\documentclass[11pt, a4paper]{article}
\usepackage{amsmath, amssymb, amsthm}
\usepackage{mathtools}
\usepackage{hyperref}
\usepackage[margin=2.5cm]{geometry}
\usepackage{graphicx}
\usepackage{mathrsfs}
\usepackage{float}
\usepackage{subcaption}
\usepackage{multirow}
\usepackage{booktabs}
\usepackage{siunitx}
\usepackage{threeparttable}
\usepackage{xcolor}
\usepackage{cleveref}
\usepackage[
  backend=biber,
  style=numeric,
  sorting=none
]{biblatex}
\numberwithin{equation}{section}
\title{Traversable Wormholes Supported by Conformally Coupled Scalar Fields in $D \geq 4$: Geometry, Geodesics, and Quasinormal Modes}
\author{Pablo Vidal\thanks{\texttt{pablo.vidal@alumnos.uach.cl}}}
\date{}
\begin{document}
\maketitle

\begin{abstract}
   In this work, we analyze a two-parameter $(\mu, b)$ asymmetric family of traversable wormhole solutions generated by a massless conformally coupled scalar field. We show that, for a suitable range of parameters, the spacetime is everywhere regular and has two asymptotically
   flat regions. We then define the wormhole throat and verify the flare-out condition, which leads to violations of the classical energy conditions in GR, thereby requiring exotic matter to hold the throat open. We further analyze the motion of photons governed by null geodesics through an effective potential. The resulting trajectories are computed numerically and visualized using embedding diagrams. Following this route, we investigate scalar perturbations governed by the massless Klein–Gordon equation and find the QNM spectrum semianalytically and numerically. Interestingly enough, we show that such perturbations are governed by only one parameter $\mu$. Therefore, in $D = 4$, we find an isospectrality between the wormhole and the BBMB/Extremal Reissner–Nordstr\"om black hole. In the spirit of the seminal work of Morris and Thorne, we derive the conditions for traversability and compute the proper time for a traveler to traverse the wormhole. Finally, we discuss possible extensions and future directions for our work.
\end{abstract}
\newpage
\section{Introduction}
\label{sec:intro}
In 1915, Albert Einstein presented the theory of general relativity to the world, describing gravity not as a force, but as a purely geometric phenomenon. Since then, it has served as the foundation for describing, with remarkable precision, a wide range of natural phenomena (for more details see \cite{Ryder:2009zz, Misner1973}), such as the perihelion precession of Mercury (which was unexplained by Newtonian gravity and remained an open question at the time), the Lense–Thirring effect, which accounts for the precession of gyroscopes in orbit around the Earth, the redshift experienced by light emitted from a source moving away from an observer, the expansion of the Universe, the existence of black holes whose first image was obtained on April 10, 2019, thanks to the Event Horizon Telescope (EHT) collaboration \cite{EventHorizonTelescope:2019dse}, gravitational waves, first detected in 2015 by the Laser Interferometer Gravitational-Wave Observatory (LIGO) \cite{Abbott2016GW150914}, and many more phenomena.

\medskip

\noindent
In this context, Einstein’s theory also predicts the existence of an object that has not yet been experimentally confirmed: the \textbf{wormhole}. A wormhole is a \textit{tunnel} connecting two distant regions of the Universe (intra-universal) or two different universes (extra-universal) \cite{Visser:1995cc}. The first exploration of the possibility of connecting two regions of spacetime dates back to 1916, when Ludwig Flamm showed that the Schwarzschild solution to the field equations admits a wormhole interpretation throught Kruskal maximally symmetric extension \cite{Flamm1916, Kruskal:1959vx} which turned out to be non traversable because the throat pinches off extremely quickly so that no signal can travel across the spacetime \cite{PhysRev.128.919}. Later, in 1935, Albert Einstein and Nathan Rosen further developed and popularized this idea in an attempt to remove the singularities present in general relativity, and such structures became known as \textbf{Einstein–Rosen bridges} \cite{PhysRev.48.73}. However, it was not until 1957 that Charles Misner and John Wheeler introduced the term wormhole in the context of geometrodynamics which essentially treats spacetime as dynamical geometry and therefore coined the popular phrase: \textit{physics is geometry} \cite{Misner1957ClassicalPA, Wheeler:1957mu}. Since then, arguably the most remarkable work dates back to 1988, when Michael Morris and Kip Thorne worked out the idea of traversable wormholes \cite{Morris:1988cz}, which involve violations of the energy conditions. In this sense, wormhole physics is particularly relevant in the context of semiclassical gravity, since such violations naturally arise in quantum field theory, with the \textbf{Casimir effect} \cite{Casimir:1948dh} as one of the most well-known examples \cite{Visser:1995cc, Kontou:2020bta}.

\medskip

\noindent
With this in mind, our work focuses on a family of traversable wormholes generated by conformally coupled scalar fields. Such systems trace back to the late 1960s and early 1970s, when conformal invariance in curved spacetime and the associated improvement of the energy–momentum tensor were systematically studied in the context of quantum field theory. In particular, the work of Callan, Coleman, and Jackiw \cite{Callan:1970ze} established the improved stress–energy tensor for scalar fields, ensuring conformal invariance in the massless case and consistency with renormalization. Furthermore, exact gravitational solutions involving conformally coupled scalars were soon constructed: the Bocharova–Bronnikov–Melnikov solution \cite{Bocharova1970} was later independently analyzed and interpreted by Bekenstein \cite{Bekenstein:1974sf}, leading to the well-known BBMB black hole, about the same time a family of wormholes was found by solving the minimally coupled system throught the JNWW metric \cite{Janis:1968zz, Wyman:1981bd, Virbhadra:1997ie} and mapping the solution throught a conformal factor \cite{Barcelo:1999hq}. More recently, a unified classification of static, spherically symmetric solutions in Einstein gravity with a conformally coupled scalar field in arbitrary spacetime dimensions has been developed, revealing a rich solution space in which black hole and wormhole branches arise as different regions of a single underlying structure \cite{Ray:2024fyx}.

\medskip

\noindent
The matter action for the massless conformally coupled scalar field is given by
\begin{equation}\label{Accion scalar field d}
    S_{\mathcal{M}} = -\frac{1}{2}\int \left[\nabla_{\mu}\phi \nabla^{\mu}\phi + \frac{(D - 2)}{4(D - 1)}R\phi^{2}\right]\sqrt{-g}\,d^{D}x.
\end{equation}
Whose stress-energy momentum tensor, along with the field equation, reads
\begin{align} 
    &T_{\mu \nu} = \nabla_{\mu}\phi\nabla_{\nu}\phi - \frac{1}{2}g_{\mu\nu}\nabla_{\lambda}\phi\nabla^{\lambda}\phi + \frac{(D - 2)}{4(D - 1)}\left[G_{\mu\nu} + \left(g_{\mu \nu}\Box - \nabla_{\mu}\nabla_{\nu}\right)\right]\phi^{2},\label{stresstensorford}\\
    &\left(\Box - \frac{(D - 2)}{4(D - 1)}R\right)\phi = 0\label{cc laplacian}.
\end{align}
Thus, the 2-parameter spherically symmetric solution claimed to represent a traversable wormhole is a subfamily of a more general class discussed in \cite{Ray:2024fyx}. 
\begin{align}
&ds^{2} = \left(1 - \frac{b \mu}{r^{D - 3}}\right)^{\frac{4}{D - 2}}\left[-\left(1 + \frac{\mu}{r^{D - 3}}\right)^{-\frac{4}{D - 2}}dt^{2} + \left(1 + \frac{\mu}{r^{D - 3}}\right)^{\frac{4}{(D - 2)(D - 3)}}\left(dr^{2} + r^2 d\Sigma_{D - 2}^2\right)\right],\label{line element RAY}\\
&\phi = \pm\frac{br^{D - 3} - \mu}{r^{D - 3} - b\mu}.\label{scalar field for d}
\end{align}
It is worth mentioning that for $b = 0$, $D > 4$, the line element corresponds to the so-called Bekenstein spaces \cite{Klimcik:1993cia}. In $D = 4$ we recover the BBMB black hole which has no higher dimensional generalization \cite{Xanthopoulos1991} and whose line element is the same as the extremal Reissner-Nordstr\"om black hole (ERN). Finally, for $D = 4$ and making use of new parameters such that $\mu \equiv \eta/2 > 0, \, b \equiv \tan(\Delta/2), \, \Delta \in (-\pi, 0)$, the solution matches the Barceló-Visser wormhole \cite{Barcelo:1999hq}.
\section{General properties of the Wormhole}

Before proceeding with the discussion, it is useful to introduce a set of definitions that will be employed throughout this work.

\medskip
\noindent
Given the line element
\begin{equation}\label{general line element}
    ds^{2} = -A(r)dt^{2} + B(r)dr^{2} + C(r)d\Sigma^{2}_{D - 2},
\end{equation}
we define $\theta^{i} \equiv (\theta^{1}, \theta^{2}, \ldots, \theta^{D-2})$ and $d\Sigma^{2}_{D-2} \equiv \gamma_{ij} d\theta^{i} d\theta^{j}$, where $\gamma_{ij}$ is the metric tensor on the $(D-2)$-sphere. 

\medskip
\noindent
The functions to be used are
\begin{equation}\label{ABC FUNCTIONS}
    \begin{aligned}
        &A  \equiv \left(1 - \frac{b\mu}{r^{D - 3}}\right)^{\frac{4}{D - 2}}\left(1 + \frac{\mu}{r^{D - 3}}\right)^{-\frac{4}{D - 2}}, \quad   B \equiv \left(1 - \frac{b\mu}{r^{D - 3}}\right)^{\frac{4}{D - 2}}\left(1 + \frac{\mu}{r^{D - 3}}\right)^{\frac{4}{(D - 2)(D - 3)}},\\
        &C \equiv r^{2}B,
    \end{aligned}
\end{equation}
we also define
\begin{equation}\label{uv functions}
    u \equiv 1 + \frac{\mu}{r^{D - 3}}, \quad v \equiv 1 - \frac{b\mu}{r^{D - 3}},
\end{equation}
thus, Eqs. \eqref{ABC FUNCTIONS} and \eqref{uv functions} are assumed to hold unless explicitly stated otherwise. 

\medskip

\noindent
Finally, it is convenient to compute
\begin{equation}
    \begin{aligned}
        &\frac{A^{\prime}}{A} = \frac{4\mu(D - 3)}{(D - 2)r^{D - 2}}\left(\frac{b}{v} + \frac{1}{u}\right), \quad  \frac{B^{\prime}}{B} = \frac{4\mu}{(D - 2)r^{D - 2}}\left(\frac{b(D - 3)}{v} - \frac{1}{u}\right),\\\\
        &\frac{d}{dr}\left(\frac{A^{\prime}}{A}\right) = -\frac{4\mu(D - 3)}{r^{D - 1}}\left(\frac{b}{v} + \frac{1}{u}\right) + \frac{4\mu^{2}(D - 3)^{2}}{(D - 2)r^{2(D - 2)}}\left(\frac{1}{u^{2}} - \frac{b^{2}}{v^{2}}\right),\\\\
        &\frac{d}{dr}\left(\frac{B^{\prime}}{B}\right) = -\frac{4\mu}{r^{D - 1}}\left(\frac{b(D - 3)}{v} - \frac{1}{u}\right) - \frac{4\mu^{2}(D - 3)}{(D - 2)r^{2(D - 2)}}\left(\frac{b^{2}(D - 3)}{v^{2}} + \frac{1}{u^{2}}\right).
    \end{aligned}
\end{equation}
\subsection{Singularities, asymptotically flat regions and throats}
We begin by examining the spacetime described by Eq.~\eqref{line element RAY}. Since a traversable wormhole must be free of horizons and curvature singularities, it is necessary to determine the range of the free parameters $\mu$ and $b$ for which the geometry remains regular. It is important to note that the choice $b = - 1$ is always forbidden because it has fundamental issues with the construction of the solution (see \cite{Ray:2024fyx} for more details). Also, the choice $\mu=0$ is immediately discarded, as we would recover Minkowski spacetime. As said in the introduction, for $D = 4$, $\mu>0$, and $b$ can take any negative value (except, again, $b = -1$), and as we shall see, this is going to be the case no matter the number of dimensions. Strictly speaking, it is useful to compute the Kretschmann invariant $\mathscr{K} = R_{\mu \nu \rho \delta}R^{\mu \nu \rho \delta}$ in order to properly justify our assumptions, because it also serves for determining the two asymptotically flat regions. Therefore, the non-vanishing Riemann components of Eq. \eqref{line element RAY} are given by
\begin{align}
    &R_{trtr} = \frac{2\mu(D - 3)A}{(D - 2)r^{D - 1}}\Bigg[-(D - 2)\left(\frac{b}{v^{2}} + \frac{1}{u^{2}}\right) + \frac{\mu}{r^{D - 3}}\left(\frac{b}{v} + \frac{1}{u}\right)^{2}\Bigg],\label{Riemann for d tr}\\
    &R_{titj} = \frac{4\mu(D - 3)A}{(D - 2)^{2}r^{D - 3}}\left(\frac{b}{v} + \frac{1}{u}\right)\Bigg[\frac{D - 2}{2} + \frac{\mu}{r^{D - 3}}\left(\frac{b(D - 3)}{v} - \frac{1}{u}\right)\Bigg]\gamma_{ij}, \label{Riemann for d tl}\\
    &R_{rirj} = \frac{2\mu(D - 3)B}{(D - 2)r^{D - 3}}\Bigg(\frac{b(D - 3)}{v^{2}} - \frac{1}{u^{2}}\Bigg)\gamma_{ij},\label{Riemann for d rl}\\
    &R_{ijkl} = -\frac{8\mu C}{(D - 2)^{2}r^{D - 3}}\left(\frac{b(D - 3)}{v} - \frac{1}{u}\right)\left(\frac{D - 3}{v} + \frac{1}{u}\right)\gamma_{i[k}\gamma_{l]j}.\label{Riemann for d kl}
\end{align}
Thus, we are left with an exact expression for $\mathscr{K}$
\begin{equation}\label{K invariant}
        \begin{aligned}
        \mathscr{K} = \frac{16\mu^{2}(D - 3)^{2}}{(D - 2)B^{2}r^{2(D - 1)}}\Bigg\{(D - 2)\left[-\left(\frac{b}{v^{2}} + \frac{1}{u^{2}}\right) + \frac{\mu}{(D - 2)r^{D - 3}}\left(\frac{b}{v} + \frac{1}{u}\right)^{2}\right]^{2}\\
    + \left(\frac{b}{v} + \frac{1}{u}\right)^{2}\left[1 + \frac{2\mu}{(D - 2)r^{D - 3}}\left(\frac{b(D - 3)}{v} - \frac{1}{u}\right)\right]^{2} + \left(\frac{b(D - 3)}{v^{2}} - \frac{1}{u^{2}}\right)^{2}\\
    +\frac{2}{(D - 2)^{2}(D - 3)}\left(\frac{b(D - 3)}{v} - \frac{1}{u}\right)^{2}\left(\frac{D - 3}{v} + \frac{1}{u}\right)^{2}
    \Bigg\}
    \end{aligned}
\end{equation}
As a consistency check, we can set $b = 0$, $D = 4$, as well as make the linear transformation $r + \mu \to r$ and recover the well-known expression for the BBMB black hole, see for example \cite{Qi_2024}
\begin{equation}
    \mathscr{K}_{\mathrm{BBMB}} = \frac{8\mu^{2}}{r^{6}}\left(6 - \frac{12\mu}{r} + \frac{7\mu^{2}}{r^{2}}\right).
\end{equation}
Now, for $\mu > 0$, $b = 0$ and $D > 4$ the Bekenstein spaces represents a naked singularity lying in $r = 0$, which was already known in \cite{Klimcik:1993cia}, in fact
\begin{equation}
    \mathscr{K}_{\mathrm{Bekenstein}} \sim \begin{dcases}&\frac{16 (D - 3)^{2}(D - 3)^{2}(D^{2} - 5D + 8)}{(D - 2)^{3}\mu^\frac{8}{(D - 2)(D - 3)}} \frac{1}{r^{\frac{4(D - 4)}{D - 2}}}, \qquad r \to 0,\\\\
     &16\mu^{2}(D - 1)(D - 3)\frac{1}{r^{2(D - 1)}}, \qquad r\to +\infty.\end{dcases}
\end{equation}
\noindent
Now, from Eq. \eqref{K invariant} we reaffirm the initial idea of restricting attention to $\mu>0$ and $b<0$, throughout the remainder of this work, because otherwise $\mathscr{K}$ will be ill-defined. Having established the admissible parameter domain, we proceed to identify the two asymptotically flat regions connected by the wormhole. In Eq. \eqref{K invariant} we are hinted that both regions lie in $r \to 0, +\infty$, in fact if we take the limit $r\to\infty$, $\mathscr{K}$ behaves as
\begin{equation}
    \mathscr{K} \sim \frac{16\mu^{2}(D - 1)(D - 3)(b^{2}(D - 3) + 1)}{r^{2(D - 1)}},
\end{equation}
meanwhile for $r\to 0$, we find
\begin{equation}
    \mathscr{K} \sim \frac{16(D - 1)(D - 3)(D - 3 + b^{2})}{\mu^{2 + \frac{8}{D - 3}}(-b)^{2 + \frac{8}{D - 2}}}r^{2(D - 1)}.
\end{equation}
So in both limits, $\mathscr{K}$ vanishes identically. On the other hand, for $r\to+\infty$ it is easy to see that the metric approaches the Minkowski line element, while Eq.~\eqref{scalar field for d} yields to $\phi\to\pm b$. Meanwhile, for $r\to0$, we have
\begin{equation}
    ds^{2} \simeq - (-b)^{\frac{4}{D - 2}}dt^{2} + (-b)^{\frac{4}{D - 2}}\mu^{\frac{4}{D - 3}}\left(\frac{dr^{2}}{r^{4}} + \frac{d\Sigma_{D - 2}^{2}}{r^{2}}\right), \quad \phi = \pm \frac{1}{b},
\end{equation}
if we perform the change of variables
\begin{equation}
    T = (-b)^{\frac{2}{D - 2}}t, \quad R = \frac{(-b)^{\frac{2}{D - 2}} \mu^{\frac{2}{D - 3}}}{r},
\end{equation}
one recovers Minkowski spacetime again
\begin{equation}
    ds^{2} \simeq -dT^{2} + dR^{2} + R^{2}d\Sigma_{D - 2}^{2}.
\end{equation}
Therefore, the coordinate $r$ is well defined for $r \in (0,+\infty)$, which physically corresponds to the two asymptotically flat regions of spacetime. Notice that the fact of the Kretschmann invariant as well as the scalar field taking different values in both regions is because the wormhole is asymmetric. This was noted by Barceló-Visser themselves, and i quote \textit{``this behavior is disappointing because we would like to have a bridge that connects two equivalent regions of spacetime''} \cite{Barcelo:1999hq}.

\medskip

\noindent
Now, the most important feature of wormholes is the existence of some minimal surface $r_T$, known as the wormhole throat, which has an analogous surface for black holes (the event horizon). In practice, the fact that some general spacetime representing a wormhole/black hole admits a throat/horizon is by no means an easy task to prove, and there has been considerable ensure the existence of such a quantity by relying on the vanishing of curvature invariants constructed from the Riemann tensor; see for example, \cite{Coley:2017woz, Abdelqader:2014vaa, McNutt:2017gjg}. However, since our wormhole is static and spherically symmetric, the problem is relatively much simpler, since we only need to compute the invariant $J$ and check whether it vanishes or not, defined in \cite{McNutt2021GeometricSurfaces}
\begin{equation}{\label{invariant SPI}}
    J \equiv 4I_{1}I_{3} - I_{5},
\end{equation}
where
\begin{equation}
    I_{1} = C_{\mu \nu \alpha \beta}C^{\mu \nu \alpha \beta}, \quad I_{3} = \nabla_{\lambda}C_{\mu \nu \alpha \beta}\nabla^{\lambda}C^{\mu \nu \alpha \beta}, \quad I_{5} = \partial_{\mu}I_{1}\partial^{\mu}I_{1},
\end{equation}
taking advantage of the fact that $R = 0$ for our system, the Weyl tensor simplifies to
\begin{equation}
    C_{\mu \nu \alpha \beta} = R_{\mu \nu \alpha \beta} - \frac{2}{D - 2}\left(R_{\mu [\alpha}g_{\beta]\nu} + R_{\nu[ \beta}g_{\alpha]\mu}\right),
\end{equation}
using Eqs. \eqref{Riemann for d tr}, \eqref{Riemann for d tl}, \eqref{Riemann for d rl} and \eqref{Riemann for d kl} we first compute the Ricci tensor, yielding
\begin{align}
    &R_{tt} = \frac{2\mu^{2}(D - 3)^{2}A}{(D - 2)Br^{2(D - 2)}}\left(\frac{b}{v} + \frac{1}{u}\right)^{2},\label{Ricci tt}\\
    &R_{rr} = \frac{2 \mu(D - 3)}{(D - 2)r^{D - 1}}\left[\frac{b(D - 2)^{2}}{v^{2}} - \frac{\mu}{r^{D - 3}}\left(\frac{b}{v} + \frac{1}{u}\right)^{2}\right],\label{Ricci rr}\\
    &R_{ij} =  \frac{2 \mu(D - 3)}{(D - 2)r^{D - 3}}\left[-\frac{b(D - 2)}{v^{2}} + \frac{\mu}{r^{D - 3}}\left(\frac{b}{v} + \frac{1}{u}\right)^{2}\right]\gamma_{ij}.\label{Ricci ij}
\end{align}
\noindent
Therefore, the non-vanishing components of the Weyl tensor are given by
\begin{align}
    &C_{t r t r} = -\frac{2\mu(D - 3) A}{u^{2}r^{D - 1}}, \quad C_{t i t j} = \frac{2\mu(D - 3) A}{(D - 2)u^{2}r^{D - 3}}\gamma_{ij},\\
    &C_{r i r j} = -\frac{2\mu(D - 3) B}{(D - 2)u^{2}r^{D - 3}}\gamma_{ij}, \quad C_{i j k l} = \frac{8\mu C}{(D - 2)u^{2}r^{D - 3}}\gamma_{i [k}  \gamma_{l]j},
\end{align}
before proceeding, we define
\begin{equation}
    \mathcal{P}(r) \equiv \frac{D - 1}{r^{D - 2}} + \frac{2\mu}{(D - 2)r^{2D - 5}}\left(\frac{2b(D - 3)}{v} - \frac{(D^{2} - 5D + 8)}{u}\right).
\end{equation}
Thus,
\begin{equation}
    I_{1} = \frac{16 \mu^{2}(D - 1)(D - 3)}{B^{2}u^{4}r^{2(D - 1)}}, \quad I_{5} = \frac{1024\mu^{4}(D - 1)^{2}(D - 3)^{2}}{B^{5}u^{8}r^{2(D + 1)}}\mathcal{P}^{2},
\end{equation}
In addition, the covariant derivative of the Weyl tensor yields
\begin{align}
    &\nabla_{r}C_{t r t r} = \frac{2\mu(D - 3)A}{u^{2}r^{2}}\mathcal{P}, \quad \nabla_{r}C_{t i t j} = -\frac{2\mu(D - 3)A}{(D - 2)u^{2}}\mathcal{P}\gamma_{ij},\\
    &\nabla_{r}C_{r i r j} = \frac{2\mu(D - 3)B}{(D - 2)u^{2}}\mathcal{P}\gamma_{ij}, \quad \nabla_{r}C_{i j k l} = -\frac{8\mu C}{(D - 2)u^{2}}\mathcal{P}\gamma_{i [k}  \gamma_{l]j},
\end{align}
hence
\begin{equation}
    I_{3} = \frac{16\mu^{2}(D - 1)(D - 3)}{B^{3}u^{4}r^{4}}\mathcal{P}^{2},
\end{equation}
which verifies the condition in Eq. \eqref{invariant SPI}, i.e., $4I_{1}I_{3} = I_{5}$. We may therefore conclude that \eqref{line element RAY} effectively admits a throat which can be computed by solving $\mathrm{Tr}(\mathcal{K}) = 0,\,$with $\,\mathcal{K}_{ab} \equiv -\partial_{n} g_{ab}/2$ and $g_{a b}$ being the $(D-2)-$dimensional metric written in Gaussian normal coordinates for a constant time slice \cite{Hochberg:1997wp}, for our metric yields
\begin{align}
    dl^{2} &= dn^{2} + g_{a b}dx^{a}dx^{b},\\
    &=dn^{2} + r^{2}B(r)d\Sigma_{D - 2}^{2}, \quad  dn \equiv \sqrt{B(r)} dr
\end{align}
therefore
\begin{align}\notag
    &\mathrm{Tr}(\mathcal{K}) = 0 \quad \Rightarrow \quad r_{T}^{2(D - 3)} + \frac{\mu(D - 4)(b + 1)}{D - 2}r_{T}^{D - 3} + b\mu^{2} = 0\notag.
\end{align}
Which admits only the positive branch of the square root, since the negative one goes beyond the range of $r$ \cite{Ray:2024fyx}
\begin{equation}\label{radial throat}
    r_{T} = \left\{\frac{\mu}{2}\left[-\frac{(D - 4)(b + 1)}{D - 2} + \sqrt{\frac{(D - 4)^{2}(b + 1)^{2}}{(D - 2)^{2}} - 4b}\right]\right\}^{\frac{1}{D - 3}},
\end{equation}
\noindent
In particular, for $D = 4$, one obtains $r_{T} = \mu\sqrt{-b}$, which agrees with \cite{Barcelo:1999hq}. Now that we have found an explicit expression for the radial throat, we finally need to check the \textit{generalized flare-out} condition, showing that in fact such a surface is indeed minimal, that is \cite{Hochberg:1997wp}
\begin{equation}\label{flare-out condition}
    \begin{aligned}
        S &\equiv \frac{\partial \,\mathrm{Tr}(\mathcal{K})}{\partial n}\Bigg|_{r = r_{T}}\Bigg.,\\
        &= \frac{1}{C}\Bigg[D - 2 - \frac{4(D - 3)}{D - 2}\left(\frac{1}{v} - \frac{1}{u}\right)^{2} + 2(D - 1)\left((D - 3)\left(\frac{1}{v^{2}} - \frac{1}{v}\right) + \frac{1}{u^{2}} - \frac{1}{u}\right)\Bigg]\Bigg.\Bigg|_{r = r_{T}}
    \end{aligned}
\end{equation}
Now, the key is to rewrite the $u$ function using the throat condition, yielding,
\begin{equation}\label{uthroat}
    \frac{1}{u(r_{T})} = \frac{D - 2}{2} - \frac{D - 3}{v(r_{T})},
\end{equation}
so that $S$ can be written as
\begin{equation}
    S(r_{T}) = \frac{(D - 2)(D - 3)}{2C(r_{T})}\left[4(D - 3)\left(\frac{1}{v(r_{T})} - \frac{1}{2}\right)^{2} - 1\right]
\end{equation}
Therefore, we are left to prove that the inside of the square bracket in $S$ is negative everywhere. For this task, recall that $r \in (0, +\infty)$ and $r_{T} > 0$; therefore, $1/u \in (0, 1)$, thus Eq. \eqref{uthroat} is sufficient to show
\begin{equation}
     4(D - 3)\left(\frac{1}{v(r_{T})} - \frac{1}{2}\right)^{2} < \frac{1}{D - 3} \leq 1,
\end{equation}
then
\begin{equation}
    4(D - 3)\left(\frac{1}{v(r_{T})} - \frac{1}{2}\right)^{2} - 1 < 0.
\end{equation}
This is proof that $S(r_{T}) < 0$ everywhere, $\forall \mu > 0, b < 0, D \geq 4$, and we can safely conclude that $r_{T} > 0$ corresponds indeed to the minimal surface representing the throat of the wormhole. 

\medskip

\noindent
One final comment before moving on, if we set $b = 0, D = 4$ as well as shifting $(r + \mu \to r)$, because of Eq. \eqref{invariant SPI} we then have proven explicitly that the BBMB/ERN black hole admits an event horizon located in Eq. \eqref{radial throat} (this time, $r_{H} = \mu$) and therefore it is easy to see that $S(r_{H}) = 0$.

\subsection{The need for exotic matter}
\noindent
To finish up this section, we need to note that if this wormhole exists in nature, it would violate the so-called energy conditions, which generalize the intuition that an inertial observer measures positive mass-energy densities. In the literature, the most commonly used (pointwise) energy conditions are the (WEC), (NEC), (SEC), and (DEC), which are extensively reviewed in \cite{Visser:1995cc, Kontou:2020bta}. In Einstein's gravity, the presence of negative energy densities and pressures in the vicinity of the throat is an intrinsic feature of wormholes \cite{Morris:1988cz, Visser:2003yf}, in particular, for the matter contribution in Eq. \eqref{Accion scalar field d}, it is well-known that such scalar field would violate in $D = 4$, the (NEC) \cite{Barcelo:2000zf}. Any material exhibiting this property is referred to as \textit{exotic matter}. At first sight, such objects may appear unphysical, exotic matter is nevertheless consistent with the framework of semi-classical gravity. A well-known example of this phenomenon is the Casimir effect \cite{Casimir:1948dh}, which predicts the existence of microscopic negative energy densities and therefore provides a physical setting in which very small chunks of exotic matter can, in principle, arise. Also, it is worth noting that this is not quite the full story; in fact, one can avoid the use of exotic matter by modifying the underlying theory of gravity; see, for instance, \cite{Churilova:2021tgn,Bronnikov:2015pha,Rosa:2022osy,Moraes:2024mxk, Shaikh:2018yku}.
\medskip

\noindent
In practice, it is useful to define an orthonormal reference frame in which our measurements become meaningful within an inertial observer (see any standard textbook \cite{Misner1973}) such that
\begin{equation}
    g_{\hat{\alpha} \hat{\beta}} = \eta_{\mu\nu}\,e^{\mu}_{\,\,\hat{\alpha}}e^{\nu}_{\,\,\hat{\beta}}.    
\end{equation}
Where $e^{\mu}_{\,\,\hat{\alpha}}$ is known as the \textit{vielbein} and for our spacetime, can be written as
\begin{equation}\label{orthonormal reference frame}
    e^{t}_{\,\,\hat{t}} = \frac{1}{\sqrt{A}},\quad e^{r}_{\,\,\hat{r}} = \frac{1}{\sqrt{B}}, \quad e^{j}_{\,\,\hat{j}} = \frac{1}{\sqrt{C}}\tilde{e}^{j}_{\,\,\hat{j}}, \quad \gamma_{i j} = \delta_{\hat{k}\hat{l}}\,\tilde{e}^{\hat{k}}_{\,\,i}\tilde{e}^{\hat{l}}_{\,\,j}.
\end{equation}
Thus, if we want to meassure some quantity $Q_{\alpha_{1} \dots \mu_{p}}^{\qquad\beta_{1} \dots \nu_{q}}$ in the observers rest frame, that would look like
\begin{equation}
    Q_{\hat{\mu}_{1}\dots\hat{\mu}_{p}}^{\qquad\hat{\nu}_{1}\dots\hat{\nu}_{q}} = Q_{\alpha_{1} \dots \alpha_{p}}^{\qquad\beta_{1} \dots \beta_{q}}\, e^{\alpha_{1}}_{\,\,\hat{\mu}_{1}}\dots e^{\alpha_{p}}_{\,\,\hat{\mu}_{p}}e^{\hat{\nu}_{1}}_{\,\,\beta_{1}}\dots e^{\hat{\nu}_{q}}_{\,\,\beta_{q}}
\end{equation}
\noindent
Now we can compute the components of $T_{\mu\nu}$ explicitly or, even better, use Einstein's field equations, since we already computed the Ricci tensor in Eqs. \eqref{Ricci tt}, \eqref{Ricci rr}, \eqref{Ricci ij}, and  make use of our orthonormal reference frame previously defined in Eq.~\eqref{orthonormal reference frame}, so every component of the stress tensor becomes the energy density and principal pressures, that is
\begin{equation}
    T_{\hat{t}\hat{t}} \equiv \mathcal{E} , \quad T_{\hat{r}\hat{r}} \equiv P_{r} , \quad T_{\hat{j}\hat{k}} \equiv P_{\theta}\gamma_{jk},
\end{equation}
therefore
\begin{align}
    &\mathcal{E} = \frac{2\mu^{2}(D - 3)^{2}}{(D - 2)Br^{2(D - 2)}}\left(\frac{b}{v} + \frac{1}{u}\right)^{2} > 0,\label{T tt}\\
    &P_{r} = \frac{2 \mu(D - 3)}{(D - 2)Br^{D - 1}}\left[\frac{b(D - 2)^{2}}{v^{2}} - \frac{\mu}{r^{D - 3}}\left(\frac{b}{v} + \frac{1}{u}\right)^{2}\right] < 0,\label{T rr}\\
    &P_{\theta} = \frac{2 \mu(D - 3)}{(D - 2)Br^{D - 1}}\left[-\frac{b(D - 2)}{v^{2}} + \frac{\mu}{r^{D - 3}}\left(\frac{b}{v} + \frac{1}{u}\right)^{2}\right] > 0.\label{T ij}
\end{align}
Notice that it is easy to see that the trace of the energy-momentum tensor vanishes identically,
i.e,
\begin{equation}\label{traceless check}
    T = -\mathcal{E} + P_{r} + \sum_{\theta}P_{\theta} = 0,
\end{equation}
meanwhile, the sum of all terms is always positive
\begin{equation}
    \mathcal{E} + P_{r} + \sum_{\theta}P_{\theta} = 2\mathcal{E} > 0.
\end{equation}
Now, we can evaluate the NEC explicitly at the throat, yielding
\begin{equation}
    \left(\mathcal{E} + P_{r}\right)_{r = r_{T}} = \frac{2(D - 3)}{C}\left[(D - 2)\left(\frac{1}{v^{2}} - \frac{1}{v}\right) + \frac{(D - 4)}{(D - 2)}\left(\frac{1}{v} - \frac{1}{u}\right)^{2}\right],
\end{equation}
by making use of Eq. \eqref{uthroat}, the NEC becomes exactly the flare-out condition found earlier 
\begin{equation}
\left(\mathcal{E} + P_{r}\right)_{r=r_{T}} = \frac{(D - 2)(D - 3)}{2C(r_{T})}\left[4(D - 3)\left(\frac{1}{v(r_{T})} - \frac{1}{2}\right)^{2} - 1\right] = S(r_{T}) < 0.
\end{equation}
 Therefore, the fact that the wormhole admits a throat is exactly why it has to be filled with exotic matter, thus violating the energy conditions of GR and the culprit is none other but the flare-out condition itself.

\subsection{Embedding diagrams}
Having established that the geometry represents a wormhole for an appropriate choice of parameters, we now turn to constructing embedding diagrams, which provide a useful tool for visualizing the spacetime geometry and will aid the discussion of particle trajectories in the next section. Exploiting spherical symmetry, we may restrict ourselves to a single angular sector. We then embed the spatial geometry associated with Eq. \eqref{line element RAY} into three-dimensional Euclidean space, yielding
\begin{align}
    ds^{2} &= B(r)\left(dr^{2} + r^2 d\varphi^{2}\right),\label{ds RAY reduced}\\
    &= d\rho^{2} + \rho^{2}d\varphi^{2} + dz^{2},\notag\\
    &= \left[\left(\frac{d \rho}{dr}\right)^{2} + \left(\frac{dz}{dr}\right)^{2}\right]dr^{2} + \rho^{2}d\varphi^{2}.\label{rho, phi}
\end{align}
Matching \eqref{ds RAY reduced} and \eqref{rho, phi}, it follows that
\begin{align}
    &\rho(r) = r \sqrt{B(r)},\label{funcion rho} \\
    & \frac{dz(r)}{dr} = \pm\sqrt{\left(\frac{\rho}{r}\right)^{2} - \left(\frac{d \rho}{dr}\right)^{2}}.
\end{align}
Therefore
\begin{equation}\label{z shape}
z(r)=
\begin{cases}
\displaystyle
+\frac{2}{D - 2}\int\limits_{r_{T}}^{\,r}
\sqrt{
-B(\tilde{r}) \mathcal{P}(\tilde{r})\left(D - 2 + \mathcal{P}(\tilde{r})\right)
}\, d\tilde r,
& r > r_{T}, \\[1.2em]
\displaystyle
-\frac{2}{D - 2}\int\limits_{r_{T}}^{\,r}
\sqrt{
-B(\tilde{r}) \mathcal{P}(\tilde{r})\left(D - 2 + \mathcal{P}(\tilde{r})\right)
}\, d\tilde r,
& 0 < r \le r_{T},
\end{cases}
\end{equation}
where, 
\begin{equation}
    \mathcal{P}(r) \equiv \frac{\mu}{r^{D - 3}}\left(\frac{b(D - 3)}{v} - \frac{1}{u}\right).
\end{equation}
Since it is clear that $\mathcal{P} < 0$, one might be worried about the sign of the factor inside the parentheses of the square root, but if we rearrange this expression differently, we get 
\begin{equation}
    \begin{aligned}
    D - 2 + \mathcal{P} &= \frac{1}{u r^{D - 3}}\left((D - 2)r^{D - 3} + \mu(D - 3)\right) + \frac{b\mu (D - 3)}{vr^{D - 3}},\\
    &= \frac{(D - 2)r^{D - 3} + \mu(D - 3) - b\mu}{uvr^{D - 3}} > 0.
    \end{aligned}
\end{equation}
\begin{figure}[H]
    \centering
    \label{fig:wormhole_pairs}
     \caption{Embedding diagrams of the surface $S(\rho(r), \varphi) = (\rho(r) \cos(\varphi), \rho(r)\sin(\varphi), z(r))$, $\varphi \in [0, 2\pi]$, $r \in (0, +\infty)$, where the throat was highlighted in red for $b = -3$, $\mu = 1$.}
    \begin{subfigure}[H]{0.24\textwidth}
        \includegraphics[width=\linewidth]{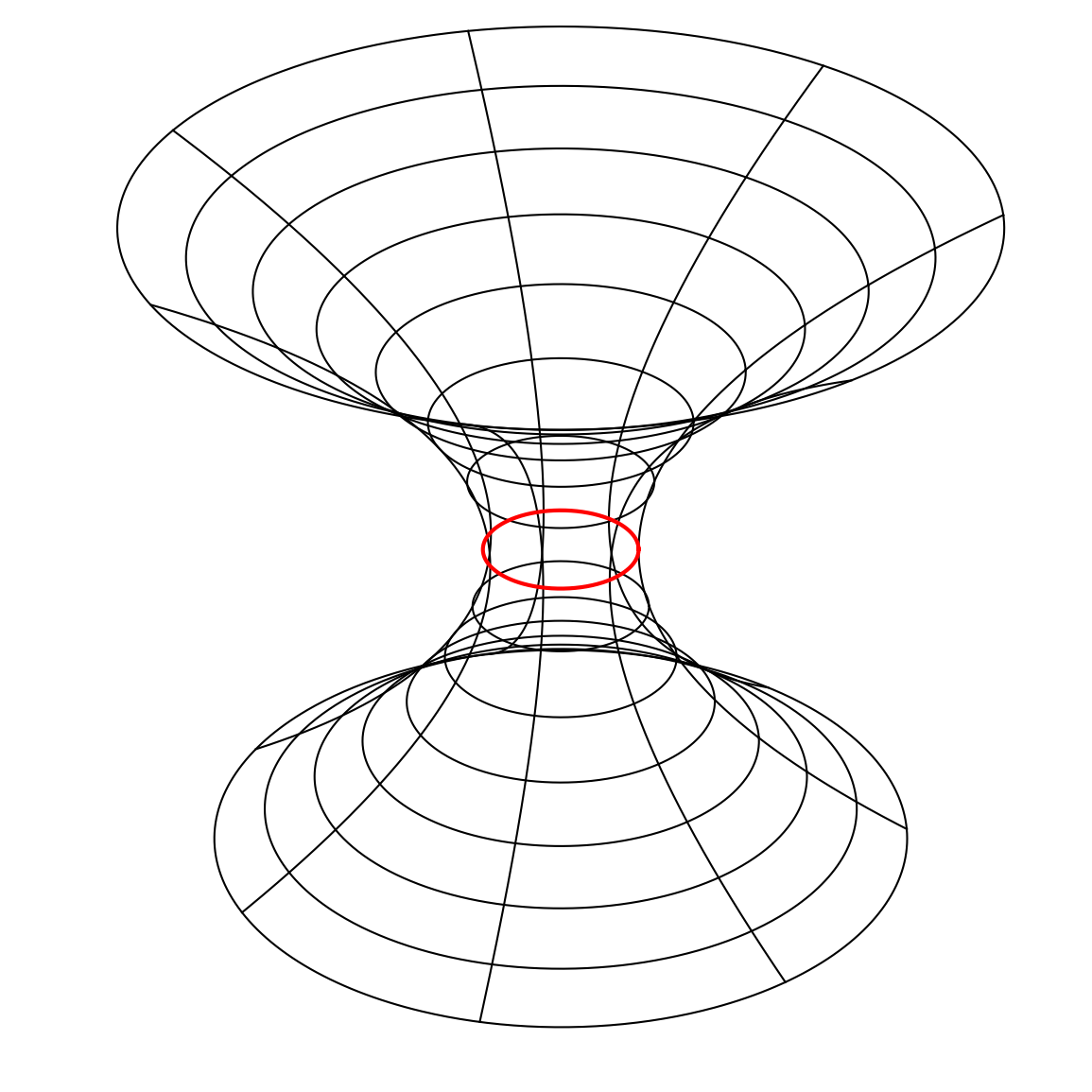}
        \caption{$D = 4$}
    \end{subfigure}
    \begin{subfigure}[H]{0.24\textwidth}
        \includegraphics[width=\linewidth]{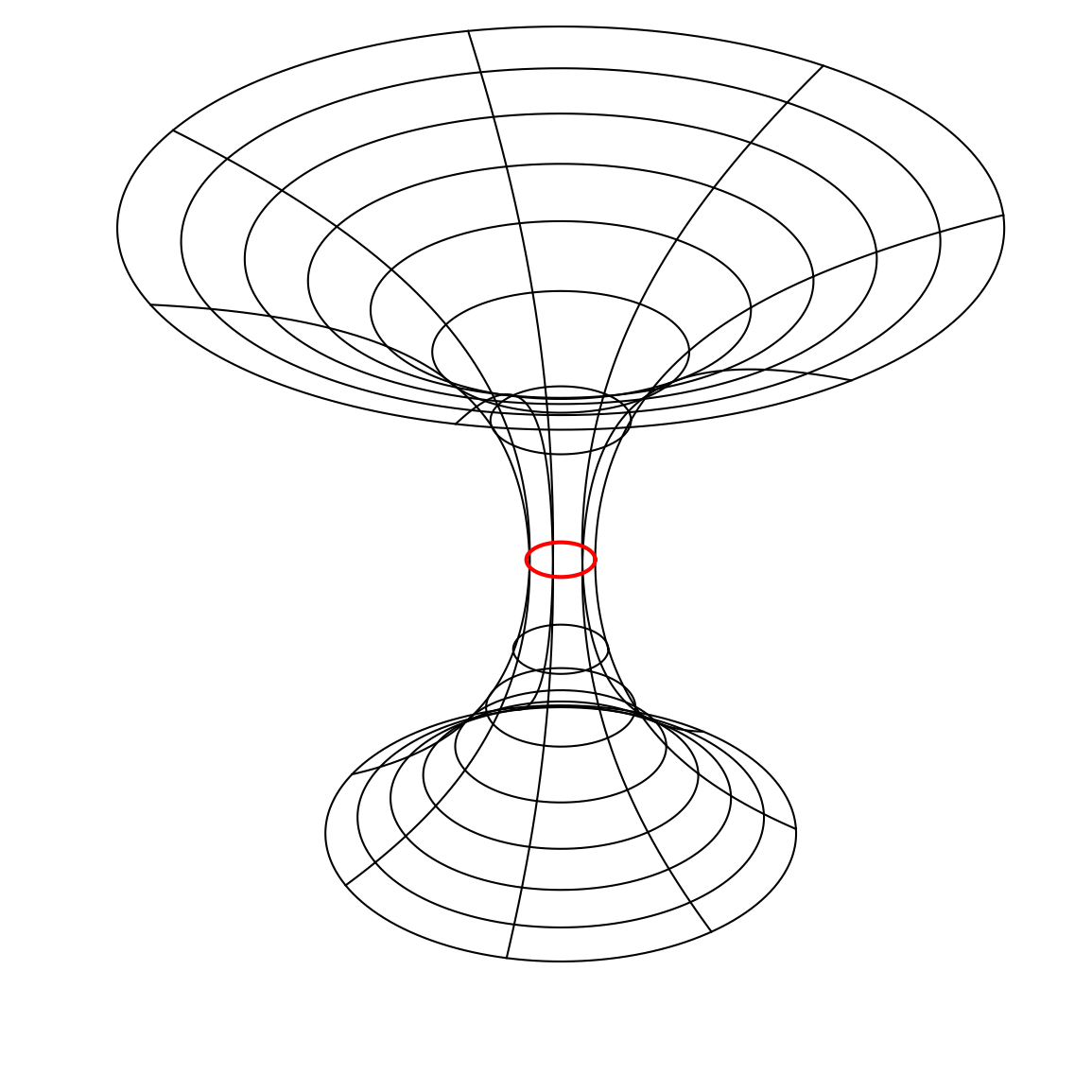}
        \caption{$D = 5$}
    \end{subfigure}
    \begin{subfigure}[H]{0.24\textwidth}
        \includegraphics[width=\linewidth]{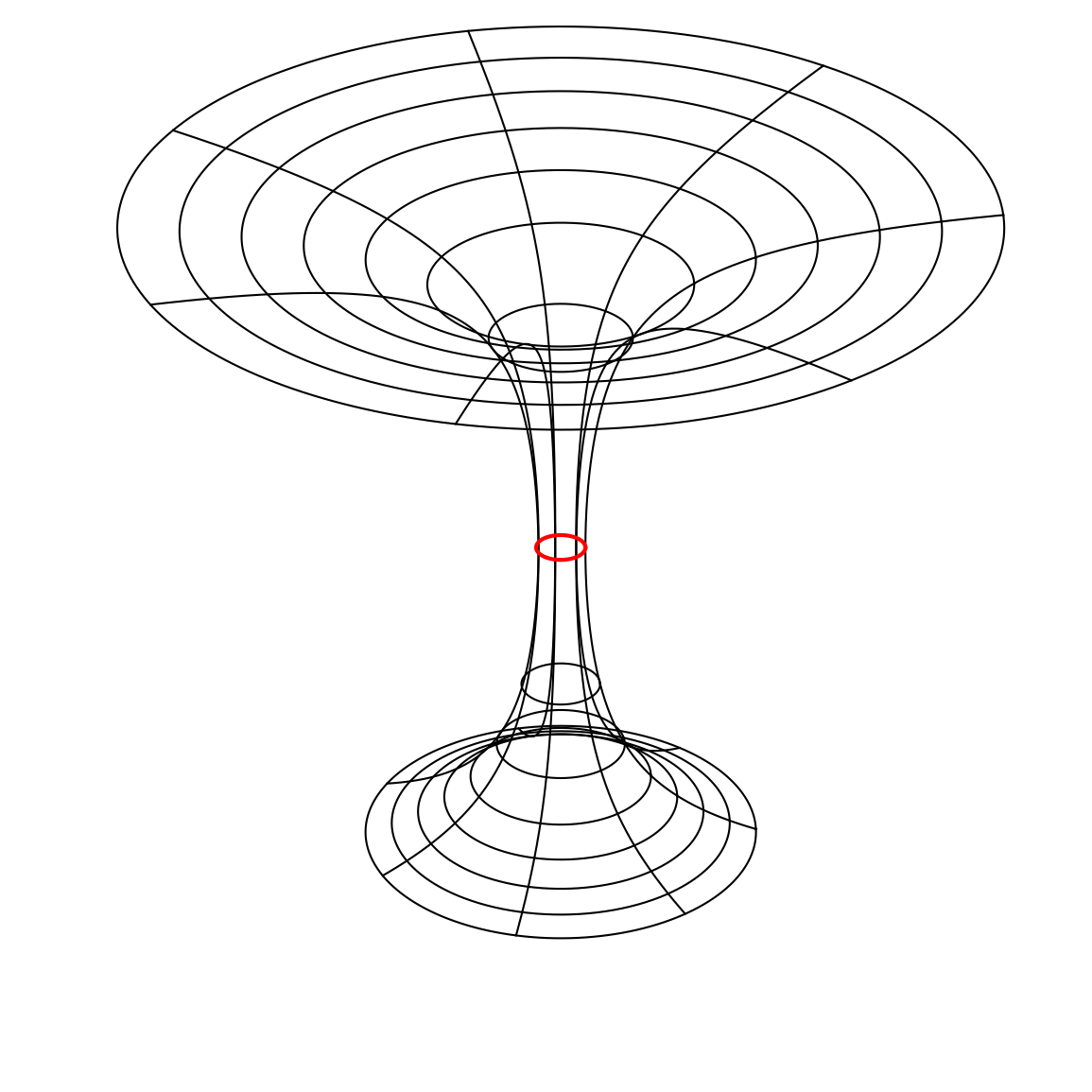}
        \caption{$D = 6$}
    \end{subfigure}
    \begin{subfigure}[H]{0.24\textwidth}
        \includegraphics[width=\linewidth]{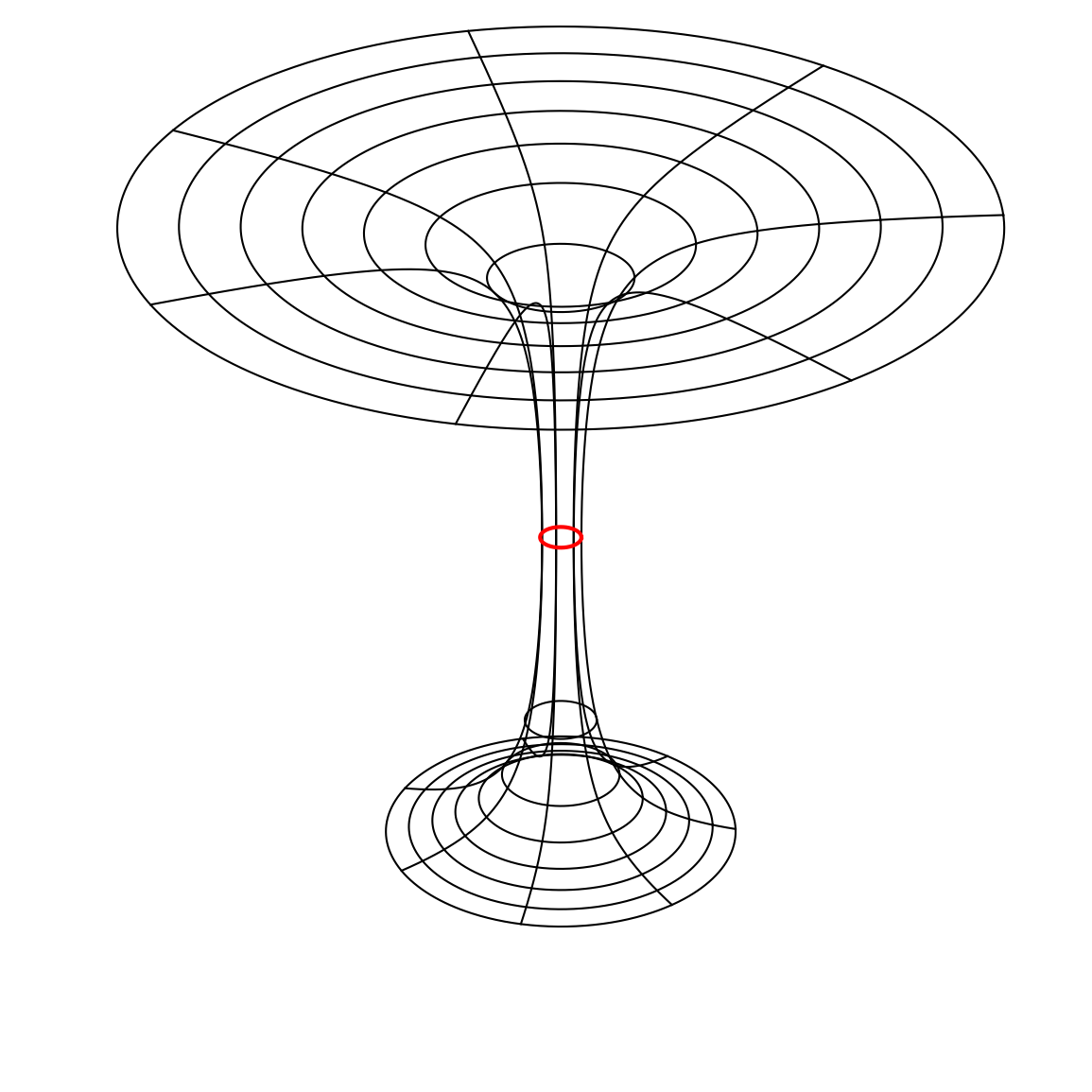}
        \caption{$D = 7$}
    \end{subfigure}
\end{figure}
\section{Equations of motion for test particles}
\noindent
In this section, we now turn our attention to one of the most important features of traversable wormholes: the geodesic motion. In the literature, there is a lot of work related in solving analyticaly/numericaly the geodesics Eqs. for traversable wormholes because they hold the key into understanding very important phenomena such as propagation of light in weak/strong gravitational lensing, multi-photons rings, shadows, quasinormal modes, accretion disks and so on, see for example \cite{Clement:2015aka, Potashov:2020zmg, Mishra:2017yrh, Huang:2023yqd, Huang:2024bbs, Errehymy:2025vvs, Macedo:2025ipc, Deligianni:2021ecz}. In particular, for the Barceló-Visser wormhole in $D=4$, the classification of trajectories has been analyzed in detail by using the Weierstrass elliptic functions \cite{Willenborg:2018zsv}. As in the previous section, we consider only one angle contribution,
\begin{equation}\label{ecuatorial line element}
    ds^{2} = -A(r)dt^{2} + B(r)(dr^{2} + r^{2}d\varphi^{2}).
\end{equation}
Because our reduced system has both translational and angular invariance, there are guaranteed to be two associated conserved quantities which can be interpreted as the angular momentum and the energy of the particle, 
\begin{equation}\label{conserved quantities}
    \dot{t} = \frac{E}{A(r)}, \quad 
    \dot{\varphi} = \frac{L}{C(r)}.
\end{equation}
In addition to these two equations, one must determine the radial geodesic, which will help distinguish between the different types of trajectories under consideration, namely $\zeta = -1$ and $\zeta = 0$, corresponding to timelike and null curves, respectively.
\begin{equation}\label{constarin what-like}
    \zeta = -A(r)\dot{t}^{2} + B(r)(\dot{r}^{2} + r^{2}\dot{\varphi}^{2}).
\end{equation}
furthermore, we can eliminate $\dot{t}, \,\dot{\varphi}$ by using Eq. \eqref{conserved quantities}, yielding
\begin{equation}
    \begin{aligned}
    &\dot{r}^{2} = \frac{1}{B}\Bigg(\zeta + \frac{E^{2}}{A} - \frac{L^{2}}{C}\Bigg).\label{r punto eqn}    
    \end{aligned}
\end{equation}
And, finally we get two differential equations associated with $t, \, \varphi$, respectively
\begin{align}
    & \left(\frac{dr}{dt}\right)^{2} = u^{-\frac{4}{D - 3}}\Bigg[\frac{1}{E^{2}}\left(A(r) \zeta - u^{-\frac{4}{D - 3}}\frac{L^{2}}{r^{2}}\right) + 1\Bigg]\label{drdt motion},\\
    & \left(\frac{dr}{d\varphi}\right)^{2} = r^{2}\left[\frac{r^{2}}{L^{2}}\left(B(r) \zeta + E^{2}u^{\frac{4}{D - 3}}\right) - 1\right].\label{drdphi motion}
\end{align}
The general strategy for classifying the different types of trajectories should be requiring that Eqs.~\eqref{drdt motion}, \eqref{drdphi motion} be mathematically well defined, i.e,
\begin{equation}
    \left(\frac{dr}{dt}\right)^2 \ge 0, \quad \left(\frac{dr}{d\varphi}\right)^2 \ge 0.    
\end{equation} 
From either equation, we can define an effective potential containing all the information stored in the different meanings of its critical points, that is, a minimum/maximum for stable/unstable orbits, respectively
\begin{equation}\label{efective potential}
    E^{2} \geq V_{\text{eff}}(r) \equiv \frac{L^{2}}{r^{2}}u^{-\frac{4}{D - 3}} - A(r) \zeta,
\end{equation}
it is worth noticing that the behavior of the effective potential in the two asymptotic regions is given by
\begin{equation}
    V_{\text{eff}}(r \to 0) = -\zeta(-b)^{\frac{4}{D - 2}}, \quad V_{\text{eff}}(r \to +\infty) = -\zeta.
\end{equation}

\subsection{Classification of orbits}
\begin{figure}[H]
    \centering
    \caption{Numerical solution (FT, UO, SU, respectively) of Eq. \eqref{drdphi motion}, for $\mu = 1, L = 7, D = 11$.}

    \begin{subfigure}[b]{0.32\textwidth}
        \centering
        \includegraphics[width=\textwidth]{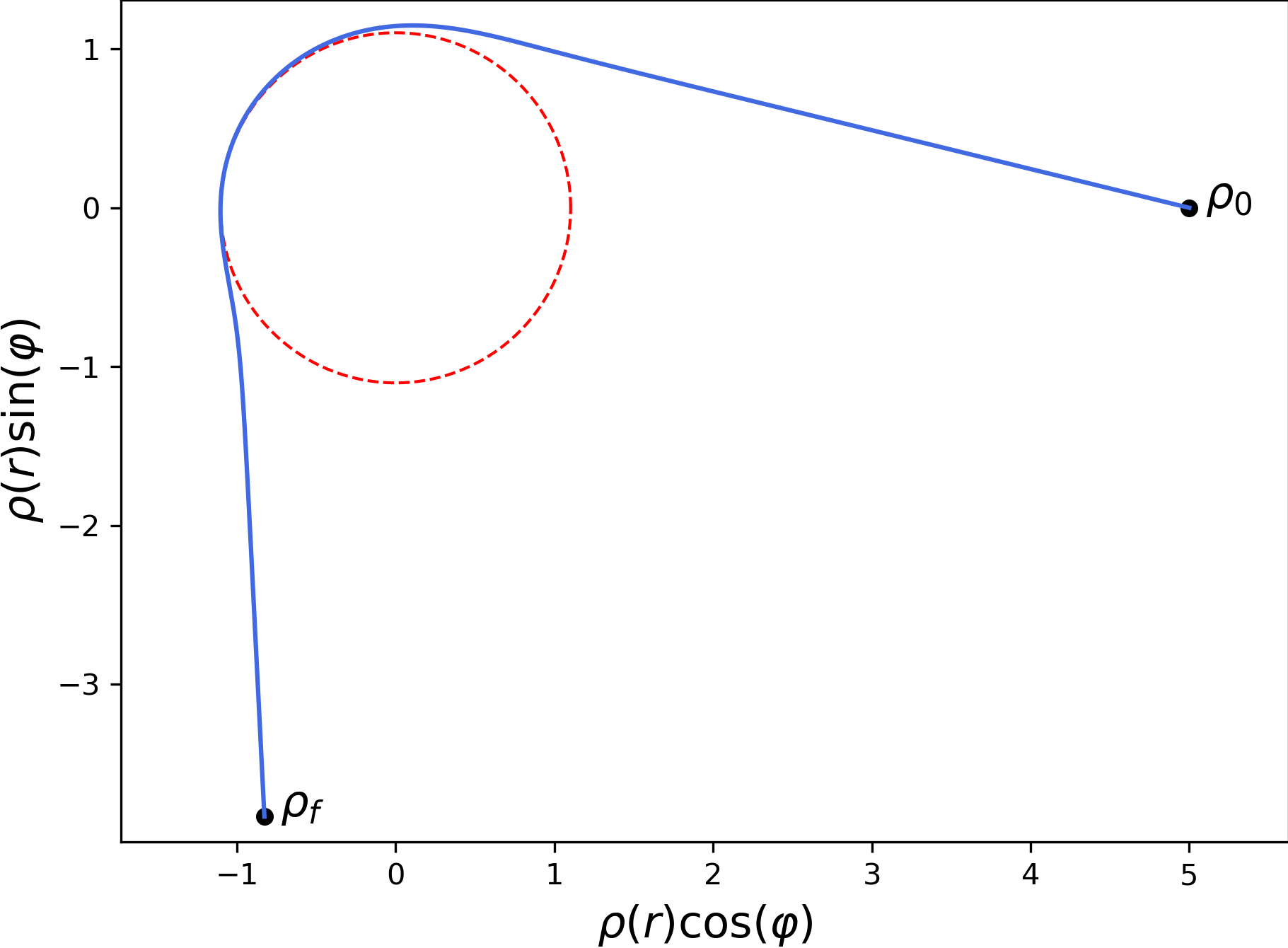}
        \caption{$E = 5\text{.}9, \, b = -0\text{.}5$}
        \label{fig:2Dnull}
    \end{subfigure}
    \begin{subfigure}[b]{0.32\textwidth}
        \centering
        \includegraphics[width=\textwidth]{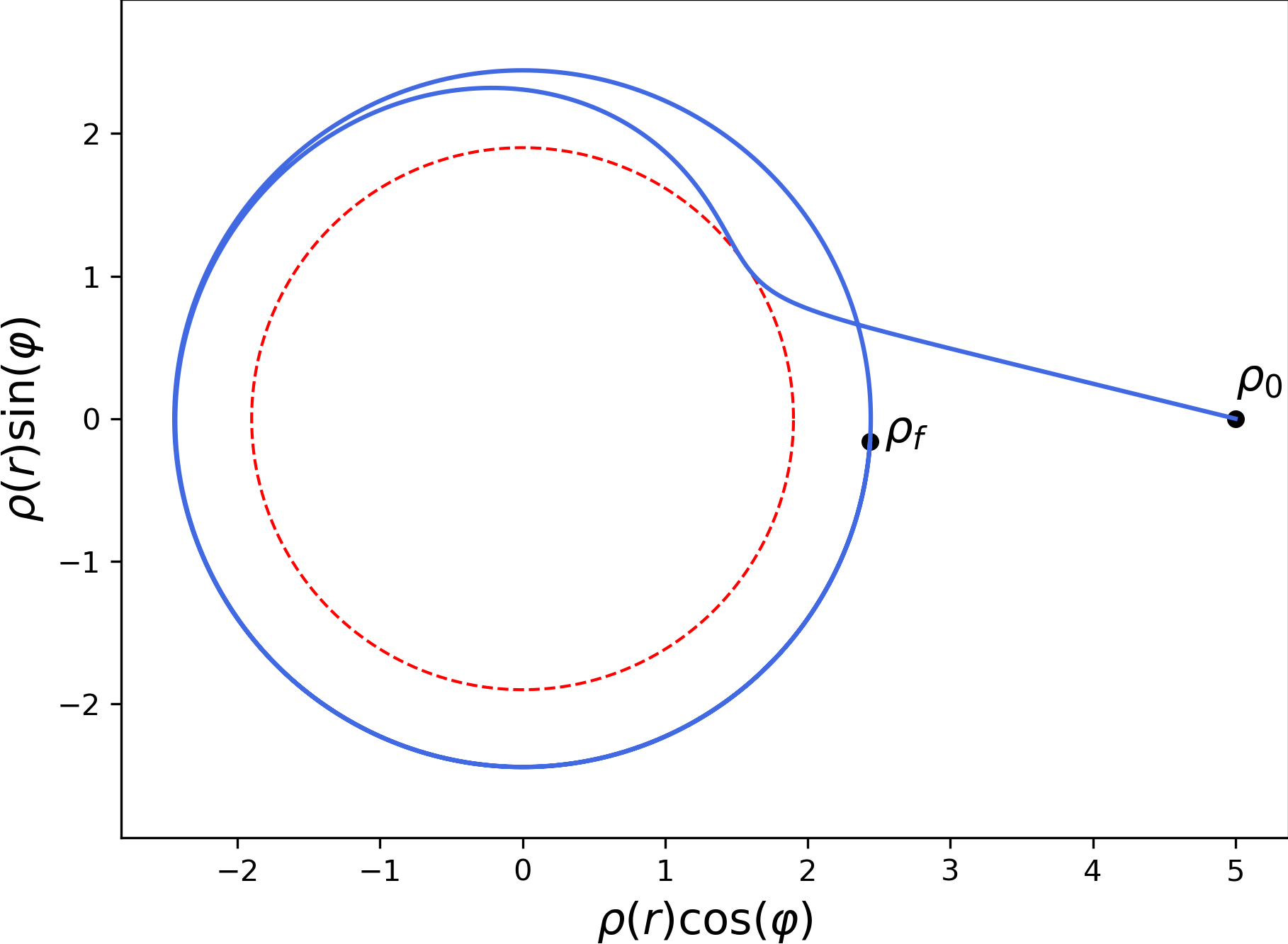}
        \caption{$E^{2} = \text{max}(V_{\text{eff}}), \, b = -50$}
    \label{fig:2Dcritico}
    \end{subfigure}
    \begin{subfigure}[b]{0.32\textwidth}
        \centering
        \includegraphics[width=\textwidth]{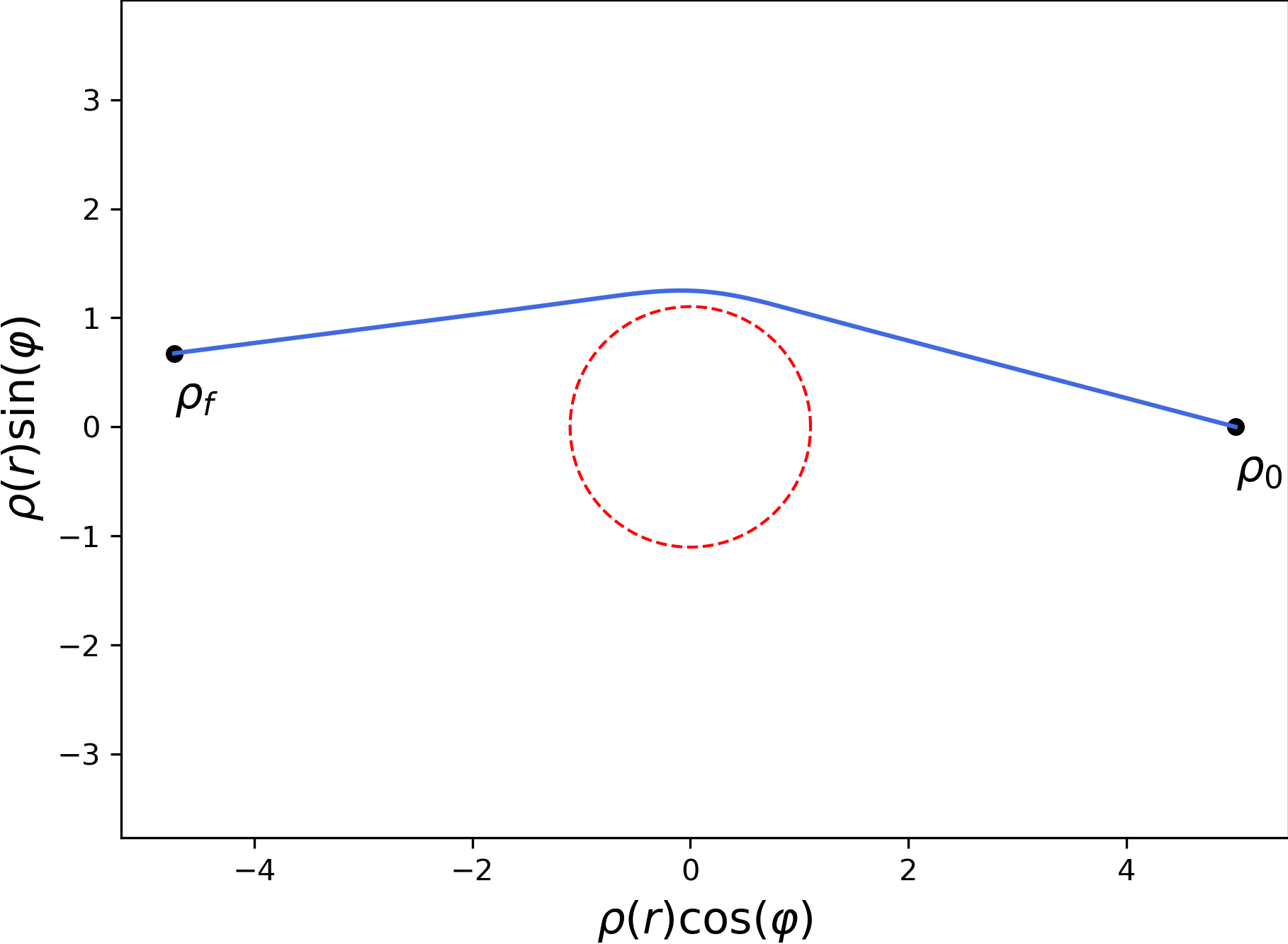}
        \caption{$E = 5\text{.}5, \, b = -0\text{.}5$}
    \label{fig:2Dsqrt}
    \end{subfigure}
    \label{fig:three_trajectories}
\end{figure}

\noindent
Once the orbital motion in Eq.~\eqref{drdphi motion} and the condition in Eq.~\eqref{efective potential} have been established, we are in a position to classify the different types of trajectories followed by particles in the spacetime described by Eq.~\eqref{line element RAY}. However, the systematic analytical determination of the extrema of $V_{\text{eff}}$ is highly non-trivial, since in practice the condition $V_{\text{eff}}^{\prime}=0$, evaluated in a critical radius $r = r_{c}$ leads to
\begin{equation}\label{dVdr}
(\mu - r_{c}^{D - 3})(r_{c}^{D - 3} - b\mu)^{(D - 6)/(D - 2)} = \frac{2\mu(b + 1)(D - 3)\zeta}{(D - 2)L^{2}} r_{c}^{D - 5}(r_{c}^{D - 3} + \mu)^{4/(D - 2)(D - 3)},
\end{equation}
which does not admit a closed-form analytical solution for general $\zeta \, \mathrm{and} \,D$. To this end, it is necessary to define the three main classes of trajectories to be considered, and they are to be displayed in the figure \ref{fig:three_trajectories}.
\begin{enumerate}
\item \textit{Fully traversable}(FT): The condition $E^{2} > V_{\text{eff}}(r)$ holds for all $r$, allowing a particle from one asymptotic region to traverse the throat and reach the other asymptotic region.

\item \textit{Stable/unstable orbits} (SO/UO): The condition $E^{2} = V_{\text{eff}}(r_c)$ is satisfied. Then, the particle moves in a circular orbit around the wormhole at $r=r_c$, which may lie in either of the two asymptotic regions.

\item \textit{Single-universe} (SU): The particle does not have the required energy, that is $E^{2} = V_{\text{eff}}(r)$. It reaches a turning point within the wormhole geometry and returns to the asymptotic region from which it originated.

\end{enumerate}
In addition, it is worth mentioning that in all figures of this section, a red dashed line will be used to denote the wormhole throat given in Eq.~\eqref{radial throat}, while the points $\rho_0$ and $\rho_f$ correspond to the initial and final positions of the particle, respectively, where $\rho(r)$ is defined in Eq.~\eqref{funcion rho}.

\subsubsection{Photon trajectories \texorpdfstring{$(\zeta = 0)$}{(zeta = 0)}}
\noindent
We now turn our attention to the trajectories followed by light, since Eqs. \eqref{efective potential}, \eqref{dVdr} are sufficiently simple to allow full analytical analysis. We begin by noticing that because of conformal invariance of the null geodesics, the $b$ parameter will not enter directly into the equations. However, it will play a role when visualizing the motion using embedding diagrams because of the $B(r)$ function, as in Eq. \eqref{funcion rho}. Now, solving Eq. \eqref{dVdr} and subsequently taking the second derivative of the effective potential, yields
\begin{align}
    &r_{c} = \mu^{\frac{1}{D - 3}},\\ 
    &V_{\text{eff}}^{\prime\prime}(r_{c}) = -\frac{L^{2}(D - 3)}{(2\mu)^{\frac{4}{D - 3}}} < 0.
\end{align}

\noindent
Thus, encountering a single critical point, corresponding to an unstable orbit (UO) commonly known as the photon sphere or Einstein ring, therefore Eq. \eqref{drdphi motion} is going to be well defined if and only if
\begin{equation}\label{real trayectory}
E^{2} \geq \frac{L^{2}}{(4\mu)^{\frac{2}{D - 3}}}.
\end{equation}
Notice that, provided $E^{2} > V_{\text{eff}}$ everywhere along the trajectory, the photon crosses the wormhole throat without being reflected and reaches the other asymptotic region, as shown in the figure below.
\begin{figure}[H]
    \centering
    \caption{FT trajectories followed by light rays in the wormhole geometry for $\mu = 3$, $b = -0.7$, and $L = 7$.
}
    \begin{minipage}{0.32\textwidth}
        \centering
        \includegraphics[width=\linewidth]{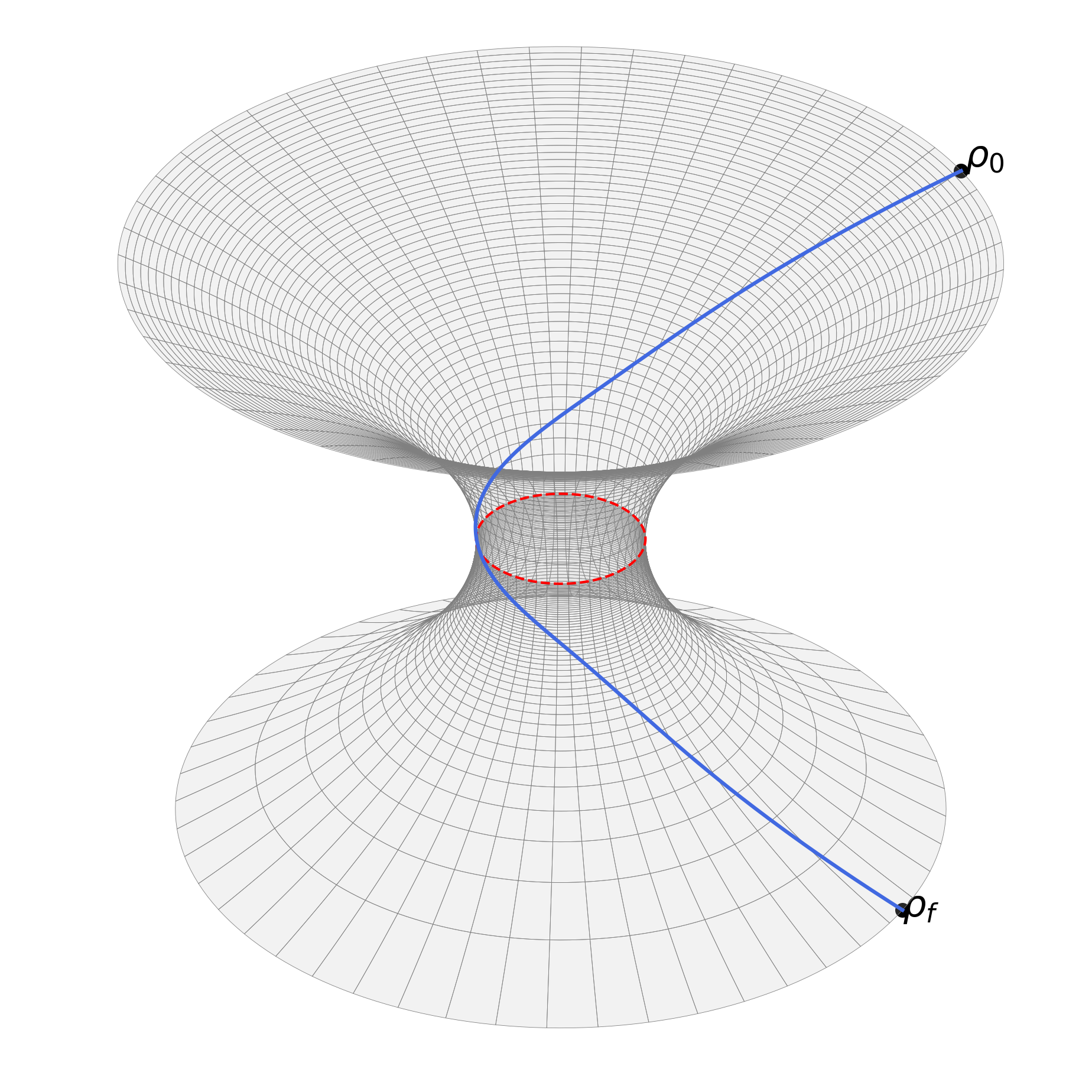}
        \caption*{(a) $D=4, E = \sqrt{0\text{.}4}$}
    \end{minipage}
    \hfill
    \begin{minipage}{0.32\textwidth}
        \centering
        \includegraphics[width=\linewidth]{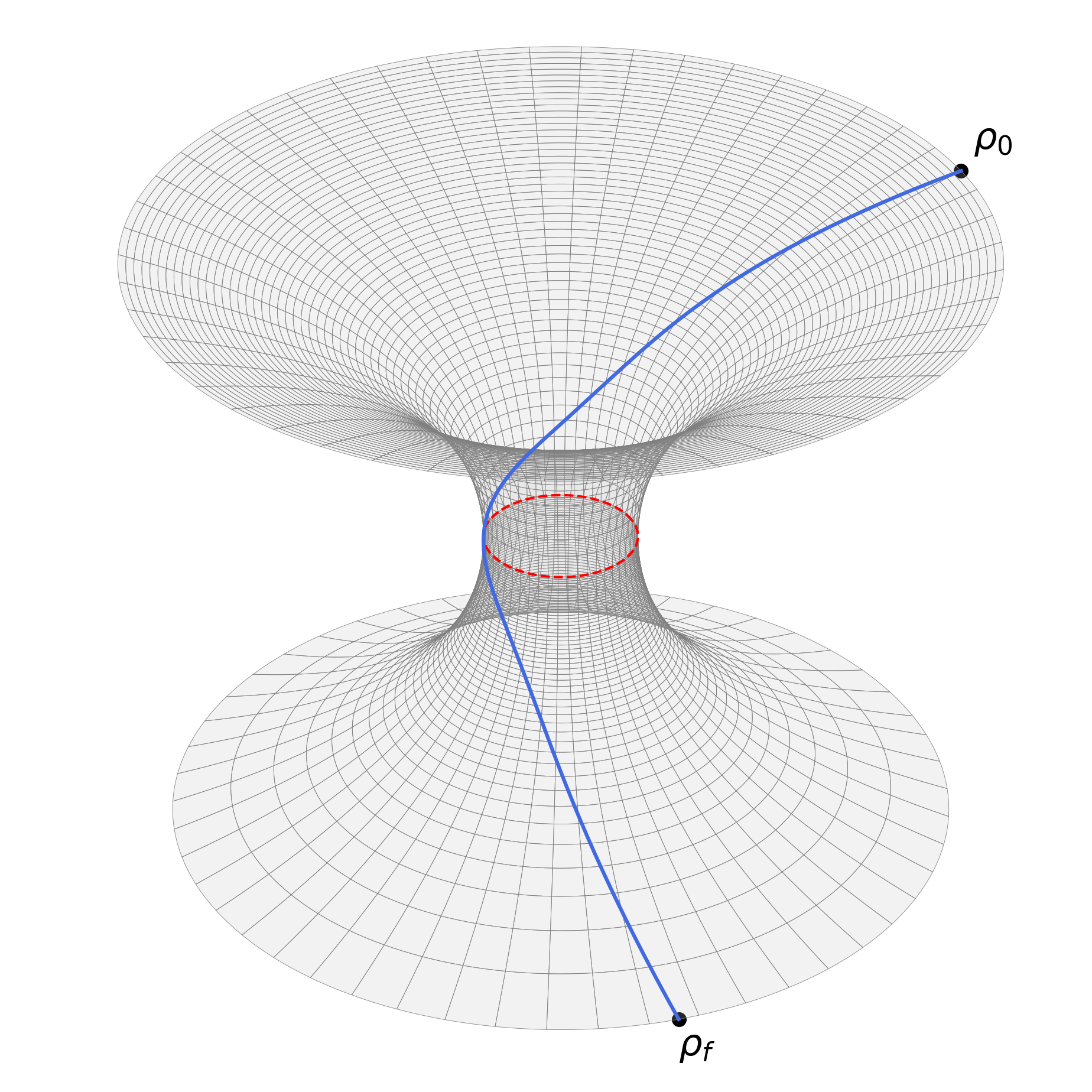}
        \caption*{(b) $D=5, E = \sqrt{4\text{.}5}$}
    \end{minipage}
    \hfill
    \begin{minipage}{0.32\textwidth}
        \centering
        \includegraphics[width=\linewidth]{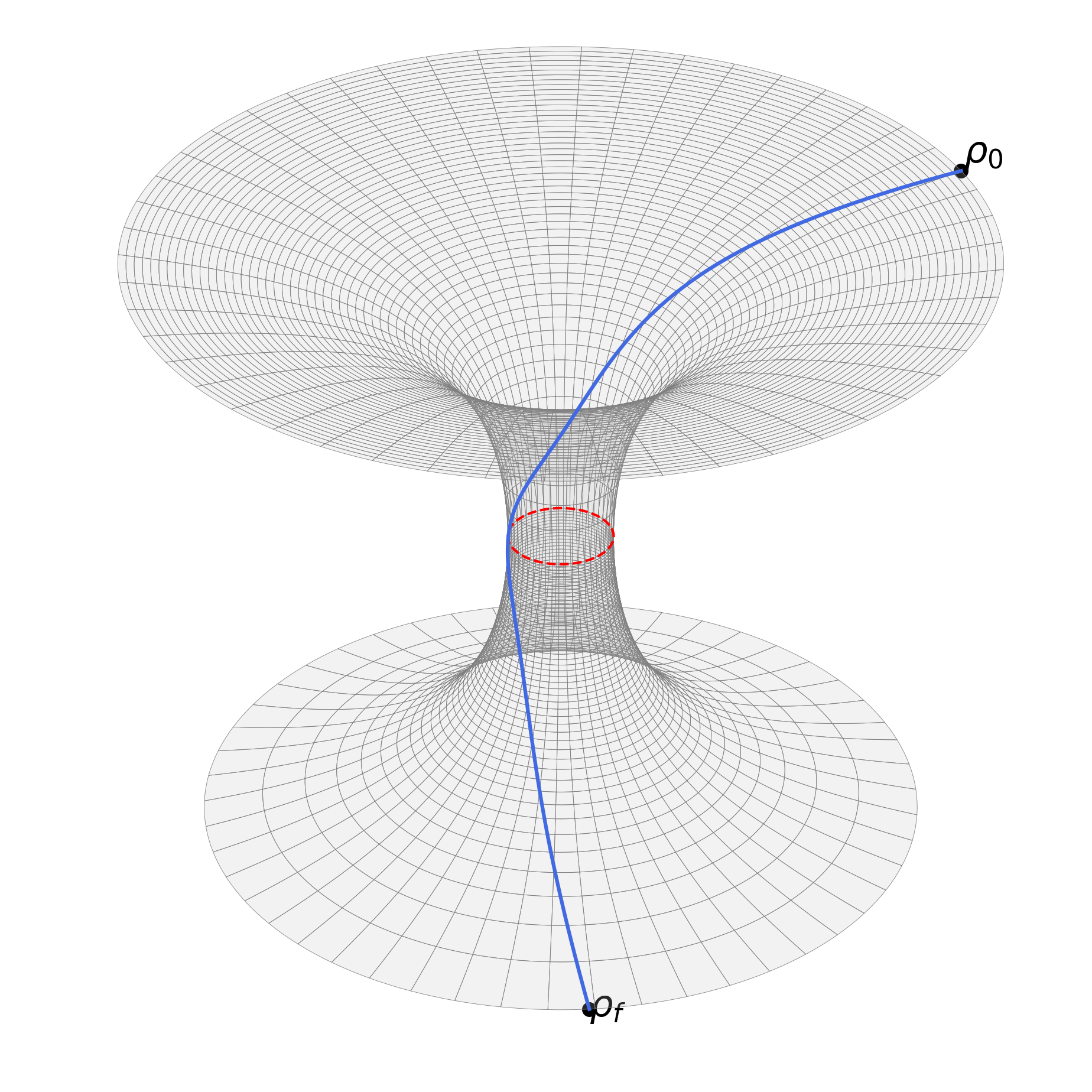}
        \caption*{(c) $D=6, E = \sqrt{10}$}
    \end{minipage}
    \begin{minipage}{0.32\textwidth}
        \centering
        \includegraphics[width=\linewidth]{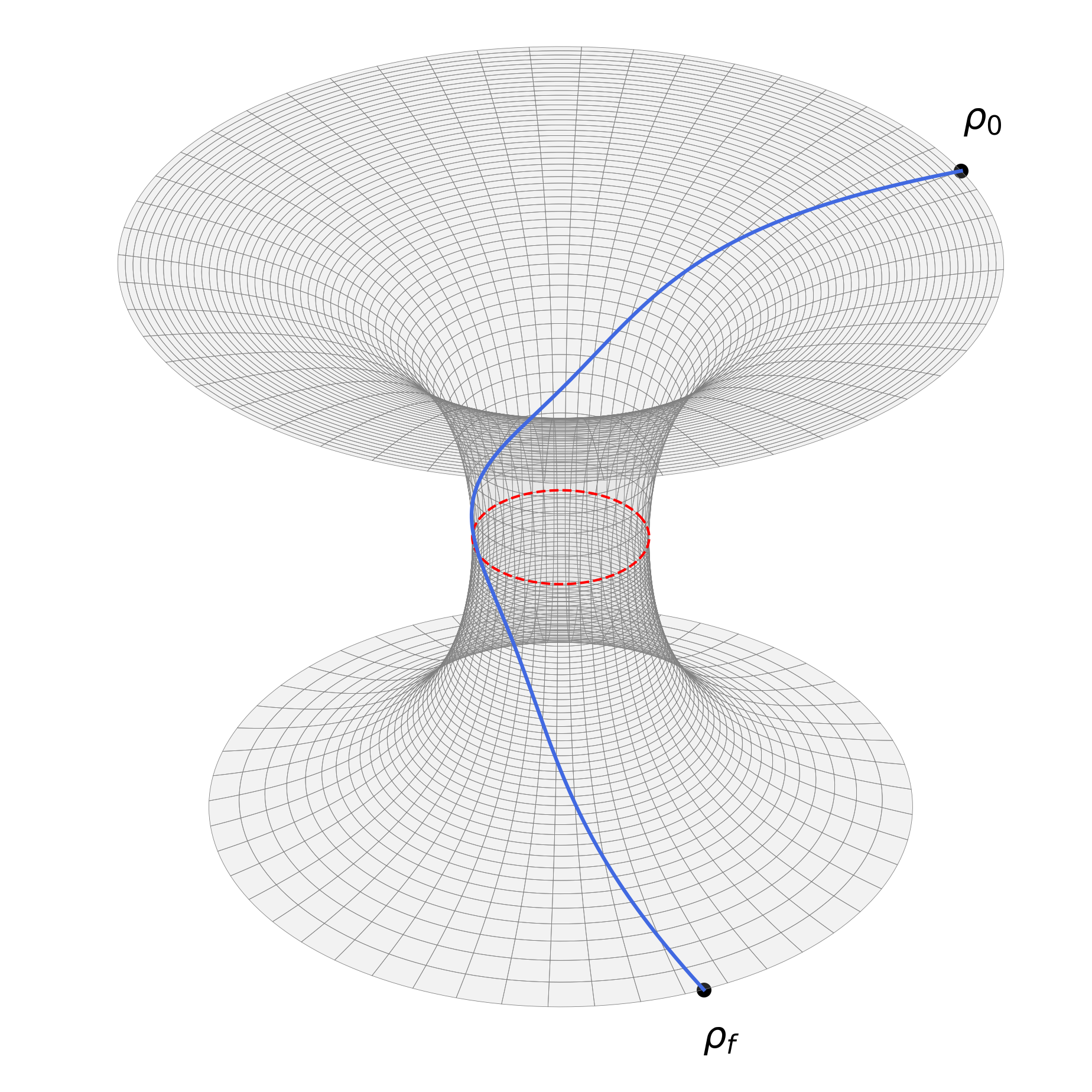}
        \caption*{(a) $D=7, E = 5$}
    \end{minipage}
    \hfill
    \begin{minipage}{0.32\textwidth}
        \centering
        \includegraphics[width=\linewidth]{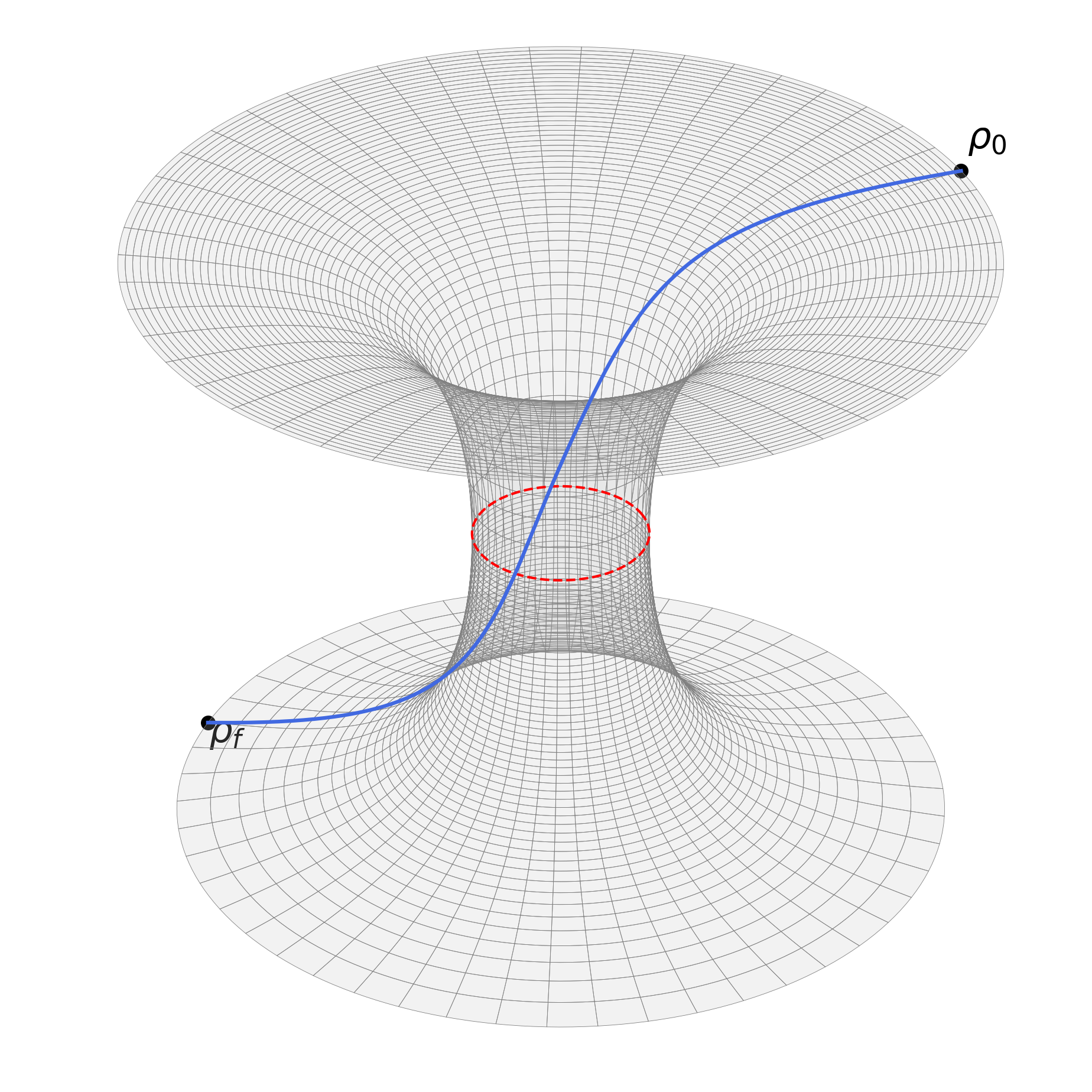}
        \caption*{(b) $D=8, E = 5$}
    \end{minipage}
    \hfill
    \begin{minipage}{0.32\textwidth}
        \centering
        \includegraphics[width=\linewidth]{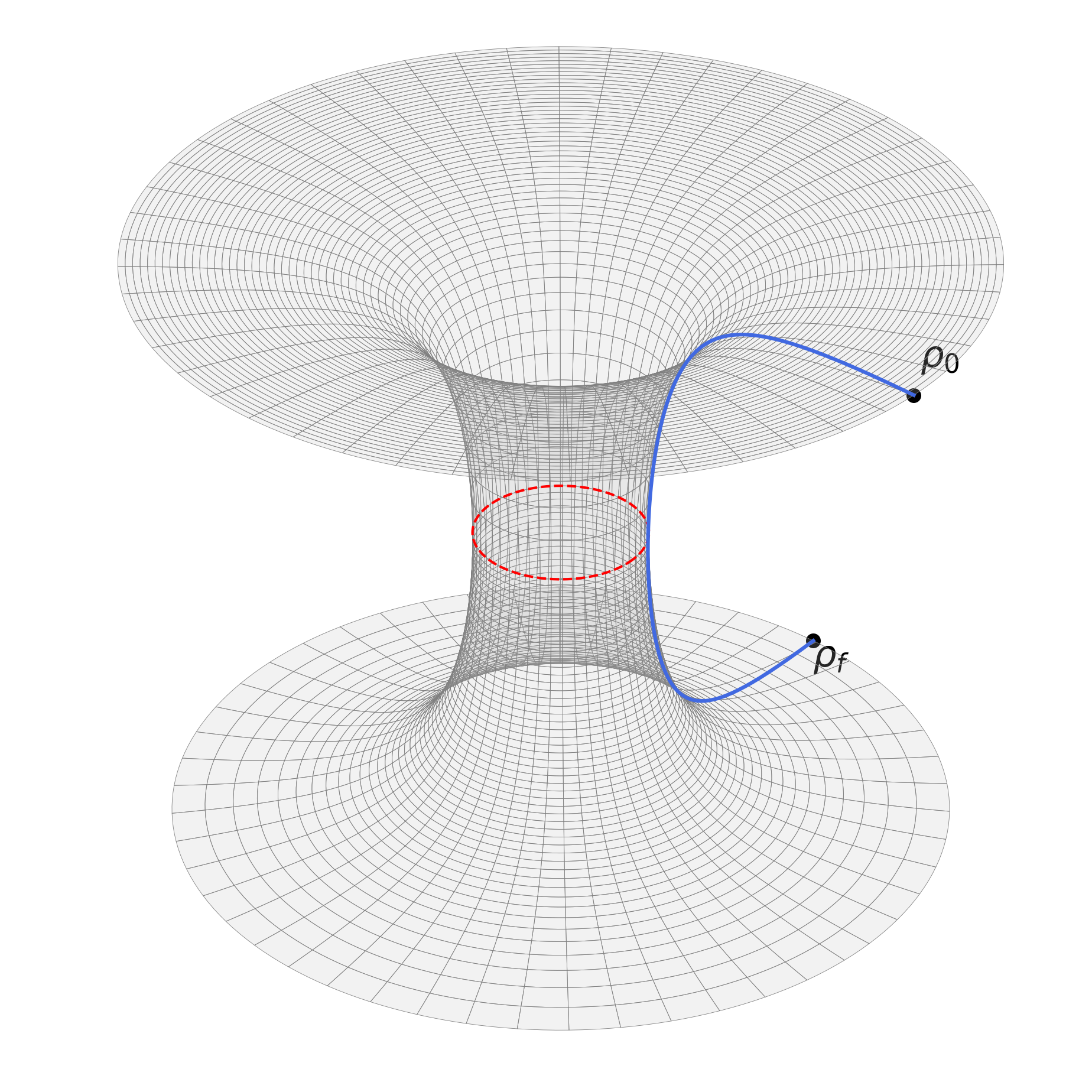}
        \caption*{(c) $D=9, E = 7$}
    \end{minipage}
    \label{fig:three_images}
\end{figure}
\noindent
On the other hand, if $E^{2}$ is exactly equal to $V_{\text{eff}}(r_c)$, then the trajectory corresponds to a circular orbit. Furthermore, it is possible to determine in which asymptotic region this orbit is located by considering
\begin{equation}\label{orbits segun b}
        r_{c} > r_{T} \quad \Rightarrow \quad \frac{(D - 3)(b + 1)}{D - 2} > 0,
\end{equation}
It follows that,

\begin{enumerate}
    \item $b \in (-1,0)$, the circular orbit lies in the same asymptotic region from which the particle was initially emitted.
    \item $b \in (-\infty,-1)$, the particle must cross the wormhole throat, and once situated in the second asymptotic region, it will move in a circular orbit.
\end{enumerate}
It is interesting to note that because $b = -1$ is forbidden, such an orbit cannot be placed exactly at the wormhole throat, as shown in the figure below.
\begin{figure}[H]
    \centering
    \caption{(UO) trayectories followed by light rays in the wormhole geometry for $\mu = 1, L = 5.$}
    \begin{minipage}{0.3\textwidth}
        \centering
        \includegraphics[width=\linewidth]{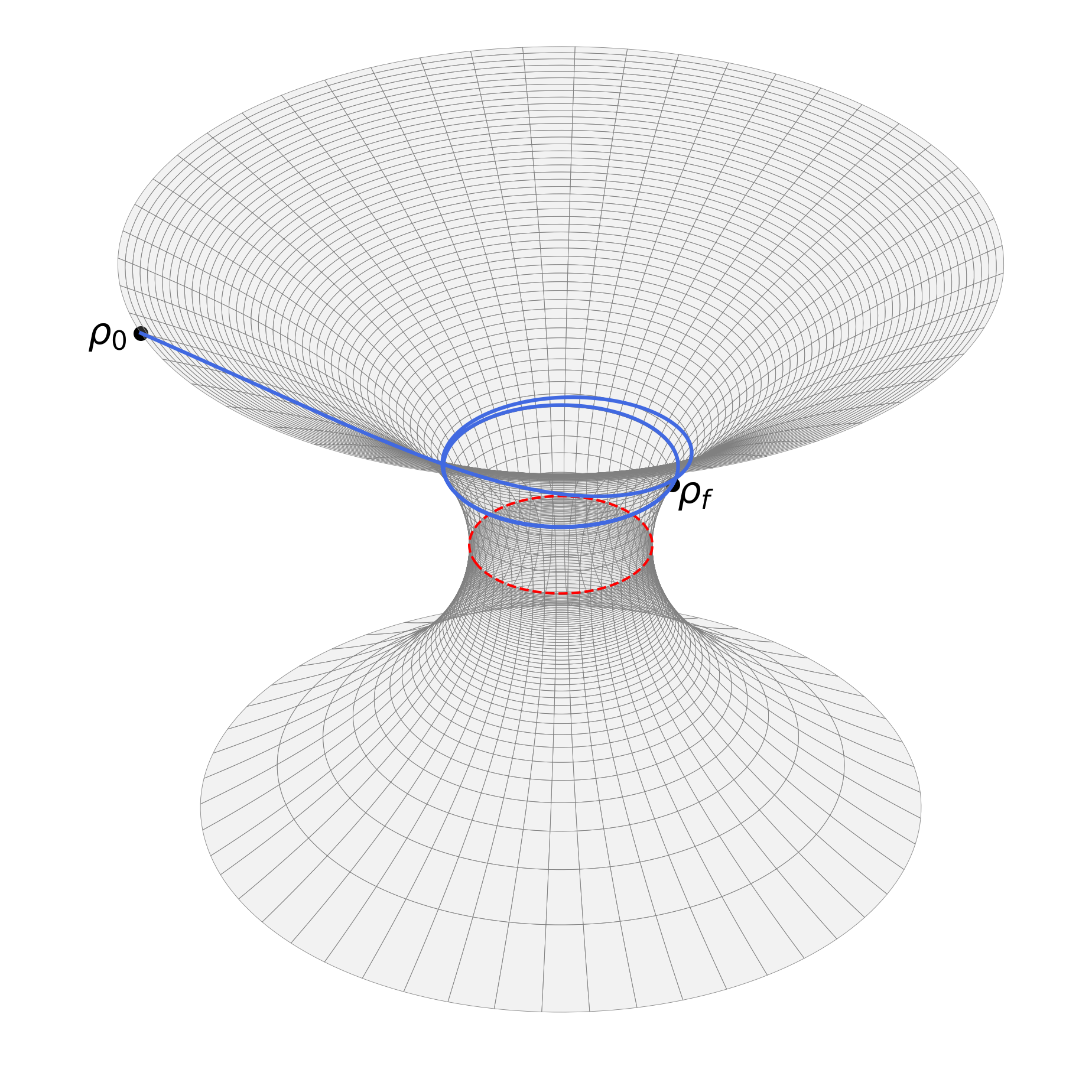}
        \caption*{(a) $D=4, b = -0.1$}
    \end{minipage}
    \hfill
    \begin{minipage}{0.3\textwidth}
        \centering
        \includegraphics[width=\linewidth]{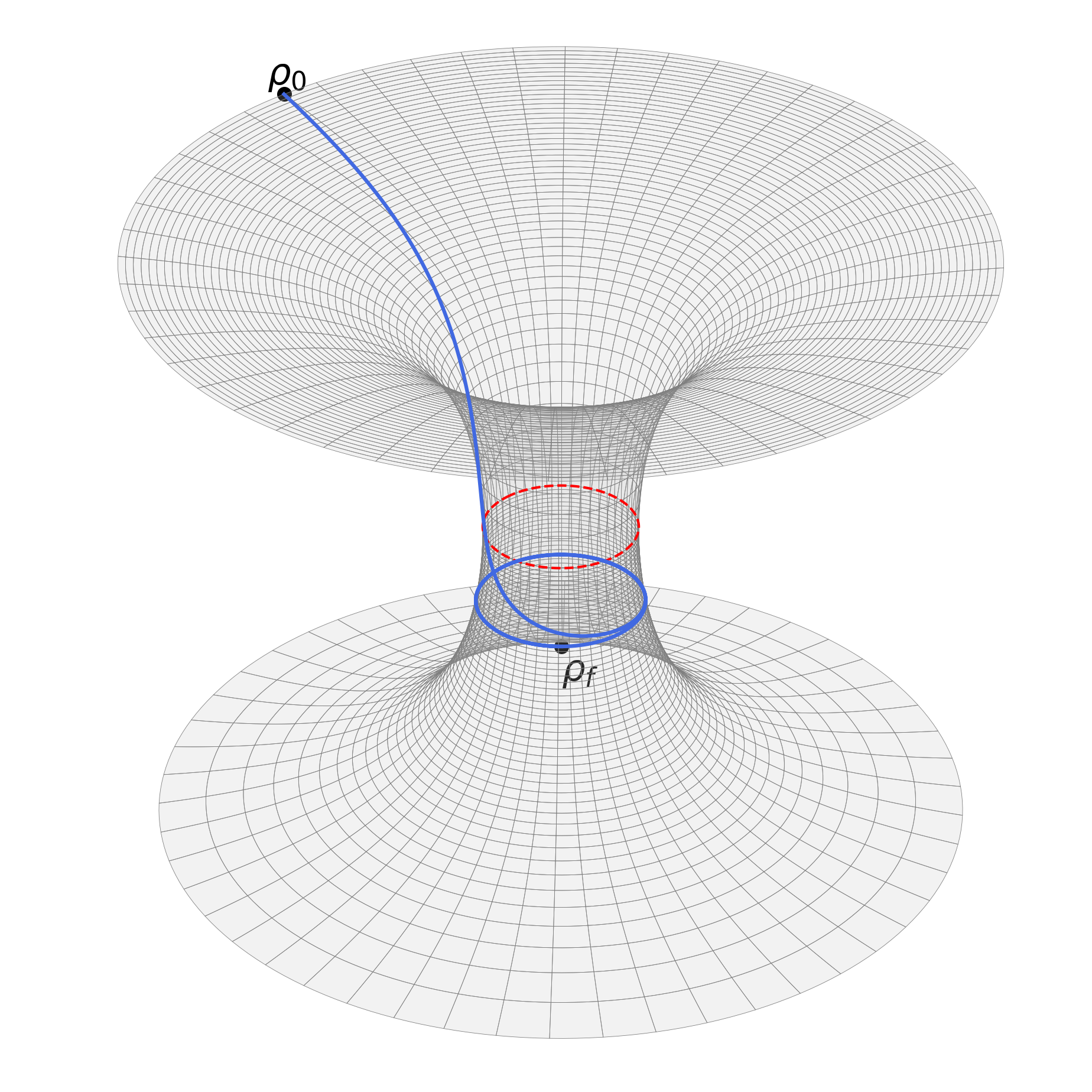}
        \caption*{(b) $D=5, b = -0.7$}
    \end{minipage}
    \hfill
    \begin{minipage}{0.3\textwidth}
        \centering
        \includegraphics[width=\linewidth]{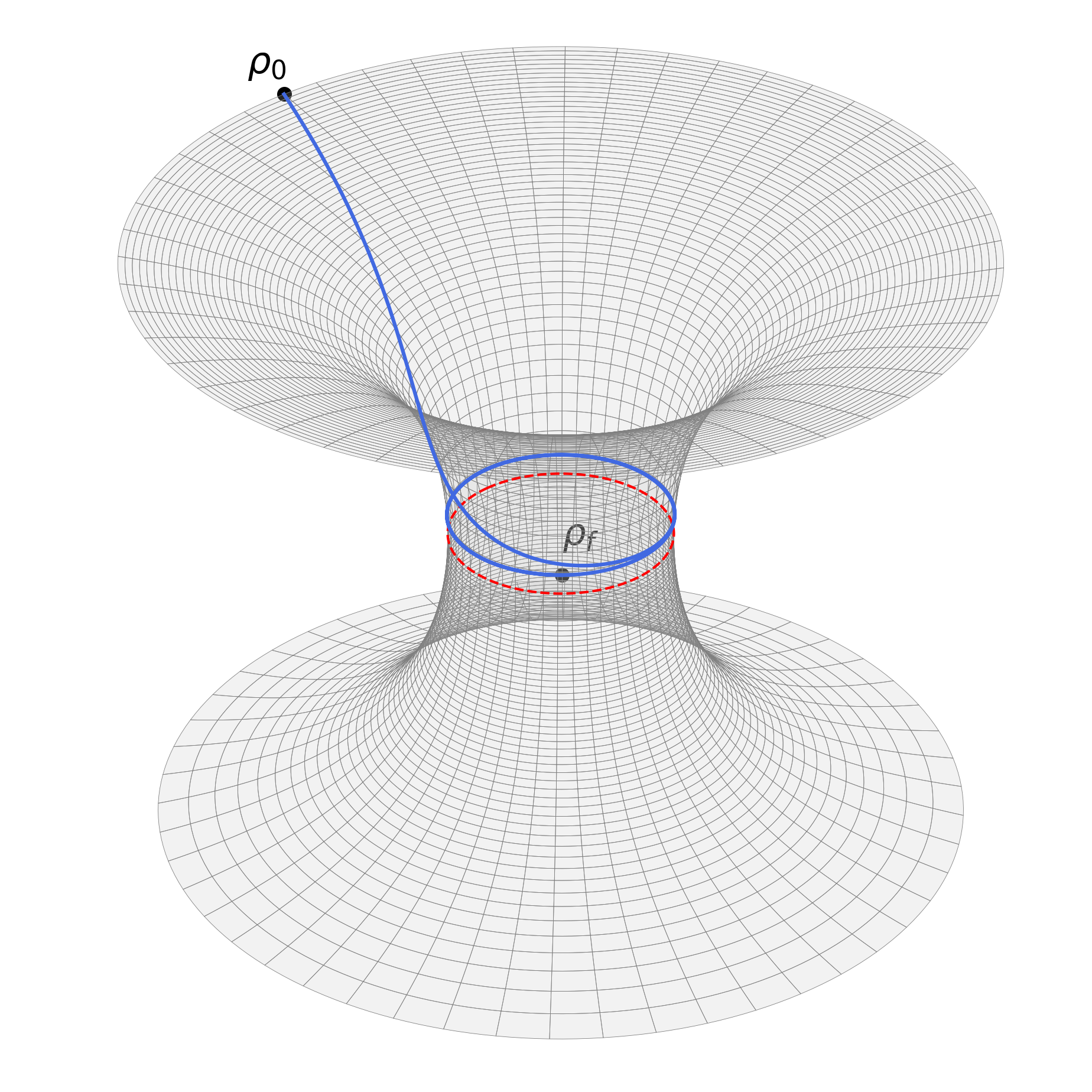}
        \caption*{(c) $D=6, b = -5$}
    \end{minipage}
    \begin{minipage}{0.3\textwidth}
        \centering
        \includegraphics[width=\linewidth]{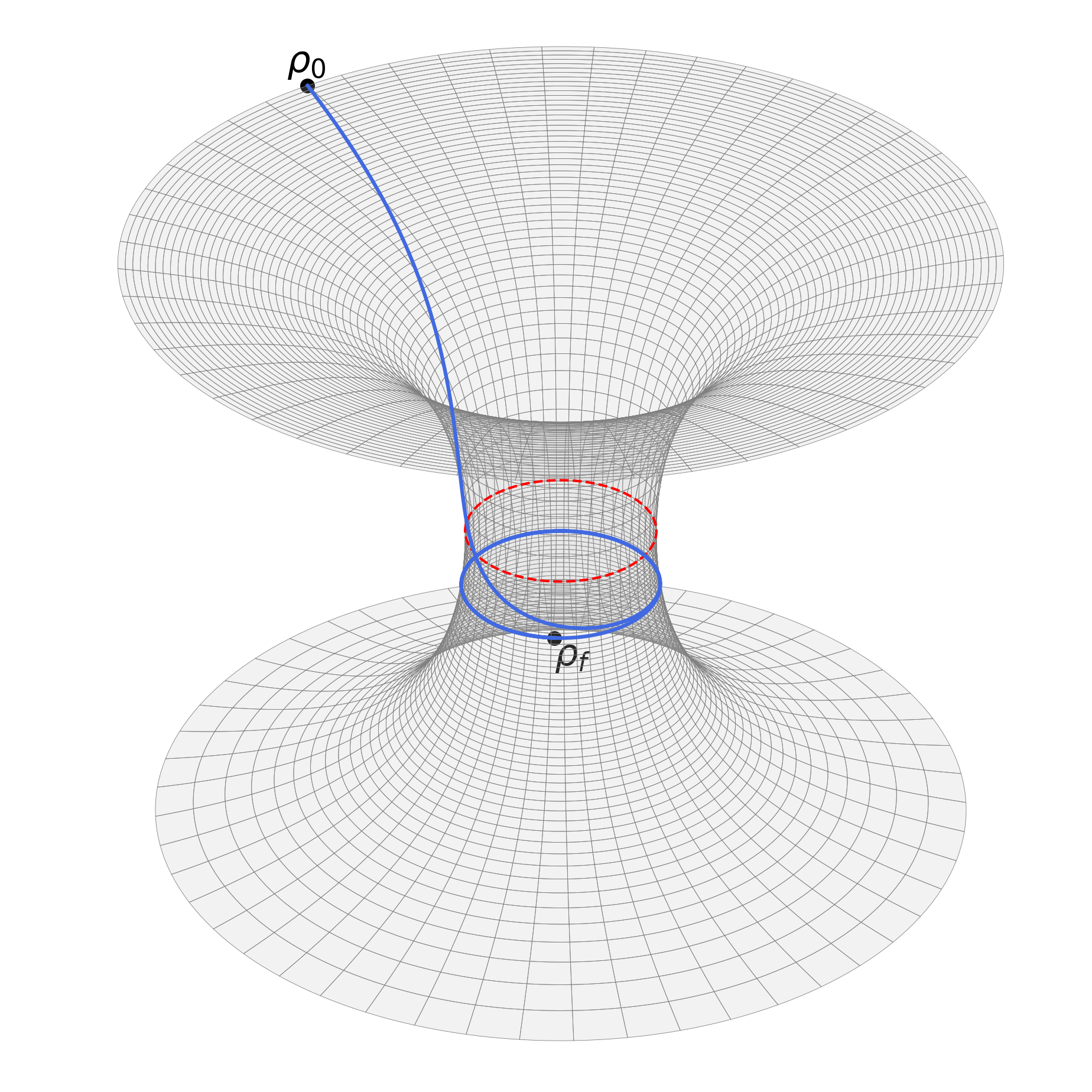}
        \caption*{(d) $D=7, b = -3$}
    \end{minipage}
    \hfill
    \begin{minipage}{0.3\textwidth}
        \centering
        \includegraphics[width=\linewidth]{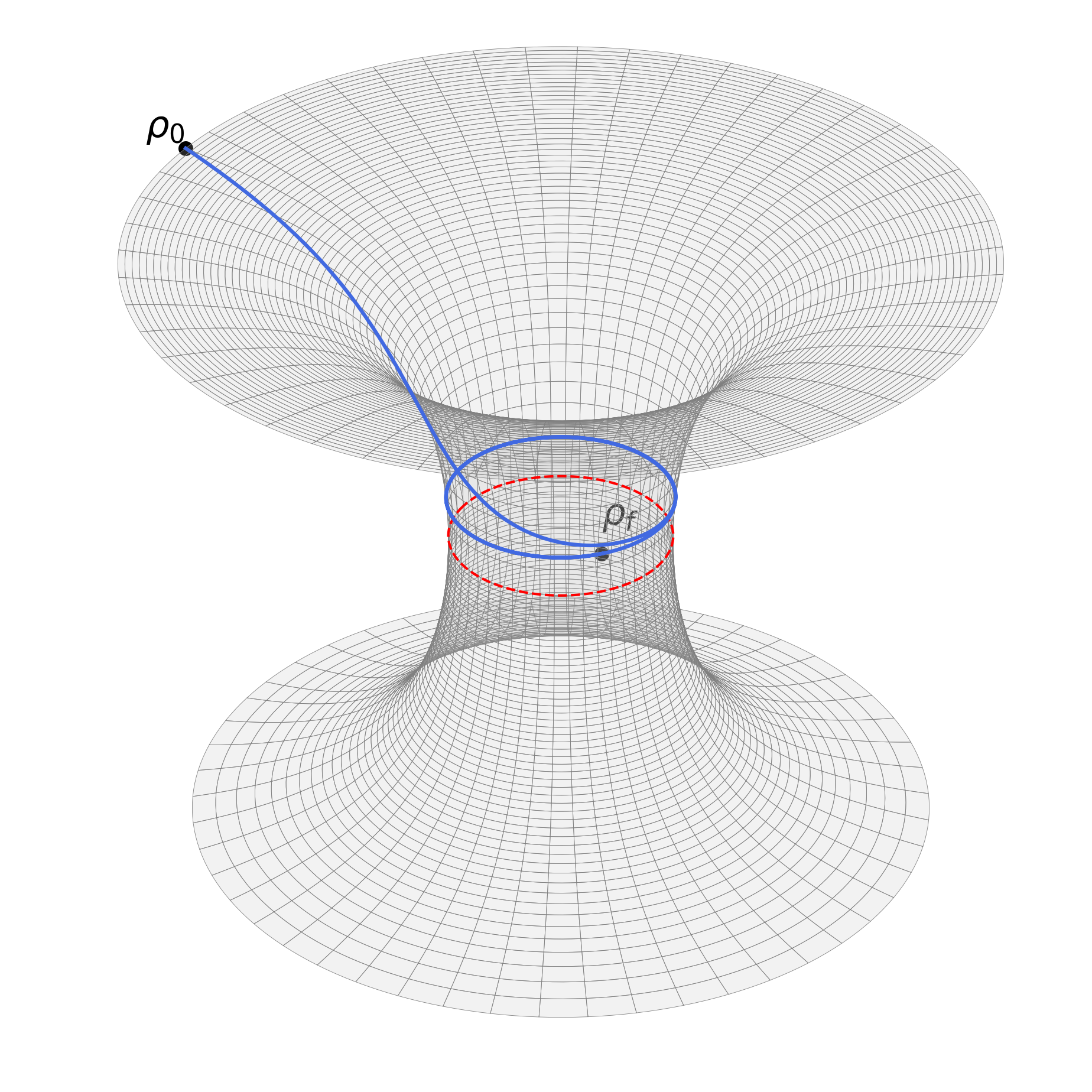}
        \caption*{(e) $D=8, b = -0.5$}
    \end{minipage}
    \hfill
    \begin{minipage}{0.3\textwidth}
        \centering
        \includegraphics[width=\linewidth]{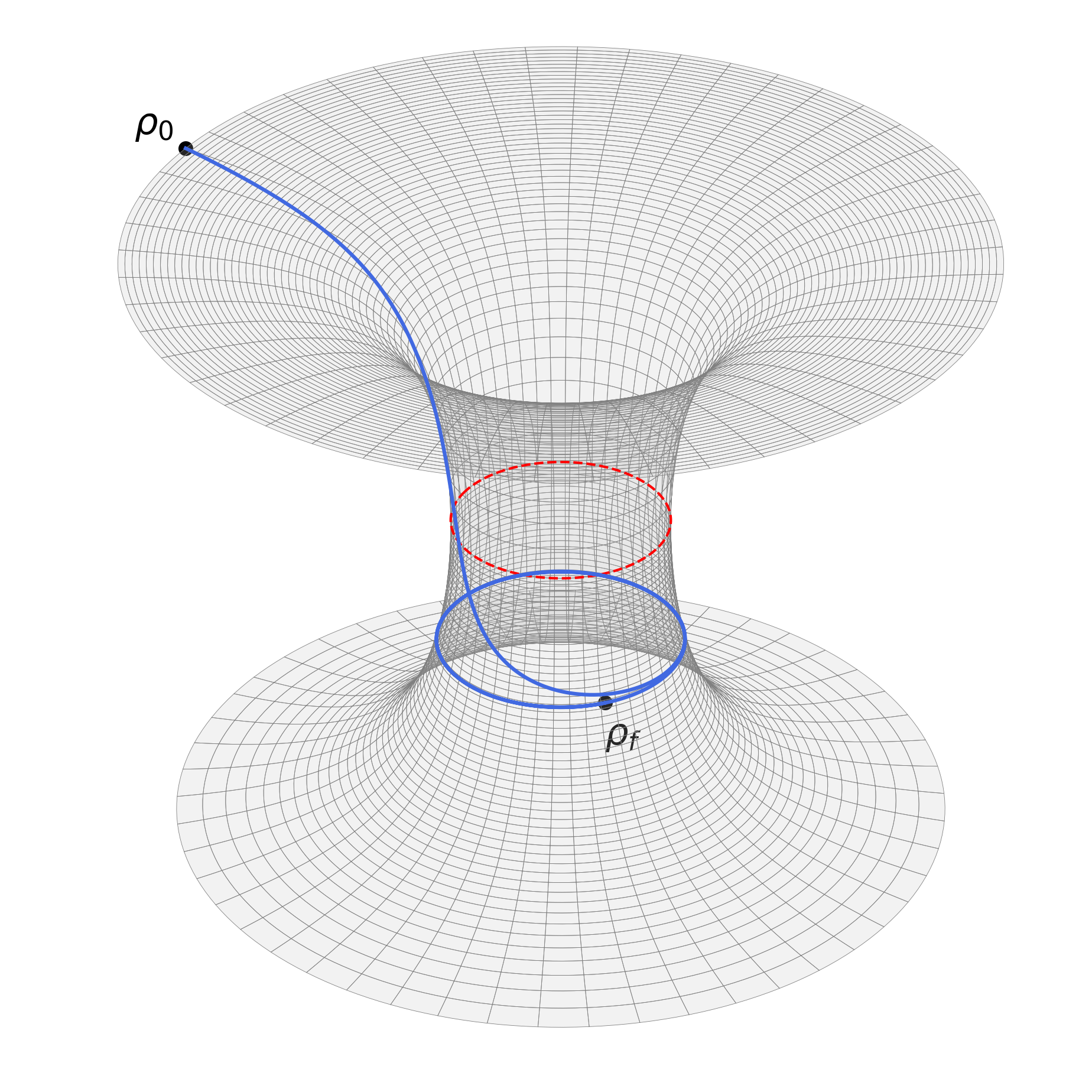}
        \caption*{(f) $D=9, b = -10$}
    \end{minipage}
    \label{fig:six_images}
\end{figure}
\noindent
The final possibility corresponds to the existence of a turning point $r_{tp}$, which is found from the condition $E^{2}=V_{\text{eff}}(r_{tp})$. Solving this equation gives
\begin{equation}
    r_{\text{tp}} = \left\{\frac{1}{2}\left(\frac{L}{E}\right)^{D - 3} + \sqrt{\left(\frac{L}{E}\right)^{D - 3}\left[\frac{1}{4}\left(\frac{L}{E}\right)^{D - 3} - \mu\right]} - \mu \right\}^{\frac{1}{D - 3}}, \quad  \mu \left(\frac{E}{L}\right)^{D - 3} \leq \frac{1}{4}.
\end{equation}
\noindent
Where the minus branch on the square root was discarded because for $E, L, \mu>0$, it is always going to be smaller than the positive branch. As before, it is possible to determine the values of $b$ for which the asymptotic region of the turning point must be situated, that is
\begin{equation}
    r_{\text{tp}} \geq r_{T} \quad \Rightarrow \quad b_{\text{tp}} \geq - \left[\frac{(D - 2)r_{\text{tp}}^{D - 3} + \mu(D - 4)}{(D - 4)r_{\text{tp}}^{D - 3} + \mu(D - 2)}\right]\frac{r_{\text{tp}}^{D - 3}}{\mu}.
\end{equation}
Therefore,
\begin{enumerate}
    \item $b \in (b_{\text{tp}},0)$, the particle never reaches the wormhole throat and instead turns around to the same asymptotic region from which it originated. 
    \item $b \in (-\infty,b_{\text{tp}})$, the particle crosses the throat and then reaches a turning point in the second asymptotic region, returning to where it came from.
    \item $b=b_{\text{tp}}$, the turning point coincides exactly with the wormhole throat and then turns toward the first asymptotic region.
\end{enumerate}

\begin{figure}[H]
    \centering
    \caption{(SU) trayectories followed by light rays in the wormhole geometry for $\mu = 10^{-2}, L = 5, E = 10$.}
    \begin{minipage}{0.3\textwidth}
        \centering
        \includegraphics[width=\linewidth]{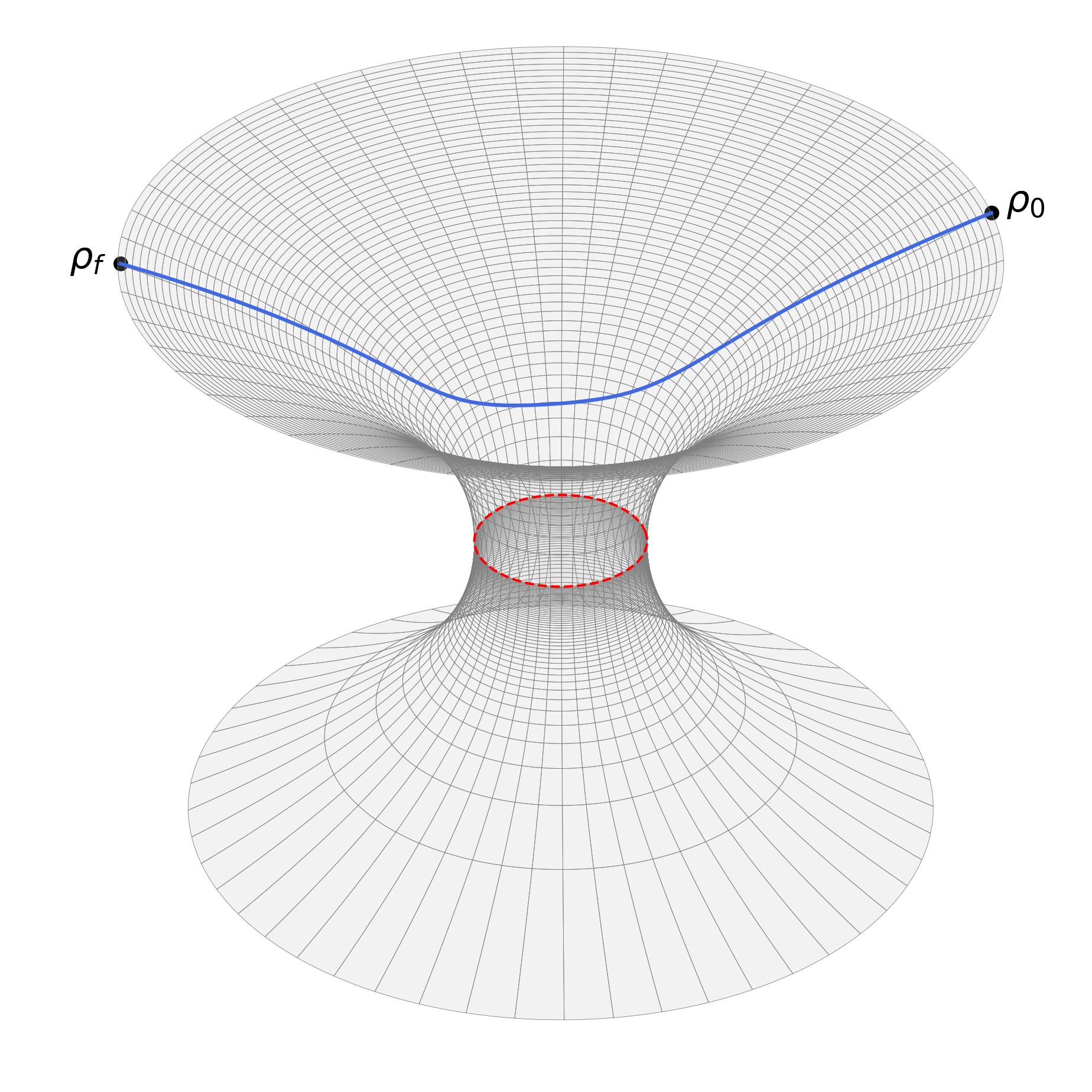}
        \caption*{(a) $D=4, b = -100 > b_{\mathrm{tp}}$}
    \end{minipage}
    \hfill
    \begin{minipage}{0.3\textwidth}
        \centering
        \includegraphics[width=\linewidth]{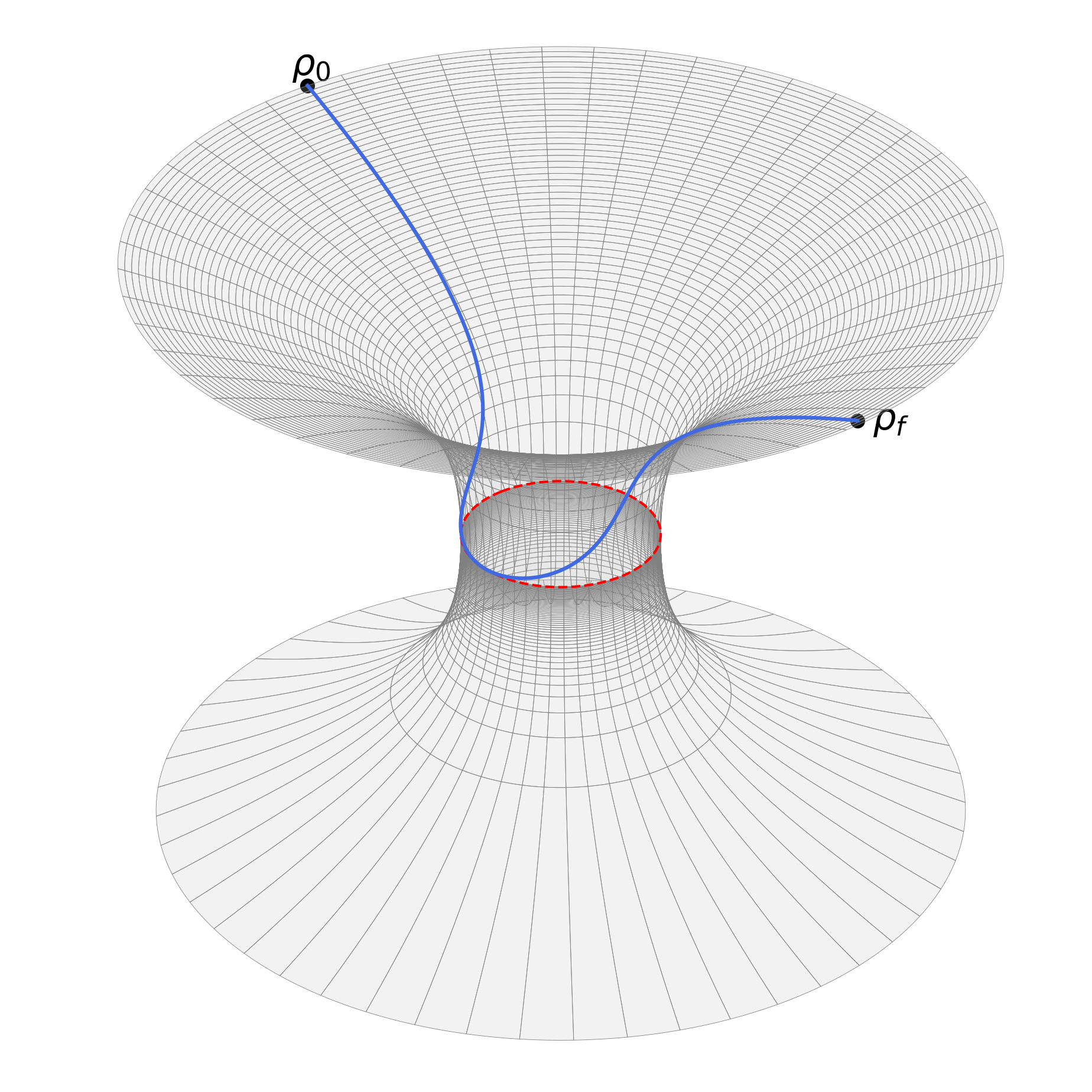}
        \caption*{(b) $D=4, b = b_{\text{tp}} \sim -2302$}
    \end{minipage}
    \hfill
    \begin{minipage}{0.3\textwidth}
        \centering
        \includegraphics[width=\linewidth]{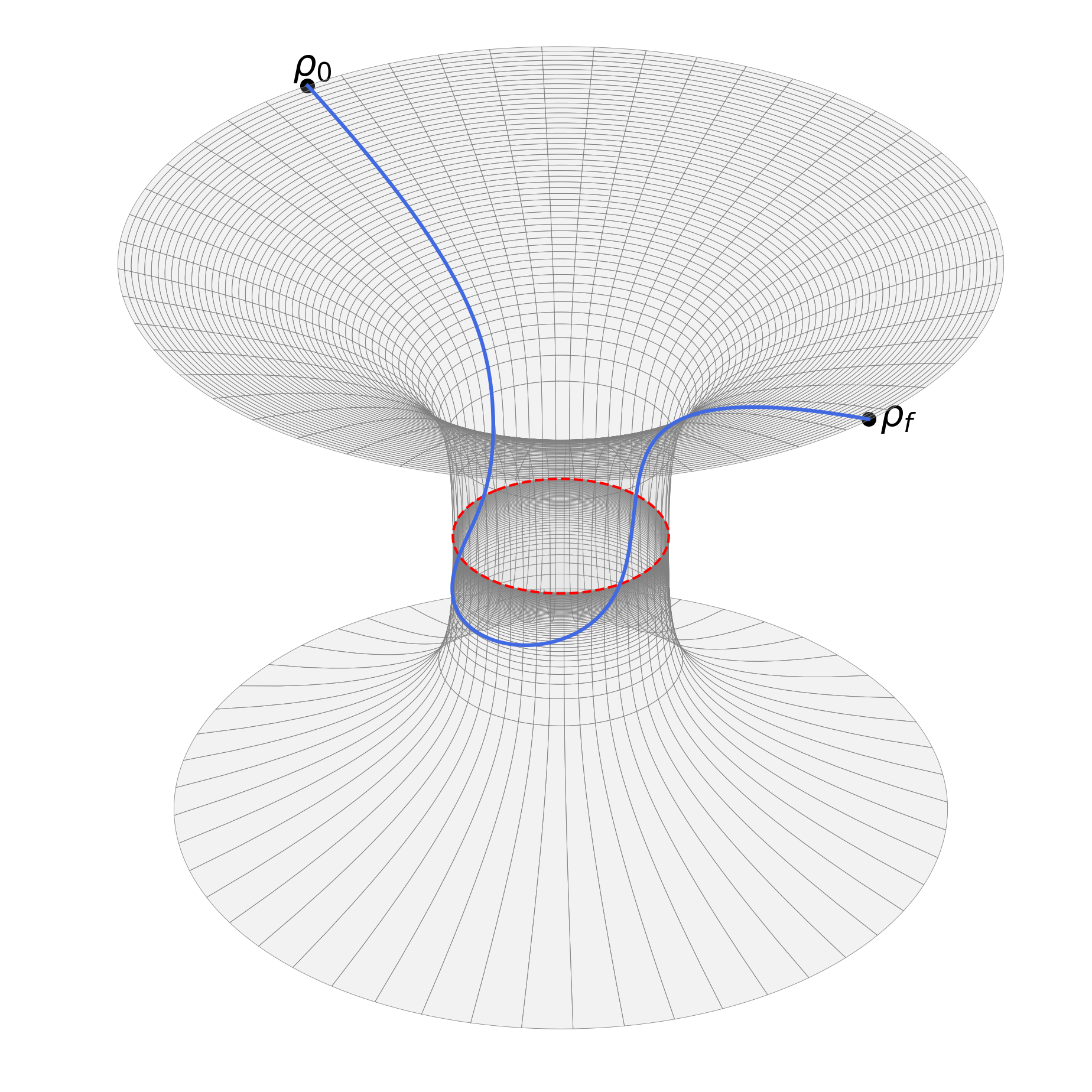}
        \caption*{(c) $D=4, b = -10^{5}$}
    \end{minipage}
    
    \begin{minipage}{0.3\textwidth}
        \centering
        \includegraphics[width=\linewidth]{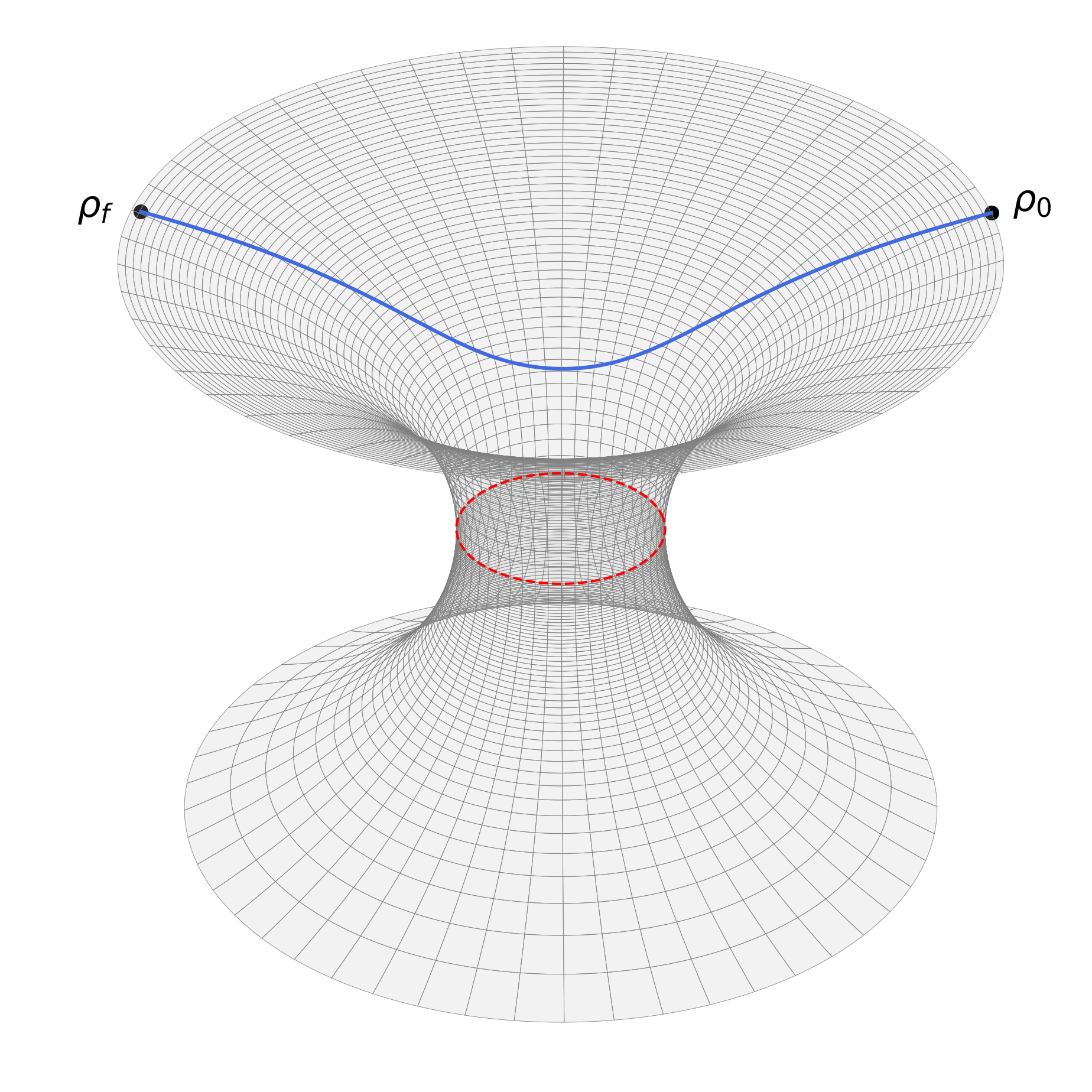}
        \caption*{(d) $D=5, b = -10 > b_{\text{tp}}$}
    \end{minipage}
    \hfill
    \begin{minipage}{0.3\textwidth}
        \centering
        \includegraphics[width=\linewidth]{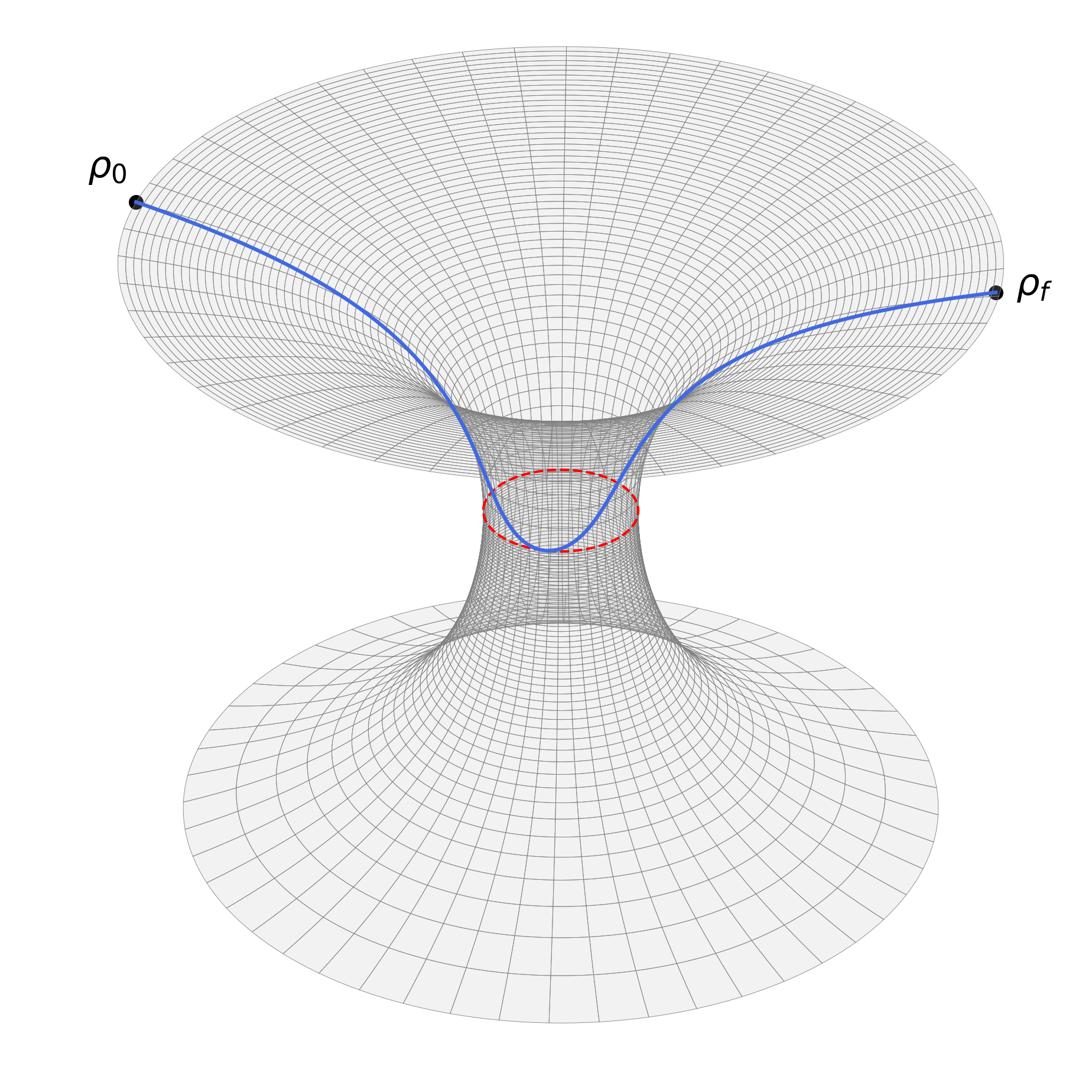}
        \caption*{(e) $D=6, b = b_{\text{tp}} \sim -18.3$}
    \end{minipage}
    \hfill
    \begin{minipage}{0.3\textwidth}
        \centering
        \includegraphics[width=\linewidth]{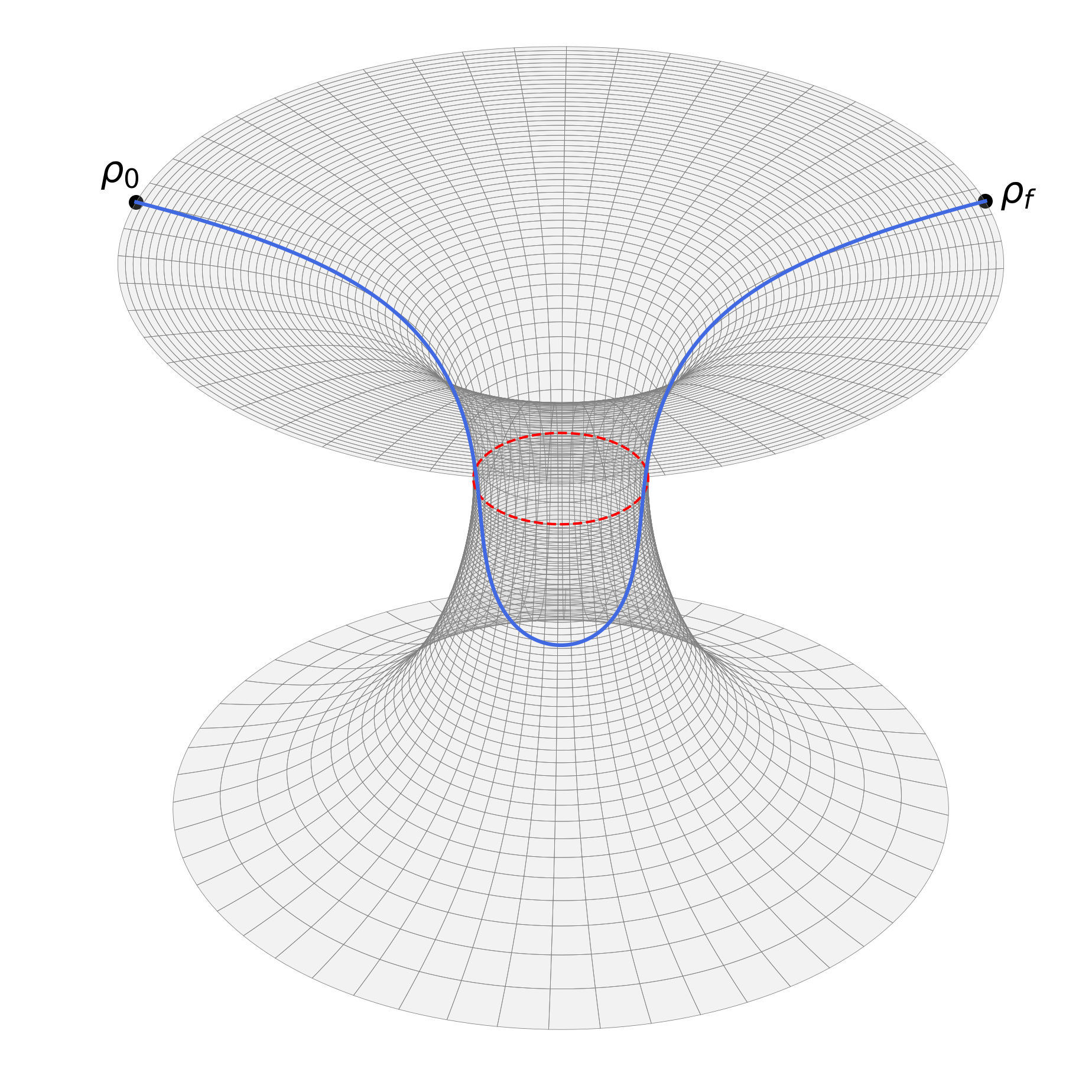}
        \caption*{(f) $D=7, b = -70 < b_{tp}$}
    \end{minipage}
    \label{fig:six_images b}
\end{figure}
\noindent
To finish this section, we can summarize our results by launching a stream of photons from some initial position $r_{0}$, each with a different angular momentum and fixed energy as shown in the figure below, where the red, blue lines correspond to reflecting and traversing trajectories as well as the yellow line which corresponds to the Einstein's ring placed at $r = r_{c}$.
\begin{figure}[H]
    \centering
    \caption{Null geodesics for $\mu = 1, b = -0.5, r_{0} = 7$ and $L \in [0.01, 7], \Delta_{L} = 100$.}
    \begin{minipage}{0.32\textwidth}
        \centering
        \includegraphics[width=\linewidth]{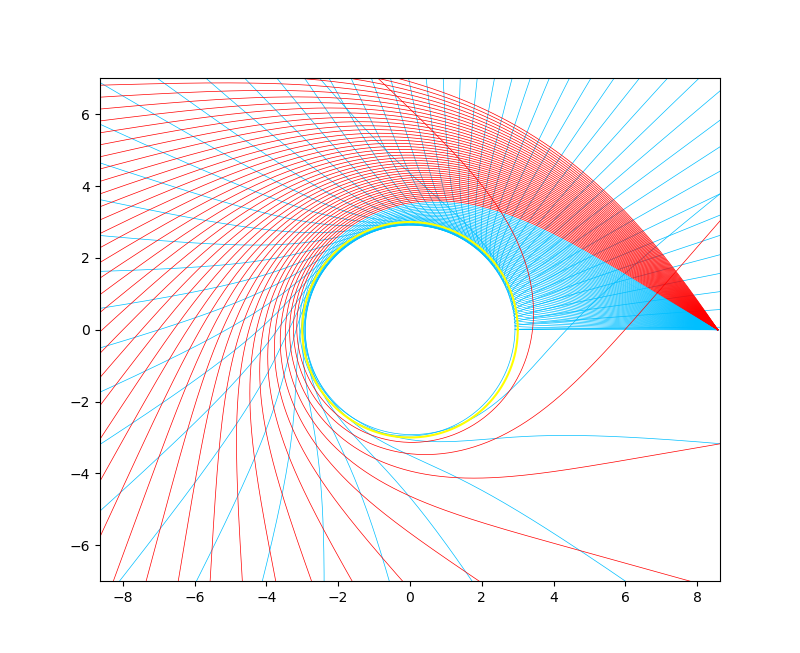}
        \caption*{(a) $D=4$}
    \end{minipage}
    \begin{minipage}{0.32\textwidth}
        \centering
        \includegraphics[width=\linewidth]{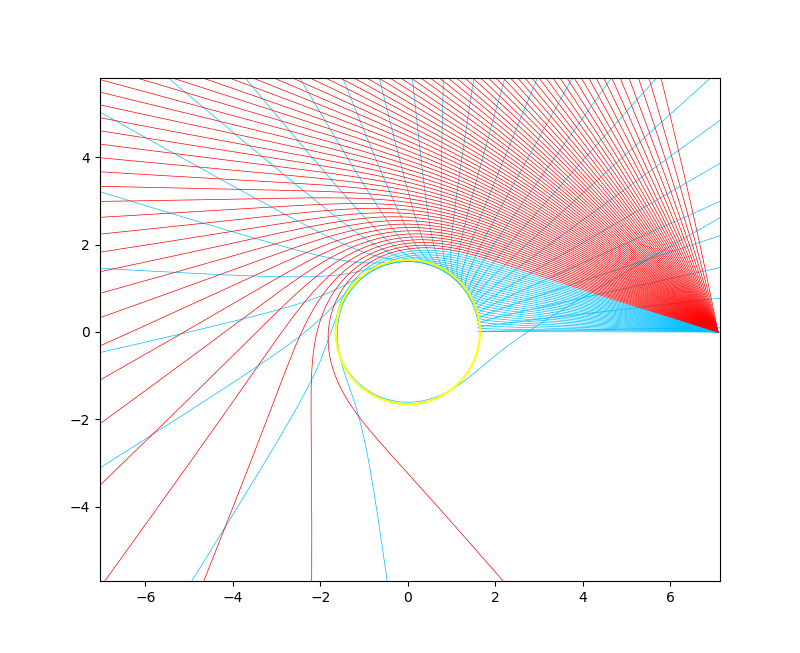}
        \caption*{(b) $D=5$}
    \end{minipage}
    \begin{minipage}{0.32\textwidth}
        \centering
        \includegraphics[width=\linewidth]{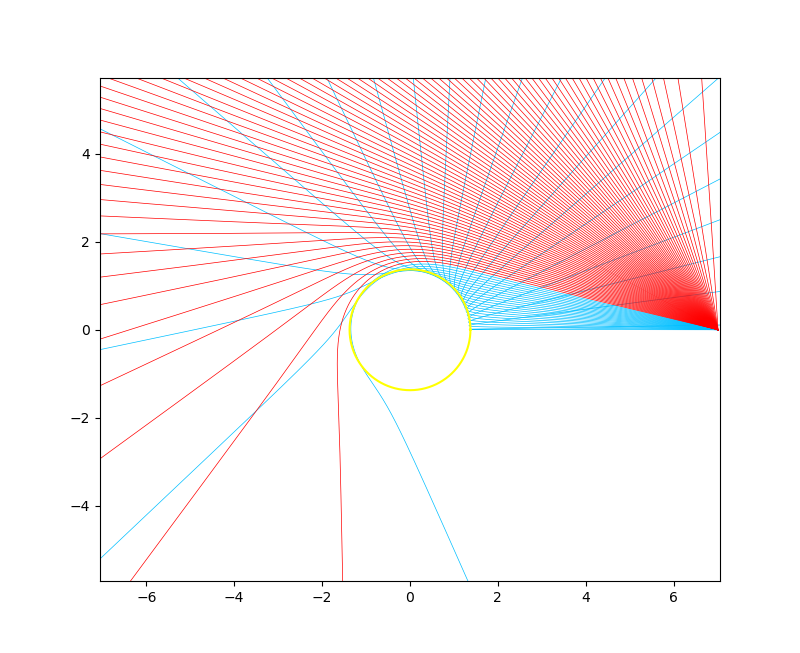}
        \caption*{(c) $D=6$}
    \end{minipage}
    \label{fig:three_ray_tracing}
\end{figure}

\section{Scalar perturbations and quasinormal modes}
Having treated the geodesic analysis for photons in the wormhole geometry, we move on to a beautiful connection with perturbation theory enunciated in \cite{Cardoso:2008bp} and find an isospectrality between the wormhole family and the ERN black hole in $D = 4$. In GR, there is a lot of interest when dealing with perturbations of some solution,whether that be a star, a black hole, a wormhole or other structure that solves Einstein's equations, because it may reveal the stability/instability of the object. For example, it is well-known that some \textit{black strings} are unstable to long wavelength perturbations, where the event horizon pinches off thus creating a naked singularity \cite{Gregory:2011kh} moreover, from an astrophysical point of view when the antennas such as LIGO, VIRGO, LISA, and others manage to detect a gravitational signal from a source, the dominating contribution to such signal will be the QNM with
lowest frequency: the fundamental mode, and hence they are the key in distinguishing black holes from wormholes \cite{Konoplya:2011qq}. 

\medskip

\noindent   
In the literature, there is a vast class of perturbations that one might consider; the most used are: scalar, vector, and tensor (gravitational) perturbations, but for simplicity, we are going to restrict ourselves to the scalar type. Mathematically, we need to analyze how a wave function $\Psi$ behaves in the wormhole background. The covariant equation for a scalar wave is the massless Klein-Gordon equation, namely
\begin{equation}\label{KleinGordonfield}
    \Box \Psi = 0.
\end{equation}
Since our spacetime is static, spherically symmetric, we can always postulate the ansatz
\begin{equation}\label{ansatz scalar perturbation}
    \Psi = e^{-i\omega t}Y_{l}(\theta)\Phi(r),
\end{equation}
where $Y_{l}(\theta)$ are the higher dimensional spherical harmonics which satifies $\nabla^{2}_{S^{D - 2}}Y_{l} = -l(l + D - 3)Y_{l}$ (an extensive review on this topic can be found in \cite{Frye:2012jj}), now inserting Eq.~\eqref{ansatz scalar perturbation} in \eqref{KleinGordonfield}, yields
\begin{equation}
    \frac{d^{2}\Phi}{dr^{2}} + \left(\frac{2\mu b(D - 3)}{vr^{D - 2}} + \frac{D - 2}{r}\right)\frac{d \Phi}{dr} + \left(\frac{B}{A}\omega^{2} - \frac{l(l + D - 3)}{r^{2}}\right)\Phi = 0.
\end{equation}
In order to remove the weight in the frequency, we define the \textit{tortoise coordinate} $dr_{*} = \sqrt{B/A}dr$, which can be solved exactly in terms of the Gauss hypergeometric function $_{2}F_{1}$
\begin{equation}
    r_{*}(r) = 
    \begin{dcases}
    &r + 2\mu \ln r - \frac{\mu^{2}}{r} + \mathrm{const}, \quad D = 4,\\
    &r \, _{2}F_{1}\left(-\frac{2}{D -3}, -\frac{1}{D - 3}; \frac{D - 4}{D - 3}; -\frac{\mu}{r^{D - 3}}\right) + \mathrm{const}, \quad D > 4.
    \end{dcases}
\end{equation}
Where now the asymptotic regions are mapped throught $(-\infty, +\infty)$, thus the new coordinate $r_{*}$ behaves as
\begin{equation}\label{asymptotic r asterisco}
\begin{aligned}
    &r_{*} \simeq r, \quad \text{as} \quad r_{*} \to +\infty,\\
    &|r_{*}| \simeq \frac{\mu^{2/(D - 3)}}{r}, \quad \text{as} \quad r_{*} \to -\infty,
\end{aligned}
\end{equation}
Then, the radial equation becomes
\begin{equation}
    \frac{d^{2}\Phi}{dr_{*}^{2}} + \frac{1}{u^{2/(D - 3)}}\left(\frac{D - 2}{r} + \frac{2\mu}{r^{D - 2}}\left(\frac{b(D - 3)}{v} - \frac{1}{u}\right)\right)\frac{d \Phi}{dr_{*}} + \left(\omega^{2} - \frac{l(l + D - 3)}{r^{2}u^{4/(D - 3)}}\right)\Phi = 0.
\end{equation}
With this in mind, we perform the following transformation
\begin{equation}
    \begin{aligned}
    \Phi &= \text{exp}\left[-\frac{1}{2} \int \frac{1}{u^{2/(D - 3)}}\left(\frac{D - 2}{r} + \frac{2\mu}{r^{D - 2}}\left(\frac{b(D - 3)}{v} - \frac{1}{u}\right)\right)dr_{*}\right]\psi\\
    &= \frac{\psi}{v u^{1/(D - 3)}r^{(D - 2)/2}},
    \end{aligned}
\end{equation}
so that the whole problem reduces to finding the solutions of a Schr\"odinger-like equation, namely
\begin{align}
    &-\frac{d^{2}\psi}{dr^{2}_{*}} + \mathcal{V}_{\text{eff}}\psi = \omega^{2}\psi,\label{shrodingers equation scalar field}\\
    &\mathcal{V}_{\text{eff}} \equiv \frac{1}{u^{\frac{4}{D - 3}}r^{2}}\left[\mathscr{L}^{2} + \frac{\mu(D - 2)}{u^{2}r^{D - 3}} - \frac{1}{4}\right], \quad \mathscr{L} \equiv l + \frac{D - 3}{2},\label{effectivepotentialKGF}
\end{align}
subject to the boundary conditions
\begin{equation}
    \Bigg\{\begin{aligned}
        &\psi \sim e^{i\omega r_{*}}, \quad r_{*} \to +\infty\\
        &\psi \sim e^{-i\omega r_{*}}, \quad r_{*} \to -\infty
    \end{aligned}\Bigg. 
\end{equation}
Immediately, we find a very useful symmetry of the potential barrier 
    \begin{equation}
        \mathcal{I}(r): r \to \frac{\mu^{2/(D - 3)}}{r}, \qquad \mathcal{V}_{\mathrm{eff}}(\mathcal{I}(r)) = \mathcal{V}_{\mathrm{eff}}(r),
    \end{equation}
indeed, if we let $r_{*} = 0$ at $r_{c} = \mu^{1/(D - 3)}$, the tortoise coordinate reads $r_{*}(\mathcal{I}(r)) = -r_{*}(r)$, then we have shown that the barrier is symmetric with respect to $r_{*}$
\begin{equation}\label{v symmetric}
    \mathcal{V}_{\mathrm{eff}}(r_{*}) = \mathcal{V}_{\mathrm{eff}}(-r_{*}).
\end{equation}
Moreover, we notice that something remarkable has happened in Eq. \eqref{shrodingers equation scalar field}: even though the parameter $b$ is carried out in $\Phi$, it has \textit{disappeared} from the canonical Schr\"odinger equation controlling the radial perturbation. Furthermore, the equation given in \eqref{shrodingers equation scalar field} matches exactly in $D = 4$ with the equation for the ERN black hole, as given, for example, in \cite{PhysRevD.103.064005}. Therefore, the quasinormal spectrum will be the same. It is also interesting to note that the extrema of $\mathcal{V}_{\mathrm{eff}}$ lie exactly at $r = r_{c}$ where the photon sphere is located, and it is the only global maximum, i.e 
\begin{equation}
    \frac{d^{2} \mathcal{V}_{\mathrm{eff}}}{dr^{2}}\Bigg|_{r = r_{c}}\Bigg. = - \frac{(D - 3)}{2^{3 + 4/(D - 3)}\mu^{4/(D - 3)}}\left(8\mathscr{L}^{2} + D(D - 3)\right) < 0.
\end{equation}
Now, we can apply the WKB approximation formulated in \cite{PhysRevD.35.3621, IyerII, Kokkotas:1988fm, Konoplya_2003}, and give the full eigenfrequency spectrum. For this work, we focus on orders I, III, and VI where the corrections $\mathcal{A}$ are supplied as an appendix.
\begin{equation}\label{qnms wkb}
    \begin{aligned}
    &\omega_{\mathrm{WKB\,I}}^2
    =
    \mathcal{A}_{0}
    -
    i\alpha\mathcal{A}_{1}
    \\
    &\omega_{\mathrm{WKB\,III}}^2=\mathcal{A}_{0} + \mathcal{A}_2-i\alpha\mathcal{A}_{1}\left(1+\mathcal{A}_3\right)\\
    &\omega_{\mathrm{WKB\,VI}}^2
    =
    \mathcal{A}_{0}
    +
    \mathcal{A}_2
    +
    \mathcal{A}_4
    +
    \mathcal{A}_6
    -
    i\alpha\mathcal{A}_{1}
    \left(
        1
        +
        \mathcal{A}_3
        +
        \mathcal{A}_5
    \right)
    \end{aligned}
    \end{equation}
$\alpha \equiv n + 1/2, \quad n = 0, 1, 2, \dots $ , where $n$ is known as the overtone number. 
\subsection{The eikonal limit}
Now, if we take the eikonal limit, that is $\mathscr{L} \gg 1$, the effective potential matches it with Eq.~\eqref{efective potential} for photons by the interchange of $\mathscr{L} \leftrightarrow L$, so
\begin{equation}
    \mathcal{V}_{\text{eff}} \simeq \frac{\mathscr{L}^{2}}{r^{2}u^{\frac{4}{D - 3}}}.
\end{equation}
In order to find the QNMs in the eikonal limit, we follow the procedure of \cite{Cardoso:2008bp}, so first we compute the angular velocity of the angular unstable photon orbit and the corresponding Lyapunov exponent
\begin{equation}\label{angular velocity and lyapunov exponent}
\Omega_{c} = \frac{1}{\mathscr{L}}\sqrt{\mathcal{V}_{\text{eff}}(r_{c})} = \frac{1}{(4\mu)^{1/(D - 3)}}, \quad \lambda_{L} = \sqrt{-\frac{1}{2\mathcal{V}_{\text{eff}}(r_{c})} \frac{d^{2}\mathcal{V}{\text{eff}}}{d r_{*}^{2}}\Bigg|_{r = r_{c}}\Bigg.} = \sqrt{\frac{D - 3}{2}}\Omega_{c}.
\end{equation}
On the other hand, we can compute those quantities by the very definition of the angular velocity of the photon, thus using Eq \eqref{conserved quantities}
\begin{equation}
    \Omega_{c} \equiv \frac{\dot{\varphi}}{\dot{t}} = \frac{L}{E r_{c}^{2}u(r_{c})^{4/(D - 3)}}
\end{equation}
along with Eq.~\eqref{real trayectory} we obtain once again \eqref{angular velocity and lyapunov exponent}. Furthermore, by looking directly at Eq.~\eqref{drdt motion} for $\zeta = 0$, we obtain
\begin{equation}\label{drdt photon}
    \left(\frac{dr}{dt}\right)^{2} = R(r) \equiv \frac{1}{u^{4/(D - 3)}}\left(1 - \frac{L^{2}}{E^{2}r^{2}u^{4/(D - 3)}}\right).
\end{equation}
Now, we can expand Eq. \eqref{drdt photon} near the critical unstable radius $r_{c}$, yielding
\begin{equation}
    R(r_{c}) =  \frac{D - 3}{2(4\mu)^{2/(D - 3)}}(r - r_{c})^{2} + \mathcal{O}((r - r_{c})^{4}),
\end{equation}
therefore, by setting $\delta r \equiv r - r_{c}$, we have
\begin{equation}
     \frac{d \delta r}{dt} \simeq \left[\sqrt{\frac{D - 3}{2}}\frac{1}{(4\mu)^{1/(D - 3)}}\right] \delta r \quad \Rightarrow \quad \delta r \propto e^{\lambda_{L}t},
\end{equation}
which is, once again, the Lyapunov exponent, and thereby measures the rate of divergence of the trajectories; one can interpret the wormhole geometry as one big barrier where particles slowly leak out to either asymptotic region. Also, note that as $D$ grows, the instability of the geodesic motion increases. Now we can compute the QNMs in the eikonal limit explicitly, given by \cite{Cardoso:2008bp}
\begin{equation}\label{qnm eikonal analytically}
    r_{c} \, \omega_{\mathrm{eikonal}} \simeq \frac{1}{4^{1/(D - 3)}}\left[\mathscr{L} - i\left(n + \frac{1}{2}\right)\sqrt{\frac{D - 3}{2}}\right].
\end{equation}
To finish up the discussion about the nature of the correspondence between QNMs and null geodesics, we can also give the transmission and reflection coefficients, $\mathcal{T}, \mathcal{R}$ respectively, in the eikonal limit as those were previously addressed in \cite{PhysRevD.35.3621}, where now $\omega$ is a real number
\begin{equation}
    \left|\mathcal{T}(\omega)\right|^{2} \simeq \left[1 + \exp\left(\frac{\pi (\omega_{R}^{2} - \omega^{2})}{2\omega_{I}\omega_{R}}\right)\right]^{-1}, \quad \left|\mathcal{R}(\omega)\right|^{2} = 1 - |\mathcal{T}(\omega)|^{2}.
\end{equation}
Therefore
\begin{equation}
    \left|\mathcal{T}(\omega)\right|^{2} = \left[1 + \exp\left\{\pi\sqrt{\frac{2}{D - 3}}\left(\mathscr{L} - \frac{\omega^{2}(4\mu)^{2/(D - 3)}}{\mathscr{L}}\right)\right\}\right]^{-1}.
\end{equation}
If we define a \textit{critical} frecuency $\omega_{\mathrm{c}} \equiv \mathscr{L}\Omega_{c}$, controlled only by the unstable, null geodesic then both amplitudes are exactly 1/2, meaning that the critical radius $r_{c}$ acts as a two-way membrane allowing or forbidding the travel through the wormhole.
\subsection{Low energy regime}
Having treated the eikonal limit, which describes short-wavelength perturbations localized near the photon-sphere, with \(\omega r_c\sim l\). We now turn on the low-energy regime, which describes long-wavelength waves with \(\omega r_c\ll1\), typically dominated by low multipoles $l$. Therefore, we would have found the two edges of the same spectral problem. In fact, in this regime, the Schr\'odinger eq found in \eqref{shrodingers equation scalar field} can be solved analytically for $\omega = 0$ as well as for $\omega \simeq 0$. This fact can be exploited once again to determine the transmission/reflection coefficients in this regime, and, as we shall see, the wormhole is not just an open window through which matter can travel; particles must have the \textit{right amount} of energy. So first of all, if we take exactly $\omega = 0$, then Eq. \eqref{shrodingers equation scalar field} reads
\begin{equation}
    \frac{d^{2}\psi_{\mathrm{exact}}}{dr^{2}} + \frac{2\mu}{r(r^{D - 3} + \mu)}\frac{d\psi_{\mathrm{exact}}}{d r} = \left[\frac{\mathscr{L}^{2} - 1/4}{r^{2}} + \frac{\mu(D - 2)r^{D - 5}}{(r^{D - 3} + \mu)^{2}}\right]\psi_{\mathrm{exact}},
\end{equation}
which has the general solution
\begin{equation}
    \psi_{\mathrm{exact}} = \mathcal{C}_{1}(r^{D - 3} + \mu)^{1/(D - 3)}r^{\mathscr{L} - 1/2} + \mathcal{C}_{2}(r^{D - 3} + \mu)^{1/(D - 3)}r^{-\mathscr{L} - 1/2},
\end{equation}
where $\mathcal{C}_{1}, \mathcal{C}_{2}$ are two arbitrary constants of integration. In the asymptotic limit given in Eq. \eqref{asymptotic r asterisco}, this solution behaves as
\begin{align}
    &\psi_{\mathrm{exact}} \simeq \mathcal{C}_{1}r_{*}^{1/2 + \mathscr{L}} + \mathcal{C}_{2}r^{1/2 - \mathscr{L}}, \quad  \quad r_{*} \to +\infty,\\
    &\psi_{\mathrm{exact}} \simeq \mathcal{C}_{1}\mu^{2\mathscr{L}/(D - 3)}r_{*}^{1/2 - \mathscr{L}} + \mathcal{C}_{2}\mu^{-2\mathscr{L}/(D - 3)}r_{*}^{1/2 + \mathscr{L}}, \quad  \quad r_{*} \to -\infty.
\end{align}
Now, we turn our attention to a more subtle analysis. We suppose that $\omega \mu \ll 1$ and write Eq. \eqref{shrodingers equation scalar field} in both asymptotic regions, yielding
\begin{equation}
    \left[\frac{d^{2}}{d r_{*}^{2}} + \omega^{2} - \frac{\mathscr{L}^{2} - 1/4}{r_{*}^{2}}\right]\psi_{\mathrm{left/right}} \simeq 0,
\end{equation}
by transforming $\psi_{\mathrm{left/right}} \to \sqrt{|r_{*}|}\psi_{\mathrm{left/right}}(\omega |r_{*}|)$ we obtain the Bessel differential equation, and solve it by using the Hankel functions because they are a natural choice for travelling waves, then
\begin{equation}
    \begin{aligned}
        &H^{(1)}_{\mathscr{L}} \equiv J_{\mathscr{L}}^{(1)} + iJ_{\mathscr{L}}^{(2)}, \quad (\mathrm{outgoing})\\
        &H^{(2)}_{\mathscr{L}} \equiv J_{\mathscr{L}}^{(1)} - iJ_{\mathscr{L}}^{(2)}, \quad (\mathrm{ingoing}).
    \end{aligned}
\end{equation}
Where $J_{\mathscr{L}}^{(1)}, J_{\mathscr{L}}^{(2)}$ are the Bessel functions of first and second kind respectively, therefore
\begin{align}
    &\psi_{\text{left}} = \mathcal{N}e^{-i\varrho}\left(H^{(2)}_{\mathscr{L}}(-\omega r_{*}) + \mathcal{R}H_{\mathscr{L}}^{(1)}(-\omega r_{*})\right)\sqrt{-r_{*}}, \quad \quad r_{*} \to -\infty,\\
    &\psi_{\text{right}} = \mathcal{N}e^{i\varrho}\mathcal{T}H^{(1)}_{\mathscr{L}}(\omega r_{*})\sqrt{r_{*}}, \quad \quad r_{*} \to +\infty.
\end{align}
Notice that $\mathcal{N} \equiv \sqrt{\pi\omega/2}$ and $\varrho \equiv \pi(\mathscr{L} + 1/2)/2$ are two normalization constants chosen carefully so that  the asymptotic Hankel modes have unit plane-wave amplitude (and hence the coeficients $|R|^{2}$ and $|T|^{2}$ are to be truly interpreted as the amplitud probabilities of the scattering analysis in the low energy limit, in this regime, the Bessel functions behave as 
\begin{equation}
    J_{\mathscr{L}}^{(1)} \simeq \frac{1}{\Gamma(\mathscr{L} + 1)}\left(\frac{\omega |r_{*}|}{2}\right)^{\mathscr{L}}, \quad J_{\mathscr{L}}^{(2)} \simeq -\frac{\Gamma(\mathscr{L})}{\pi}\left(\frac{2}{\omega |r_{*}|}\right)^{\mathscr{L}}.
\end{equation}
It is useful to define the quantity $\mathscr{A}$ as
\begin{equation}
    \mathscr{A} \equiv \frac{\pi \mu^{2\mathscr{L}/(D - 3)}}{\Gamma(\mathscr{L})\Gamma(\mathscr{L} + 1)}\left(\frac{\omega}{2}\right)^{2\mathscr{L}}
\end{equation}
We are now ready to match both solutions in the two asymptotic regions, which is the core of all the analysis done so far. This process is illustrated in the figure below.
\begin{figure}[H]
    \centering
    \includegraphics[width=\linewidth]{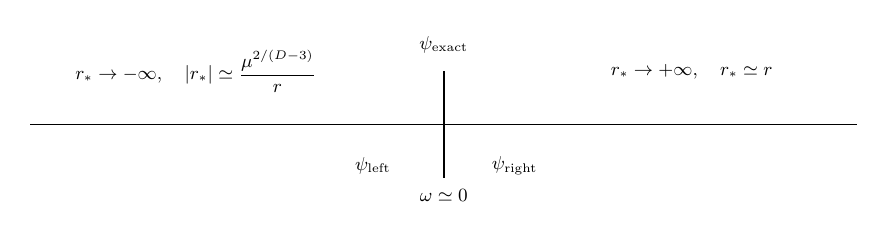}
    \caption{Schematic representation of the matching between the asymptotic solutions and the exact solution in the low-frequency regime.}
    \label{fig:matching_simple_brace}
\end{figure}
\noindent
By doing this, we find
\begin{equation}
    |\mathcal{T}|^{2} = \frac{4 \mathscr{A}^{2}}{(1 + \mathscr{A}^{2})^{2}}, \quad |\mathcal{R}|^{2} \equiv \left(\frac{1 - \mathscr{A}^{2}}{1 + \mathscr{A}^{2}}\right)^{2}.
\end{equation}
Therefore, in the low energy limit, we expand both amplitudes as $\mathscr{A} \ll 1$, then we get the desired probability coefficients
\begin{equation}
    |\mathcal{T}|^{2} \simeq 4\mathscr{A}^{2}, \quad |\mathscr{R}|^{2} \simeq 1 - 4\mathscr{A}^{2}.
\end{equation}
Notice that as the frequency decreases, the scattering process approaches pure reflection, as it should be.
\subsection{Numerical approach: The Chebyshev discretization}
Once we have covered the problem semi-analytically, we can tackle it numerically and show the consistency of the analysis so far. For this task, we can employ several methods in the literature, such as finite differences, continued fractions, Prony, Padé approximant, and many others; see, for example, \cite{Berti_2009, Leaver:1985ax, jansen2017overdampedmodesschwarzschilddesitter}. However, we find that the Chebyshev pseudospectral method is incredibly accurate, easy to implement, and a highly reliable tool for this specific problem \cite{Trefethen2000}, \cite{article}. Because of the exact symmetry in Eq. \eqref{v symmetric}, we can exploit this symmetry and only need to solve one half of the full domain; that is, the behavior as $r_{*} \to +\infty$ is the only one that will matter. With this being said, we can extract its behavior at infinity by making the following change
\begin{equation}
\begin{dcases}
    &\xi = e^{i\omega r_{*}}\psi,\\
    &s = 1 - \frac{r_{c}}{r}, \quad s \in [0, 1].
\end{dcases}
\end{equation}
Thus, one can show that Eq. \eqref{shrodingers equation scalar field} becomes a generalized eigenvalue problem, and the corresponding eigenfrequency is now linear instead of quadratic, that is
\begin{equation}\label{eigenequation}
    \left(M_{0} + \Omega M_{1}\right)\xi = 0,
\end{equation}
where
\begin{equation}
    \begin{aligned}
        &M_{0} = (1 - s)^{2}\frac{d^{2}}{ds^{2}} - \frac{2(1 - S)}{u(s)}\frac{d}{ds} - \mathscr{L}^{2} + \frac{1}{4} - \frac{(D - 2)(1 - s)^{D - 3}}{u(s)^{2}}\\
        &M_{1} = 2iu^{2/(D - 3)}\frac{d}{ds},\\
        &u(s) = 1 + (1 - s)^{D - 3}, \quad \Omega \equiv r_{c} \, \omega.
    \end{aligned}
\end{equation}
Now, the boundary condition for $s = 1$ becomes
\begin{equation}\label{boundary 1}
    2i\Omega\frac{d\xi}{ds}\Bigg|_{s = 1}\Bigg. =\left(\mathscr{L}^{2} - \frac{1}{4}\right)\xi(1),
\end{equation}
meanwhile, for $s = 0$, the modes are naturally separated for parity, that is
\begin{equation}\label{boundary 2}
    \begin{dcases}
        &2^{-2/(D - 3)}\frac{d\xi}{ds}\Bigg|_{s = 0}\Bigg. + i\Omega\xi(0) = 0, \quad (\mathrm{even})\\\\
        &\xi(0) = 0 \quad (\mathrm{odd}).
    \end{dcases}
\end{equation}
Then, we use the Chebyshev points $x_{j}$ which naturally lies in the interval $[-1, 1]$ and store all the values of the function $\xi$ in an array, such that
\begin{equation}
    x_{j} =  \cos\left(\frac{j \pi}{N}\right), \quad s_{j} = \frac{1}{2}\left(1 + x_{j}\right), \quad     \vec{\xi} = \begin{pmatrix}
        \xi(s_{0})\\
        \xi(s_{1})\\
        \vdots\\
        \xi_{s_(N)}
    \end{pmatrix}, \quad j = 0, 1, \dots N.
\end{equation}
Finally, we need to use the Chebyshev matrix for differentiation, which is defined over the Chebyshev points as
\begin{equation}
\left(D_{ij}\right)_{x}
=
\begin{dcases}
\dfrac{2N^{2}+1}{6},
& i=j=0,\\[2mm]
-\dfrac{x_i}{2(1-x_i^{2})},
& i=j,\quad 1\leq i\leq N-1,\\[2mm]
-\dfrac{2N^{2}+1}{6},
& i=j=N,\\[2mm]
\dfrac{c_i}{c_j}
\dfrac{(-1)^{i+j}}{x_i-x_j},
& i\neq j,
\end{dcases}
\end{equation}
where
\begin{equation}
    c_{i} = 
    \begin{dcases}
    &0, \quad i = 0, N\\
    &i, \quad \mathrm{otherwise.}
    \end{dcases}
\end{equation}
 Then, for the $s$ variable, we are left with
 \begin{equation}
    D_{s} = 2D_{x}.    
 \end{equation}
 Now, we can solve Eq. \eqref{eigenequation}, supply the boundary conditions in \eqref{boundary 1}, \eqref{boundary 2}, and recover the quasinormal spectrum. Notice that we are supplying $N$ equations, and therefore $N$ will be the number of eigenvalues. However, some of them will not be physical, so we need to choose those that converge to approximately the same value as $N$ changes. A good approach is to choose $N = 10, 20, 30, \dots, 80$ and select the converged frequencies.
\medskip

\noindent
Because we have found an isospectrality between the ERN black hole, we can give explicitly the results of Eqs. \eqref{qnms wkb}, \eqref{qnm eikonal analytically} as well as the Chebyshev QNMs in the table shown below and see that it matches exactly with the black hole, see for example a recent work in \cite{Wang_2026}.
\begin{table}[H]
    \centering
    \caption{
        Fundamental QNMs for several mmultipole numbers in $D=4, \mu=1$.
    }
    \label{tab:qnm_D4_n0}
    
    \resizebox{\textwidth}{!}{%
    \begin{tabular}{c c c c c c}
        \toprule
        $l$
        & $\omega_{\mathrm{eik}}$
        & $\omega_{\mathrm{WKB\,I}}$
        & $\omega_{\mathrm{WKB\,III}}$
        & $\omega_{\mathrm{WKB\,VI}}$
        & $\omega_{\mathrm{Cheb}}$
        \\
        \midrule

        0
        & $0.125000000 - 0.088388348\,i$
        & $0.200824038 - 0.095290578\,i$
        & $0.121088871 - 0.103712414\,i$
        & $0.134402040 - 0.087167948\,i$
        & $0.133458894 - 0.095843842\,i$
        \\

        1
        & $0.375000000 - 0.088388348\,i$
        & $0.405483517 - 0.090370808\,i$
        & $0.375703385 - 0.089361177\,i$
        & $0.377705583 - 0.089342281\,i$
        & $0.377641863 - 0.089384314\,i$
        \\

        2
        & $0.625000000 - 0.088388348\,i$
        & $0.643589253 - 0.089202726\,i$
        & $0.626088902 - 0.088728406\,i$
        & $0.626573862 - 0.088750805\,i$
        & $0.626572776 - 0.088748323\,i$
        \\

        3
        & $0.875000000 - 0.088388348\,i$
        & $0.888335008 - 0.088820526\,i$
        & $0.875936474 - 0.088564952\,i$
        & $0.876120158 - 0.088572522\,i$
        & $0.876120131 - 0.088571891\,i$
        \\

        4
        & $1.125000000 - 0.088388348\,i$
        & $1.135389608 - 0.088654168\,i$
        & $1.125782158 - 0.088496479\,i$
        & $1.125870031 - 0.088499489\,i$
        & $1.125870034 - 0.088499316\,i$
        \\

        5
        & $1.375000000 - 0.088388348\,i$
        & $1.383507954 - 0.088567814\,i$
        & $1.375662789 - 0.088461263\,i$
        & $1.375711329 - 0.088462662\,i$
        & $1.375711331 - 0.088462605\,i$
        \\

        6
        & $1.625000000 - 0.088388348\,i$
        & $1.632202605 - 0.088517475\,i$
        & $1.625572085 - 0.088440792\,i$
        & $1.625601637 - 0.088441524\,i$
        & $1.625601637 - 0.088441502\,i$
        \\

        7
        & $1.875000000 - 0.088388348\,i$
        & $1.881244192 - 0.088485638\,i$
        & $1.875501979 - 0.088427858\,i$
        & $1.875521276 - 0.088428276\,i$
        & $1.875521276 - 0.088428266\,i$
        \\

        8
        & $2.125000000 - 0.088388348\,i$
        & $2.130510719 - 0.088464250\,i$
        & $2.125446581 - 0.088419172\,i$
        & $2.125459864 - 0.088419427\,i$
        & $2.125459864 - 0.088419422\,i$
        \\

        9
        & $2.375000000 - 0.088388348\,i$
        & $2.379931356 - 0.088449200\,i$
        & $2.375401875 - 0.088413061\,i$
        & $2.375411404 - 0.088413225\,i$
        & $2.375411404 - 0.088413223\,i$
        \\

        10
        & $2.625000000 - 0.088388348\,i$
        & $2.629462173 - 0.088438214\,i$
        & $2.625365122 - 0.088408600\,i$
        & $2.625372187 - 0.088408710\,i$
        & $2.625372187 - 0.088408709\,i$
        \\

        \bottomrule
    \end{tabular}%
    }
\end{table}

\begin{table}[H]
    \centering
    \caption{
        QNMs for several overtones in $D=4$, $l=10$, and $\mu=1$.
    }
    \label{tab:qnm_comparison_D4_l10}
    
    \resizebox{\textwidth}{!}{%
    \begin{tabular}{c c c c c c}
        \toprule
        $n$
        & $\omega_{\mathrm{eik}}$
        & $\omega_{\mathrm{WKB\,I}}$
        & $\omega_{\mathrm{WKB\,III}}$
        & $\omega_{\mathrm{WKB\,VI}}$
        & $\omega_{\mathrm{Cheb}}$
        \\
        \midrule
        
        0
        & $2.625000000 - 0.088388348\,i$
        & $2.629462173 - 0.088438214\,i$
        & $2.625365122 - 0.088408600\,i$
        & $2.625372187 - 0.088408710\,i$
        & $2.625372187 - 0.088408709\,i$
        \\
        
        1
        & $2.625000000 - 0.265165043\,i$
        & $2.641215023 - 0.264134048\,i$
        & $2.620889176 - 0.265387164\,i$
        & $2.620912333 - 0.265385963\,i$
        & $2.620912333 - 0.265385957\,i$
        \\
        
        2
        & $2.625000000 - 0.441941738\,i$
        & $2.663972542 - 0.436462717\,i$
        & $2.612000425 - 0.442841883\,i$
        & $2.612007961 - 0.442843843\,i$
        & $2.612007963 - 0.442843842\,i$
        \\
        
        3
        & $2.625000000 - 0.618718434\,i$
        & $2.696422917 - 0.603694086\,i$
        & $2.598820801 - 0.621064347\,i$
        & $2.598690724 - 0.621106317\,i$
        & $2.598690794 - 0.621106508\,i$
        \\
        
        4
        & $2.625000000 - 0.795495129\,i$
        & $2.736966565 - 0.764680312\,i$
        & $2.581522634 - 0.800309536\,i$
        & $2.581010366 - 0.800501917\,i$
        & $2.581010993 - 0.800503301\,i$
        \\
        
        5
        & $2.625000000 - 0.972271824\,i$
        & $2.783971467 - 0.918829217\,i$
        & $2.560316910 - 0.980785408\,i$
        & $2.559036971 - 0.981364177\,i$
        & $2.559040525 - 0.981370281\,i$
        \\
        
        \bottomrule
    \end{tabular}%
    }
\end{table}
\noindent
As we can see, the WKB VI and the Chebyshev method become almost the same thing as $l$ increases, or if $l \gg n$. Also, we supply similar tables in $D > 4$, for the fundamental mode as an appendix. 
\section{Traversability Conditions}
Since the work of Morris-Thorne \cite{Morris:1988cz}, there have been many considerations aimed at ensuring human traversability; see, for instance, \cite{Lobo:2005us, Anchordoqui:1997du, Cataldo:2017ard, Lobo:2004rp, Delgaty:1994vp, Kuhfittig:2002ur}. Thus, we can perform the following test experiment: let us consider a traveller (initially at rest or with a given velocity) meassured from thw astronouts that live in some \textit{station} at a position $r=r_{1s}$, this traveller wishes to cross the wormhole throat and arrive safely at another station in $r=r_{2s}$, located in the other universe where other astronauts patiently await the traveller's arrival. Where is the minimum radial distance from the wormhole so that the forces experienced by both crews in their respective stations are at most comparable to those experienced on Earth?. What would be a \textit{good choice} of parameters $\mu, b$ so that the tidal forces experienced in the wormhole would not rip apart our traveler? What would be the total traversal time?. In this section, we shall impose several constraints required for the journey to be completed successfully. 
\subsection{Gravitational redshift measured in the stations}
\noindent
The curvature of spacetime affects virtually every measurement performed by one observer relative to another. It is therefore natural to expect that the wavelength of a signal transmitted from one point to another will undergo a gravitational redshift (or, depending on the circumstances, a blueshift in the electromagnetic spectrum). Therefore, any signal sent from either station to an observer located far from the wormhole should be measured with approximately the same wavelength. Consequently, for the \textit{first} region
\begin{equation}\label{redshift grav}
    \frac{\Delta \lambda^{(1)}}{\lambda^{(1)}} \equiv \frac{\lambda^{(1)}_{+\infty} -  \lambda^{(1)}}{\lambda^{(1)}} = \sqrt{\frac{g_{00}(r \to +\infty)}{g_{00}(r)}}  - 1,
\end{equation}
expanding in the weak field limit, we obtain
\begin{equation}\label{redshift felt}
    \left|\frac{\Delta \lambda^{(1)}}{\lambda^{(1)}}\right| \ll 1 \quad \Rightarrow \quad \left|\frac{2(1 + b)}{(D - 2)}\frac{\mu}{r_{1s}^{D - 3}}\right| \ll 1,
\end{equation}
meanwhile, for the other asymptotically flat region, we find 
\begin{equation}
    \begin{aligned}
    &\frac{\Delta \lambda^{(2)}}{\lambda^{(2)}} \equiv \sqrt{\frac{g_{00}(r \to 0)}{g_{00}(r)}}  - 1,\\
    &\left|\frac{\Delta \lambda^{(2)}}{\lambda^{(2)}}\right| \ll 1 \quad \Rightarrow \quad \left|\frac{2(1 + b)}{(D - 2)}\frac{r_{2s}^{D - 3}}{b\mu}\right| \ll 1.
    \end{aligned}
\end{equation}
\subsection{The traveler's journey}
\noindent
As discussed previously, the gravitational acceleration experienced at the stations must be less than or equal to that on Earth, namely $g_{\oplus} \approx 9.8\mathrm{m/s^{2}}$. Since the stations are at rest with respect to the wormhole geometry, the four-velocity of an observer located at either station is given by
\begin{equation}
    U_{s}^{\mu} = (\dot{t}, 0, 0, \ldots, 0),
\end{equation}
where we shall use the normalization condition for a massive particle, i.e.
\begin{equation}
    g_{\mu \nu}U^{\mu}_{s}U^{\nu}_{s} = -1 \quad \Rightarrow \quad \dot{t} = \frac{1}{\sqrt{-g_{00}}},
\end{equation}
from which we obtain an expression for the acceleration experienced by observers located at either station
\begin{align}
&g^{\mu}_{s}  = U^{\nu}_{s}\nabla_{\nu}U^{\mu}_{s}, \quad \quad U^{\mu}_{s} = \frac{1}{\sqrt{A(r)}}(1, 0, 0, \ldots, 0), \notag\\
&g_{s}^{\mu} = \frac{2 \mu (D - 3)}{(D - 2)B r^{D - 2}}\left(\frac{b}{v} + \frac{1}{u}\right)(0, 1, 0, \ldots, 0).\label{aceleracion estaciones}
\end{align}
As before, we expand in the weak field limit, and for simplicity, it is sufficient to consider only the leading-order term. We therefore define $r_{1s}$ as the minimum \textit{safe} distance from the wormhole at which a control station may be constructed in the first asymptotic region. Thus the invariant acceleration $\displaystyle g_{s} = \sqrt{g_{\mu \, s} \,  g^{\mu}_{s}}$, gives
\begin{equation}\label{gfelt}    
    g_{s} \simeq \frac{2 \mu (D - 3)|b + 1|}{(D - 2)r^{D - 2}}\lesssim g_\oplus \quad \Rightarrow \quad r_{1s} \gtrsim \Bigg[\frac{2 \mu (D - 3)|b + 1|c^2}{(D - 2)g_\oplus}\Bigg]^{\frac{1}{D - 2}}.
\end{equation}
For the second asymptotic region, \eqref{aceleracion estaciones} must instead be approximated in the limit $r \to 0$, yielding
\begin{equation}
    r_{2s} \lesssim \Bigg[\frac{g_{\oplus}(D - 2)(-b)^{\frac{D}{D- 2}} \mu^{\frac{D - 1}{D - 3}}}{2(D - 3)|b + 1|c^2}\Bigg]^{\frac{1}{D - 2}}.
\end{equation}
We now turn to the question of how long the complete journey would take for our traveler. For simplicity, let us consider a radial geodesic and assume that the journey begins at $r=r_{1s}$ with an initial velocity $v_{1s}$ measured at station 1 in the traveler's local inertial frame. Therefore, from Eqs. \eqref{conserved quantities} and \eqref{r punto eqn} the 4-velocity of the traveller is given by
\begin{equation}
    U^{\mu}_{\text{Traveller}} = \frac{1}{A(r)}\left(E, \left(1 + \frac{\mu}{r^{D - 3}}\right)^{-\frac{2}{(D - 3)}}\sqrt{E^{2} - A(r)}, 0, \ldots, 0\right), \quad E^{2} > A(r).
\end{equation}
Notice that the proper acceleration experienced by the astronaut during the journey towards the wormhole vanishes identically, suggesting that the astronaut is indeed in free fall, i.e,
\begin{equation}
    a_{\text{Traveller}}^{\mu} = U^{\nu}_{\text{Traveller}}\nabla_{\nu}U^{\mu}_{\text{Traveller}} = 0.
\end{equation}
Now, we can define the Lorentz factor that will link the energy of the traveler to its initial local velocity, yielding
\begin{equation}\label{gamma lorentz}
    \gamma \equiv -g_{\mu \nu}U_{\mathrm{s}}^{\mu}U_{\mathrm{Traveller}}^{\nu} = \frac{E}{\sqrt{A(r)}}, \quad v_{\text{Traveller}} = c\sqrt{1 - \frac{A(r)}{E^{2}}},
\end{equation}
thus
\begin{equation}
    E = \sqrt{\frac{A(r_{1s})}{1 - \frac{v_{1s}^{2}}{c^{2}}}}.
\end{equation}
It is important to mention that the traversability constrain in Eq. \eqref{efective potential} has been replaced by $E^{2} > A$, this restriction will allow us to define a minimum velocity $v_{\mathrm{min}}$ useful for determining the (human) traversability of the wormhole, since the function $A$ is monotonic for all $r$, then the full geodesic will be well defined if and only if
\begin{equation}
\Bigg\{
\begin{aligned}
    &b \in (-1, 0), \quad \Rightarrow \quad E^{2} > A(r_{1s}) \simeq 1, \quad \mathrm{and} \quad v_{\mathrm{min}} = 0,\\
    &b < - 1 \quad \Rightarrow \quad E^{2} > A(r_{2s}) \simeq (-b)^{4/(D - 2)} \quad \mathrm{and} \quad v_{\mathrm{min}} = c\sqrt{1 - (-b)^{-4/(D - 2)}}.
\end{aligned}
\Bigg.
\end{equation}
It may not seem like it, but we have uncover a really important aspect of the wormhole: Even thought at first it was deemed to be traversable for \textit{any negative values of b}, in practice the choice $b < - 1$ is highly dangerous from a human standpoint because as $|b|$ becomes large, then the traveller would require to have tremendous velocities meassured in station 1, for instance let $D = 4$, $b = -1.1$, then $v_{\mathrm{min}} \sim 0.4 c$, almost half the speed of light, which is obiously impossible for our current technology. However, things get more comfortable when dealing with higher-dimensional gravity, as shown in the table below.
\begin{table}[H]
    \centering
    \begin{threeparttable}
        \caption{Minimum initial radial velocity required for the journey.}
        \label{tab:min_initial_velocity}

        \renewcommand{\arraystretch}{1.20}
        \setlength{\tabcolsep}{9pt}

        \begin{tabular}{
            c
            S[table-format=1.3, exponent-mode=input,
              round-mode=places, round-precision=3]
            S[table-format=1.3, exponent-mode=input,
              round-mode=places, round-precision=3]
            S[table-format=1.3, exponent-mode=input,
              round-mode=places, round-precision=3]
            S[table-format=1.3, exponent-mode=input,
              round-mode=places, round-precision=3]
            S[table-format=1.3, exponent-mode=input,
              round-mode=places, round-precision=3]
            S[table-format=1.3, exponent-mode=input,
              round-mode=places, round-precision=3]
        }
            \toprule
            & \multicolumn{6}{c}{$v_{\mathrm{min}}/c$} \\
            \cmidrule(lr){2-7}

            {$D$}
            & {$b=-1.1$}
            & {$b=-1.3$}
            & {$b=-1.5$}
            & {$b=-2$}
            & {$b=-5$}
            & {$b=-10$} \\

            \midrule
            4 & 0.417 & 0.639 & 0.745 & 0.866 & 0.980 & 0.995 \\
            5 & 0.345 & 0.543 & 0.646 & 0.777 & 0.940 & 0.977 \\
            6 & 0.302 & 0.480 &0.577 & 0.707 & 0.894 & 0.949 \\
            7 & 0.271 & 0.435 & 0.526 & 0.652 & 0.851 & 0.917 \\
            8 & 0.248 & 0.400 &0.487&0.608&0.811&0.886\\
            9 &0.230 & 0.373 & 0.455 & 0.572&0.775&0.855\\
            10 &0.216 & 0.350 & 0.428 &0.541&0.743&0.827\\
            \bottomrule
        \end{tabular}
    \end{threeparttable}
\end{table}
\noindent
Now, the time measured by observers at both stations, as well as the traveler's own clock, is given by
\begin{align}\label{t_estacion 1}
    &\Delta \tau_{1s} = \frac{1}{c}\int\limits^{r_{1s}}_{r_{2s}}\left(1 + \frac{\mu}{r^{D - 3}}\right)^{\frac{2}{(D - 3)}}\frac{E\sqrt{A(r_{1s})}}{\sqrt{E^{2} - A(r)}}dr, \qquad \Delta\tau_{2s} = \Delta\tau_{1s}\sqrt{\frac{A(r_{2s})}{A(r_{1s})}},\\
    \label{t_viajero}
    &\Delta\tau^{\text{Traveller}} = \frac{1}{c}\int\limits^{r_{1s}}_{r_{2s}}\left(1 + \frac{\mu}{r^{D - 3}}\right)^{\frac{2}{(D - 3)}}\frac{A(r)}{\sqrt{E^{2} - A(r)}}\, dr.
\end{align}
It is worth pointing out that in $D = 4$, these integrals can be solved exactly in terms of elementary functions via the substitution
\begin{equation}
    x = \frac{r - b\mu}{r + \mu}, \quad x_{1s} \equiv x(r_{1s}) > 0,
\end{equation}
with this in mind, all integrals reduce to a single master integral of the form
\begin{equation}
    \mathcal{W}_{1}(x; a) \equiv \int \frac{dx}{(x - a)\sqrt{E^{2} - x^{2}}}, \quad a = \Bigg\{
    \begin{aligned}
    1\\-b
    \end{aligned}
    \Bigg.
\end{equation}
With the solution
\begin{equation}
    \mathcal{W}_{1}(x;a)=
    \begin{dcases}
        \frac{1}{\sqrt{E^{2}-a^{2}}}
        \ln\left|
        \frac{
        x\sqrt{E^{2}-a^{2}}
        -a\left(\sqrt{E^{2}-x^{2}}-E\right)}
        {
        x\sqrt{E^{2}-a^{2}}
        + a\left(\sqrt{E^{2}-x^{2}}-E\right)}
        \right|
        +\text{const},
        & E^{2}>a^{2},
        \\[1ex]
        -\frac{2}{\sqrt{a^{2}-E^{2}}}
        \tan^{-1}\left(
        \frac{
        E(a-x)-a\sqrt{E^{2}-x^{2}}
        }{
        x\sqrt{a^{2}-E^{2}}
        }
        \right)
        +\text{const},
        & a^{2}>E^{2},
        \\[1ex]
        \frac{2x}{
        E\left(
        E-x-\sqrt{E^{2}-x^{2}}
        \right)}
        +\text{const},
        & a^{2}=E^{2}.
    \end{dcases}
\end{equation}
Before moving on, we also notice that
\begin{equation}
    \frac{d}{dx}\left(\frac{\sqrt{E^{2} - x^{2}}}{x - a}\right) = \frac{a}{(x - a)\sqrt{E^{2} - x^{2}}} + \frac{a^{2} - E^{2}}{(x - a)^{2}\sqrt{E^{2} - x^{2}}}
\end{equation}
Therefore 
\begin{equation}
    \mathcal{W}_{2}(x; a) \equiv \int \frac{dx}{(x - a)^{2}\sqrt{E^{2} - x^{2}}} = \frac{\sqrt{E^{2} - x^{2}}}{(a^{2} - E^{2})(x - a)} - \frac{a}{a^{2} - E^{2}}\mathcal{W}_{1}(x; a) + \text{const,}
\end{equation}
valid if $a^{2} < E^{2}$ and $a^{2} > E^{2}$, for the special case that $a^{2} = E^{2}$, we find
\begin{equation}
     \mathcal{W}_{2} = -\frac{2}{E^{2}f}\left(1 + \frac{1}{f} + \frac{2}{3f^{2}}\right) + \text{const,} \quad f \equiv \frac{E - \sqrt{E^{2} - x^{2}}}{x} - 1.
\end{equation}
So, the proper time becomes
\begin{equation}
    \begin{aligned}
    &\Delta\tau_{1s} = \frac{\mu x_{1s}E}{c} \Bigg[2\Bigg(\mathcal{W}_{1}(x; -b) - \mathcal{W}_{1}(x; 1)\Bigg) + (b + 1)\Bigg(\mathcal{W}_{2}(x; -b) + \mathcal{W}_{2}(x;1)\Bigg)\Bigg]_{x_{2s}}^{x_{1s}},\\
    &\Delta\tau^{\text{Traveller}} = \frac{\mu}{c} \Bigg[2b\Bigg(\mathcal{W}_{1}(x; 1) - \mathcal{W}_{1}(x; -b)\Bigg) + (b + 1)\Bigg(b^{2}\mathcal{W}_{2}(x;-b) + \mathcal{W}_{2}(x;1)\Bigg)\Bigg]_{x_{2s}}^{x_{1s}}.
    \end{aligned}
\end{equation}
Finally, in addition to the inequalities imposed in Eqs.~\eqref{redshift felt}, \eqref{gfelt}, \eqref{t_estacion 1} and \eqref{t_viajero}, a further constraint must be imposed: the gravitational pull between the head and feet of our traveler must remain tolerable for a human being \cite{Morris:1988cz}. Such a condition arises naturally from the geodesic deviation and is evaluated in the traveler’s local frame, reducing to 
\begin{equation}
    \left|R_{\hat{0}^{\prime}\hat{1}^{\prime}\hat{0}^{\prime}\hat{1}^{\prime}}\right| \lesssim \frac{g_{\oplus}}{c^{2}\ell}, \quad \left|R_{\hat{0}^{\prime}\hat{i}^{\prime}\hat{0}^{\prime}\hat{j}^{\prime}}\right| \lesssim \frac{g_{\oplus}}{c^{2}\ell}.
\end{equation}
Taking a Lorentz boost and using Eq. \eqref{gamma lorentz}, we find
\begin{equation}
    R_{\hat{0}^{\prime}\hat{1}^{\prime}\hat{0}^{\prime}\hat{1}^{\prime}} = R_{\hat{t}\hat{r}\hat{t}\hat{r}}, \quad R_{\hat{0}^{\prime}\hat{i}^{\prime}\hat{0}^{\prime}\hat{j}^{\prime}} = \gamma^{2}\left(R_{\hat{t}\hat{i}\hat{t}\hat{j}} + \frac{v^{2}_{\mathrm{Traveller}}}{c^{2}}R_{\hat{r}\hat{i}\hat{r}\hat{j}}\right).
\end{equation}
Therefore
\begin{align}
    &\mathcal{F}_{\|} \equiv \frac{2\mu(D - 3)}{B(D - 2)r^{D - 1}}\Bigg|-(D - 2)\left(\frac{b}{v^{2}} + \frac{1}{u^{2}}\right) + \frac{\mu}{r^{D - 3}}\left(\frac{b}{v} + \frac{1}{u}\right)^{2}\Bigg|\frac{c^{2}\ell}{g_{\oplus}} \lesssim 1, \\ \notag\\
    &\mathcal{F}_{\perp} \equiv \frac{2\mu(D - 3)E^{2}}{AC(D - 2)r^{D - 3}}\Bigg|\left(\frac{b}{v} + \frac{1}{u}\right)\left[1 + \frac{2\mu}{(D - 2)r^{D - 3}}\left(\frac{b(D - 3)}{v} - \frac{1}{u}\right)\right] +\\
    &\left(1 - \frac{A}{E^{2}}\right)\left(\frac{b(D - 3)}{v^{2}} - \frac{1}{u^{2}}\right)\Bigg|\frac{c^{2}\ell}{g_{\oplus}} \lesssim 1.
\end{align}
\begin{table}[H]
    \centering
    \caption{Maximum radial and transverse tidal constraints experienced
    by a radial traveler for different choices of $D$ and $b$.
    We fix $\mu^{1/(D-3)}=5\times10^{7}\,\mathrm{m}$ and
    $\ell=2\,\mathrm{m}$. For $b \in (-1, 0)$, the traveler is
    released from rest at Station 1, whereas for $b=-1.1$ we take
    $v_{1s}/c=0.417$.}
    \label{tab:tidal_maxima}
    
    \renewcommand{\arraystretch}{1.2}
    \setlength{\tabcolsep}{5.5pt}

    \begin{tabular}{c cc cc cc cc}
        \toprule
        & \multicolumn{2}{c}{$b=-0.2$}
        & \multicolumn{2}{c}{$b=-0.5$}
        & \multicolumn{2}{c}{$b=-0.8$}
        & \multicolumn{2}{c}{$b=-1.1$} \\
        
        \cmidrule(lr){2-3}
        \cmidrule(lr){4-5}
        \cmidrule(lr){6-7}
        \cmidrule(lr){8-9}

        $D$
        & {$\max(\mathcal F_{\parallel})$}
        & {$\max(\mathcal F_{\perp})$}
        & {$\max(\mathcal F_{\parallel})$}
        & {$\max(\mathcal F_{\perp})$}
        & {$\max(\mathcal F_{\parallel})$}
        & {$\max(\mathcal F_{\perp})$}
        & {$\max(\mathcal F_{\parallel})$}
        & {$\max(\mathcal F_{\perp})$} \\
        \midrule

        4
        & 0.437 & 5.250
        & 0.084 & 0.551
        & 0.016 & 0.085
        & 0.005 & 0.024 \\

        5
        & 2.658 & 15.424
        & 0.639 & 2.286
        & 0.142 & 0.428
        & 0.048 & 0.252 \\

        6
        & 5.211 & 19.060
        & 1.442 & 3.537
        & 0.350 & 0.742
        & 0.125 & 0.670 \\

        7
        & 7.670 & 20.220
        & 2.331 & 4.341
        & 0.599 & 0.979
        & 0.222 & 1.201 \\
        8
        & 10.025 & 20.558
        & 3.254 & 4.879
        & 0.869 & 1.156
        & 0.329 & 1.800 \\
        9
        & 12.304& 20.602
        & 4.193 & 5.260
        & 1.151 & 1.292
        & 0.444 & 2.441 \\
        10
        & 14.531 & 20.534
        & 5.141 & 5.541
        & 1.442 & 1.399
        & 0.563 & 3.111 \\

        \bottomrule
    \end{tabular}
\end{table}
\noindent
Also notice that if we somehow manage to build a spacecraft that travells at 40-50 \% the speed of light (or more if we are extremely optimistic and very unrealistic), as in table \ref{tab:min_initial_velocity}, for allowing larger values of $|b|$, we find that for a fixed scale parameter $\mu$, the tidal forces becomes considerable more \textit{gentle}, on the other hand we observe that in higher dimensional gravity, the forces are larger than in the familiar $D = 4$. We finish this section by summarizing our results for $D = 4$ and $b \in (-1, 0)$ in the table below, where it is shown that all the traversability conditions are satisfied and therefore the wormhole can be humanly traversable.
\begin{table}[H]
    \centering
    \caption{Physical properties of the wormhole configuration, for $D = 4$, $\mu = 3.235\times10^{7}$ m, $b = -0.8$, also $r_{T} = 2.893\times10^{7}$.}
    \label{tab:wormhole_properties}
    \renewcommand{\arraystretch}{1.5}
    \begin{tabular}{lccc}
        \hline
        \textbf{Property}
        & \textbf{Station 1}
        & \textbf{Station 2}
        & \textbf{Traveller}
        \\
        \hline

        $r_{s}$
        &
        $2.438 \times 10^{11}\,\mathrm{m}$
        &
        $2.893 \times 10^{3} \, \mathrm{m}$
        &
        ---
        \\

        $\Delta\lambda/\lambda$
        &
        $2.654\times10^{-5}$
        &
        $2.654 \times 10^{-5}$
        &
        ---
        \\

        $\text{max}(\mathcal{F}_{\|})$
        &
        ---
        &
        ---
        &
        $0.038
$
        \\

        $\text{max}(\mathcal{F_{\perp}})$
        &
        ---
        &
        ---
        &
        $0.202$
        \\

        $\Delta \tau$
        &
        $1.038 \, \mathrm{days}$
        &
        $0.830 \, \mathrm{days}$
        &
        $1.030 \, \mathrm{days}$
        \\

        $v_{\mathrm{Traveller}}$
        &
        0.054 c
        &
        0.600 c
        &
        ---
        \\

        \hline
    \end{tabular}
\end{table}

\section{Conclusions}
In the present work, we have analyzed a static, spherically symmetric solution of the conformally coupled equations of gravity that has black holes and naked singularities as a subfamily. It was shown that by a suitable choice of the free parameters $\mu$ and $b$, the line element represents a family of traversable wormholes which matches in $D = 4$ to the Barceló-Visser wormhole. This spacetime connects two asymptotically flat regions of spacetime bounded together by a minimal surface known as the wormhole throat. Naturally, we show that such a spacetime allows violations of the energy conditions and, therefore, the matter content threading the throat is exotic.

\medskip

\noindent
In section three, we have analytically classified all types of null trajectories using an effective potential. Furthermore, we have integrated them numerically and visualized them using the embedding diagrams constructed in section two. Furthermore, we have found that the photon sphere lies at $r = r_{c}$, which serves as a two-way membrane allowing/forbidding the wormhole travel. This analysis paves the way for interpreting the wormhole geometry as a potential barrier where particles can be trapped, travel through, and bounce back, depending on their energy and angular momentum. Motivated by this reason, in section four, we make use of a probe scalar field, satisfying the massless Klein-Gordon equation. By a suitable change of coordinates, we derived a Schr\"odinger-like equation whose potential is manifestly symmetric with respect to the tortoise coordinate and found an isospectrality between the wormhole and the ERN black hole, thus making in $D = 4$ the Barceló-Visser wormhole indistinguishable from the BBMB/ERN black hole for a scalar probe. It would be interesting to see whether this feature breaks when other fields are allowed to perturb the wormhole geometry. Following this route, we can observe that every semianalytical/numerical method for obtaining the QNMs converges to the eikonal limit for large $\mathscr{L}$ and the WKB approximation of sixth order becomes indistinguishable from the Chebyshev pseudospectral method when the multipole number is greater than the overtone.

\medskip
\noindent
At first, the wormhole was deemed to be traversable for any negative values of $b$, in practice the forbidden value $b = -1$ acts as a threshold for human traversability, because we have found that for $b < -1$, in order to traverse the wormhole and not being reflected, we would require travelling at considerable fractions of the speed of light, which seems unlikely even with the most optimistic aspirations.

\medskip

\noindent
Even though we have no experimental evidence for the existence of such objects to date, they are incredibly rich structures that are compatible with the laws of physics as we know them and therefore offer many interesting properties worth studying. 

\section*{Acknowledgments}
The author want to thank Dr. Sourya Ray for useful discussions and insights throughout this work.

\newpage
\section*{A. WKB corrections and QNMs for $D > 4$}
In this appendix, we explicitly give the analytical corrections for the WKB series. Recalling the definitions,
\begin{equation}
    q \equiv D - 3,
    \qquad
    \alpha \equiv n + \frac{1}{2},
    \qquad
    \mathscr{L} \equiv l + \frac{D-3}{2}.
\end{equation}
We define the derivatives of the effective potential for the scalar perturbation evaluated at
the maximum of the potential barrier as
\begin{equation}
    V_k \equiv
    \left.
    \frac{d^k \mathcal{V}_{\mathrm{eff}}}
    {d r_*^k}
    \right|_{r=r_c},
    \qquad
    r_c = \mu^{1/(D-3)}.
\end{equation}
Owing to the exact symmetry found in Eq. \eqref{v symmetric}, all odd derivatives vanish identically at the maximum,
\begin{equation}
    V_{2j+1}=0,
    \qquad
    j=0,1,2,\ldots.
\end{equation}
Meanwhile, the even derivatives are given by
\begin{equation}
\begin{aligned}
    V_{0} = &\frac{\mathscr{L}^{2} + \frac{q}{4}}{(4\mu)^{2/q}}\\
    V_{2} = &- \frac{q \left(8 \mathscr{L}^{2} + q^{2} + 3 q\right)}{2^{3 + \frac{8}{q}} \mu^{\frac{4}{q}}}\\
    V_{4} = &\frac{q^{2}\left(4 \mathscr{L}^{2} q + 40 \mathscr{L}^{2} + 2 q^{3} + 11 q^{2} + 18 q\right)}{2^{3 + \frac{12}{q}} \mu^{\frac{6}{q}}}\\
    V_{6} = &- \frac{q^{3}\left(16 \mathscr{L}^{2} q^{2} + 224 \mathscr{L}^{2} q + 928 \mathscr{L}^{2} + 17 q^{4} + 139 q^{3} + 410 q^{2} + 468 q\right)}{2^{4 + \frac{16}{q}} \mu^{\frac{8}{q}}}\\
    V_{8} = &\frac{q^{4}}{2^{4 + \frac{20}{q}} \mu^{\frac{10}{q}}}\left(68 \mathscr{L}^{2} q^{3} + 1212 \mathscr{L}^{2} q^{2} + 7824 \mathscr{L}^{2} q + 19280 \mathscr{L}^{2} + 124 q^{5} + 1353 q^{4} + 5811 q^{3}\right.\\
    +&\left.11980 q^{2} + 10548 q\right)\\
    V_{10} = &- \frac{q^{5}}{2^{4 + \frac{24}{q}} \mu^{ \frac{12}{q}}}\left(496 \mathscr{L}^{2} q^{4} + 10648 \mathscr{L}^{2} q^{3} + 90192 \mathscr{L}^{2} q^{2} + 366784 \mathscr{L}^{2} q + 626944 \mathscr{L}^{2} + 1382 q^{6}\right.\\
    &\left.+ 18887 q^{5} + 106917 q^{4} + 317870 q^{3} + 508592 q^{2} + 365184 q\right)\\
    V_{12} = &\frac{q^{6}}{2^{5 + \frac{28}{q}} \mu^{\frac{14}{q}}} \left(11056 \mathscr{L}^{2} q^{5} + 276068 \mathscr{L}^{2} q^{4} + 2856772 \mathscr{L}^{2} q^{3} + 15554344 \mathscr{L}^{2} q^{2} + 45475184 \mathscr{L}^{2}q \right.\\
    &+ \left.58708000 \mathscr{L}^{2} + 43688 q^{7} + 717778 q^{6} + 5050625 q^{5} + 19643771 q^{4} + 45124642 q^{3} +\right.\\
    &\left.+ 59278964 q^{2} + 35970408 q\right)
\end{aligned}
\end{equation}
As well as the corrections
\begin{equation}
    \begin{aligned}
        &\mathcal{A}_{0} = V_{0}, \quad \mathcal{A}_{1} = \sqrt{-2V_{2}}, \quad \mathcal{A}_{2}=\frac{1+4\alpha^2}{32}\frac{V_4}{V_2}\quad \mathcal{A}_3 =-\frac{67+68\alpha^2}{4608}\frac{V_4^2}{V_2^3}+\frac{5+4\alpha^2}{576}\frac{V_6}{V_{2}^2}\\
        &\mathcal{A}_4 = \frac{2000\alpha^{4} + 4552\alpha^{2} + 513}{294912}\frac{V_4^{3}}{V_2{^4}} -\frac{176\alpha^4+472\alpha^{2} + 63}{36864}
        \frac{V_4V_6}{V_2^3} +\frac{16\alpha^4+56\alpha^2+9}{36864}\frac{V_8}{V_2^2}\\
        &\mathcal{A}_{5}
        =\frac{16\alpha^{4}+120\alpha^{2}+89}{1843200}\frac{V_{10}}{V_{2}^{3}}-\frac{171024\alpha^4+713320\alpha^2+305141}{84934656}\frac{V_4^4}{V_2^6} +\frac{50064\alpha^4+236440\alpha^2+117281}{26542080}\frac{V_4^2V_6}{V_2^5}\nonumber\\
        &-\frac{6288\alpha^{4}+33160\alpha^{2}+19277}{33177600}\frac{V_{6}^{2}}{V_{2}^{4}}-\frac{48\alpha^{4}+280\alpha^{2}+167}{245760}\frac{V_{4}V_{8}}{V_{2}^{4}}
    \end{aligned}
\end{equation}

\begin{align}
    \mathcal{A}_6=&\frac{(4\alpha^2+9)\left(16\alpha^4+184\alpha^2+25\right)}{265420800}\frac{V_{12}}{V_2^3}-\frac{256\alpha^6+2800\alpha^4+4504\alpha^2+495}{26542080}
    \frac{V_{10}V_4}{V_2^4}
    \nonumber\\
    &+
    \frac{
        933856\alpha^6
        +6264160\alpha^4
        +6211966\alpha^2
        +485523
    }{679477248}
    \frac{V_4^5}{V_2^7}
    \nonumber\\
    &-
    \frac{
        170576\alpha^6
        +1263960\alpha^4
        +1419769\alpha^2
        +122535
    }{106168320}
    \frac{V_4^3V_6}{V_2^6}
    \nonumber\\
    &+
    \frac{
        89824\alpha^6
        +728480\alpha^4
        +927246\alpha^2
        +89775
    }{265420800}
    \frac{V_4V_6^2}{V_2^5}
    \nonumber\\
    &+
    \frac{
        36992\alpha^6
        +326320\alpha^4
        +433728\alpha^2
        +41715
    }{212336640}
    \frac{V_4^2V_8}{V_2^5}
    \nonumber\\
    &-
    \frac{
        8128\alpha^6
        +77360\alpha^4
        +117012\alpha^2
        +12825
    }{265420800}
    \frac{V_{6}V_{8}}{V_2^4}
\end{align}

\begin{table}[H]
    \centering
    \caption{
        Comparison of the fundamental QNMs for $D=5$ and $\mu=1$, obtained using the eikonal
        approximation, first, third, and sixth order WKB methods,
        and the Chebyshev pseudospectral method.
    }
    \label{tab:qnm_D5_n0}

    \resizebox{\textwidth}{!}{%
    \begin{tabular}{c c c c c c}
        \toprule
        $l$
        & $\omega_{\mathrm{eik}}$
        & $\omega_{\mathrm{WKB\,I}}$
        & $\omega_{\mathrm{WKB\,III}}$
        & $\omega_{\mathrm{WKB\,VI}}$
        & $\omega_{\mathrm{Cheb}}$
        \\
        \midrule

        0  & $0.500000000 - 0.250000000\,i$
           & $0.672803867 - 0.278684486\,i$
           & $0.446270570 - 0.272318602\,i$
           & $0.521785858 - 0.229929289\,i$
           & $0.498509197 - 0.263182695\,i$ \\

        1  & $1.000000000 - 0.250000000\,i$
           & $1.092574509 - 0.262143203\,i$
           & $0.986949088 - 0.251702636\,i$
           & $1.000178538 - 0.254492365\,i$
           & $0.999643361 - 0.253669035\,i$ \\

        2  & $1.500000000 - 0.250000000\,i$
           & $1.562248733 - 0.256166164\,i$
           & $1.495576571 - 0.250935830\,i$
           & $1.499736171 - 0.252060685\,i$
           & $1.499876191 - 0.251682071\,i$ \\

        3  & $2.000000000 - 0.250000000\,i$
           & $2.046787102 - 0.253648262\,i$
           & $1.998044083 - 0.250663785\,i$
           & $1.999901770 - 0.251061879\,i$
           & $1.999944343 - 0.250958370\,i$ \\

        4  & $2.500000000 - 0.250000000\,i$
           & $2.537459694 - 0.252391959\,i$
           & $2.498976771 - 0.250483626\,i$
           & $2.499958097 - 0.250650015\,i$
           & $2.499970591 - 0.250617306\,i$ \\

        5  & $3.000000000 - 0.250000000\,i$
           & $3.031228219 - 0.251683368\,i$
           & $2.999400941 - 0.250362082\,i$
           & $2.999978518 - 0.250442340\,i$
           & $2.999982674 - 0.250430248\,i$ \\

        6  & $3.500000000 - 0.250000000\,i$
           & $3.526772600 - 0.251246828\,i$
           & $3.499620115 - 0.250278758\,i$
           & $3.499987401 - 0.250321899\,i$
           & $3.499988968 - 0.250316814\,i$ \\

        7  & $4.000000000 - 0.250000000\,i$
           & $4.023428980 - 0.250959668\,i$
           & $3.999744359 - 0.250220120\,i$
           & $3.999991897 - 0.250245292\,i$
           & $3.999992556 - 0.250242921\,i$ \\

        8  & $4.500000000 - 0.250000000\,i$
           & $4.520827479 - 0.250761023\,i$
           & $4.499819902 - 0.250177686\,i$
           & $4.499994443 - 0.250193336\,i$
           & $4.499994745 - 0.250192136\,i$ \\

        9  & $5.000000000 - 0.250000000\,i$
           & $5.018745800 - 0.250618040\,i$
           & $4.999868420 - 0.250146162\,i$
           & $4.999996006 - 0.250156395\,i$
           & $4.999996155 - 0.250155746\,i$ \\

        10 & $5.500000000 - 0.250000000\,i$
           & $5.517042337 - 0.250511767\,i$
           & $5.499900981 - 0.250122189\,i$
           & $5.499997025 - 0.250129158\,i$
           & $5.499997103 - 0.250128787\,i$ \\

        \bottomrule
    \end{tabular}%
    }
\end{table}

\begin{table}[H]
    \centering
    \caption{
        Comparison of the fundamental QNMs for $D=6$ and $\mu=1$, obtained using the eikonal
        approximation, first, third, and sixth order WKB methods,
        and the Chebyshev pseudospectral method.
    }
    \label{tab:qnm_D6_n0}

    \resizebox{\textwidth}{!}{%
    \begin{tabular}{c c c c c c}
        \toprule
        $l$
        & $\omega_{\mathrm{eik}}$
        & $\omega_{\mathrm{WKB\,I}}$
        & $\omega_{\mathrm{WKB\,III}}$
        & $\omega_{\mathrm{WKB\,VI}}$
        & $\omega_{\mathrm{Cheb}}$
        \\
        \midrule

        0  & $0.944940787 - 0.385770461\,i$
           & $1.175904215 - 0.438406128\,i$
           & $0.820147795 - 0.389584953\,i$
           & $0.995547224 - 0.339068170\,i$
           & $0.922980662 - 0.396614915\,i$ \\

        1  & $1.574901312 - 0.385770461\,i$
           & $1.717039250 - 0.412640211\,i$
           & $1.530022470 - 0.381000311\,i$
           & $1.563309203 - 0.396363189\,i$
           & $1.562360267 - 0.389996644\,i$ \\

        2  & $2.204861837 - 0.385770461\,i$
           & $2.306517138 - 0.401208035\,i$
           & $2.182943785 - 0.384165145\,i$
           & $2.194366503 - 0.391232972\,i$
           & $2.196109372 - 0.388012180\,i$ \\

        3  & $2.834822362 - 0.385770461\,i$
           & $2.913823098 - 0.395612845\,i$
           & $2.821619530 - 0.385418240\,i$
           & $2.827417004 - 0.388277941\,i$
           & $2.828097523 - 0.387155855\,i$ \\

        4  & $3.464782887 - 0.385770461\,i$
           & $3.529369335 - 0.392542676\,i$
           & $3.455692925 - 0.385841685\,i$
           & $3.459075418 - 0.387128329\,i$
           & $3.459319702 - 0.386709864\,i$ \\

        5  & $4.094743412 - 0.385770461\,i$
           & $4.149362130 - 0.390697806\,i$
           & $4.087916948 - 0.385974475\,i$
           & $4.090047423 - 0.386622117\,i$
           & $4.090141282 - 0.386448630\,i$ \\

        6  & $4.724703937 - 0.385770461\,i$
           & $4.772020803 - 0.389509371\,i$
           & $4.719268325 - 0.386004270\,i$
           & $4.720687901 - 0.386361952\,i$
           & $4.720727233 - 0.386282700\,i$ \\

        7  & $5.354664462 - 0.385770461\,i$
           & $5.396402319 - 0.388701416\,i$
           & $5.350155668 - 0.385997479\,i$
           & $5.351144964 - 0.386210144\,i$
           & $5.351162841 - 0.386170842\,i$ \\

        8  & $5.984624987 - 0.385770461\,i$
           & $6.021961262 - 0.388128215\,i$
           & $5.980772728 - 0.385978689\,i$
           & $5.981487901 - 0.386112786\,i$
           & $5.981496625 - 0.386091914\,i$ \\

        9  & $6.614585512 - 0.385770461\,i$
           & $6.648360412 - 0.387707331\,i$
           & $6.611220764 - 0.385957247\,i$
           & $6.611753703 - 0.386045910\,i$
           & $6.611758230 - 0.386034172\,i$ \\

        10 & $7.244546037 - 0.385770461\,i$
           & $7.275380096 - 0.387389425\,i$
           & $7.241556945 - 0.385936653\,i$
           & $7.241964317 - 0.385997598\,i$
           & $7.241966792 - 0.385990670\,i$ \\

        \bottomrule
    \end{tabular}%
    }
\end{table}

\begin{table}[H]
    \centering
    \caption{
        Comparison of the fundamental QNMs for $D=7$ and $\mu=1$, obtained using the eikonal
        approximation, first, third, and sixth order WKB methods,
        and the Chebyshev pseudospectral method.
    }
    \label{tab:qnm_D7_n0}

    \resizebox{\textwidth}{!}{%
    \begin{tabular}{c c c c c c}
        \toprule
        $l$
        & $\omega_{\mathrm{eik}}$
        & $\omega_{\mathrm{WKB\,I}}$
        & $\omega_{\mathrm{WKB\,III}}$
        & $\omega_{\mathrm{WKB\,VI}}$
        & $\omega_{\mathrm{Cheb}}$
        \\
        \midrule

        0  & $1.414213562 - 0.500000000\,i$
           & $1.682598832 - 0.575446635\,i$
           & $1.211362295 - 0.466260238\,i$
           & $1.516624693 - 0.429530587\,i$
           & $1.369686734 - 0.503996334\,i$ \\

        1  & $2.121320344 - 0.500000000\,i$
           & $2.301105163 - 0.543217242\,i$
           & $2.032061167 - 0.478824169\,i$
           & $2.092213307 - 0.522614973\,i$
           & $2.091757010 - 0.501753279\,i$ \\

        2  & $2.828427125 - 0.500000000\,i$
           & $2.962717092 - 0.526965434\,i$
           & $2.778657807 - 0.490071398\,i$
           & $2.798373593 - 0.513387718\,i$
           & $2.806287305 - 0.500981917\,i$ \\

        3  & $3.535533906 - 0.500000000\,i$
           & $3.642594293 - 0.518163281\,i$
           & $3.503070606 - 0.495164499\,i$
           & $3.513962014 - 0.505846165\,i$
           & $3.517834281 - 0.500627178\,i$ \\

        4  & $4.242640687 - 0.500000000\,i$
           & $4.331645045 - 0.512980307\,i$
           & $4.219157305 - 0.497484565\,i$
           & $4.226245542 - 0.502667005\,i$
           & $4.227896565 - 0.500435074\,i$ \\

        5  & $4.949747468 - 0.500000000\,i$
           & $5.025912868 - 0.509705950\,i$
           & $4.931535448 - 0.498611370\,i$
           & $4.936392431 - 0.501341302\,i$
           & $4.937112535 - 0.500319441\,i$ \\

        6  & $5.656854249 - 0.500000000\,i$
           & $5.723423278 - 0.507517506\,i$
           & $5.642032429 - 0.499195945\,i$
           & $5.645468406 - 0.500747117\,i$
           & $5.645800325 - 0.500244470\,i$ \\

        7  & $6.363961031 - 0.500000000\,i$
           & $6.423085179 - 0.505987374\,i$
           & $6.351474904 - 0.499517245\,i$
           & $6.353973973 - 0.500456939\,i$
           & $6.354136322 - 0.500193107\,i$ \\

        8  & $7.071067812 - 0.500000000\,i$
           & $7.124247433 - 0.504877698\,i$
           & $7.060278237 - 0.499702611\,i$
           & $7.062142268 - 0.500302910\,i$
           & $7.062226219 - 0.500156385\,i$ \\

        9  & $7.778174593 - 0.500000000\,i$
           & $7.826497606 - 0.504048383\,i$
           & $7.768669358 - 0.499813928\,i$
           & $7.770091579 - 0.500214656\,i$
           & $7.770137216 - 0.500129224\,i$ \\

        10 & $8.485281374 - 0.500000000\,i$
           & $8.529561800 - 0.503412849\,i$
           & $8.476780948 - 0.499883016\,i$
           & $8.477888151 - 0.500160514\,i$
           & $8.477914082 - 0.500108572\,i$ \\

        \bottomrule
    \end{tabular}%
    }
\end{table}

\begin{table}[H]
    \centering
    \caption{
        Comparison of the fundamental QNMs for $D=8$ and $\mu=1$, obtained using the eikonal
        approximation, first, third, and sixth order WKB methods,
        and the Chebyshev pseudospectral method.
    }
    \label{tab:qnm_D8_n0}

    \resizebox{\textwidth}{!}{%
    \begin{tabular}{c c c c c c}
        \toprule
        $l$
        & $\omega_{\mathrm{eik}}$
        & $\omega_{\mathrm{WKB\,I}}$
        & $\omega_{\mathrm{WKB\,III}}$
        & $\omega_{\mathrm{WKB\,VI}}$
        & $\omega_{\mathrm{Cheb}}$
        \\
        \midrule

        0  & $1.894645708 - 0.599139580\,i$
           & $2.188989824 - 0.695742496\,i$
           & $1.611368931 - 0.513046097\,i$
           & $2.072495211 - 0.513663523\,i$
           & $1.827730220 - 0.593738320\,i$ \\

        1  & $2.652503991 - 0.599139580\,i$
           & $2.861479332 - 0.659052404\,i$
           & $2.510297384 - 0.551091599\,i$
           & $2.602221606 - 0.643900721\,i$
           & $2.603793794 - 0.595748997\,i$ \\

        2  & $3.410362275 - 0.599139580\,i$
           & $3.571638236 - 0.638821082\,i$
           & $3.324400673 - 0.573717702\,i$
           & $3.349326960 - 0.629364351\,i$
           & $3.372043265 - 0.596829252\,i$ \\

        3  & $4.168220558 - 0.599139580\,i$
           & $4.299436925 - 0.627023750\,i$
           & $4.109444927 - 0.584941860\,i$
           & $4.123341624 - 0.613288754\,i$
           & $4.136638660 - 0.597471002\,i$ \\

        4  & $4.926078841 - 0.599139580\,i$
           & $5.036685485 - 0.619686573\,i$
           & $4.882354153 - 0.590672357\,i$
           & $4.892694448 - 0.605445973\,i$
           & $4.899222957 - 0.597881477\,i$ \\

        5  & $5.683937124 - 0.599139580\,i$
           & $5.779544031 - 0.614858926\,i$
           & $5.649436456 - 0.593774450\,i$
           & $5.657407697 - 0.601929215\,i$
           & $5.660580342 - 0.598159047\,i$ \\

        6  & $6.441795408 - 0.599139580\,i$
           & $6.525996826 - 0.611530888\,i$
           & $6.413402910 - 0.595556756\,i$
           & $6.419543238 - 0.600335022\,i$
           & $6.421133616 - 0.598355055\,i$ \\

        7  & $7.199653691 - 0.599139580\,i$
           & $7.274888992 - 0.609147033\,i$
           & $7.175555662 - 0.596637570\,i$
           & $7.180299959 - 0.599592556\,i$
           & $7.181131181 - 0.598498353\,i$ \\

        8  & $7.957511974 - 0.599139580\,i$
           & $8.025512397 - 0.607384506\,i$
           & $7.936578529 - 0.597324738\,i$
           & $7.940275254 - 0.599239506\,i$
           & $7.940728656 - 0.598606137\,i$ \\

        9  & $8.715370257 - 0.599139580\,i$
           & $8.777408923 - 0.606046374\,i$
           & $8.696857699 - 0.597779950\,i$
           & $8.699771036 - 0.599071157\,i$
           & $8.700028523 - 0.598689154\,i$ \\

        10 & $9.473228541 - 0.599139580\,i$
           & $9.530269117 - 0.605007432\,i$
           & $9.456624748 - 0.598092447\,i$
           & $9.458949313 - 0.598993287\,i$
           & $9.459101034 - 0.598754396\,i$ \\

        \bottomrule
    \end{tabular}%
    }
\end{table}

\begin{table}[H]
    \centering
    \caption{
        Comparison of the fundamental QNMs for $D=9$ and $\mu=1$, obtained using the eikonal
        approximation, first, third, and sixth order WKB methods,
        and the Chebyshev pseudospectral method.
    }
    \label{tab:qnm_D9_n0}

    \resizebox{\textwidth}{!}{%
    \begin{tabular}{c c c c c c}
        \toprule
        $l$
        & $\omega_{\mathrm{eik}}$
        & $\omega_{\mathrm{WKB\,I}}$
        & $\omega_{\mathrm{WKB\,III}}$
        & $\omega_{\mathrm{WKB\,VI}}$
        & $\omega_{\mathrm{Cheb}}$
        \\
        \midrule

        0  & $2.381101578 - 0.687364818\,i$
           & $2.694487647 - 0.803541017\,i$
           & $2.016894265 - 0.536748847\,i$
           & $2.657365866 - 0.599296416\,i$
           & $2.292623903 - 0.671041481\,i$ \\

        1  & $3.174802104 - 0.687364818\,i$
           & $3.407000648 - 0.763769748\,i$
           & $2.973698000 - 0.602197606\,i$
           & $3.100750593 - 0.768365337\,i$
           & $3.106237650 - 0.676733581\,i$ \\

        2  & $3.968502630 - 0.687364818\,i$
           & $4.152352701 - 0.740323606\,i$
           & $3.839687468 - 0.638653013\,i$
           & $3.862282164 - 0.748133192\,i$
           & $3.912519918 - 0.679885544\,i$ \\

        3  & $4.762203156 - 0.687364818\,i$
           & $4.914304169 - 0.725855203\,i$
           & $4.671026799 - 0.658019242\,i$
           & $4.680994050 - 0.718945405\,i$
           & $4.714909431 - 0.681820204\,i$ \\

        4  & $5.555903682 - 0.687364818\,i$
           & $5.685623348 - 0.716454010\,i$
           & $5.486646419 - 0.668670589\,i$
           & $5.496277077 - 0.702544041\,i$
           & $5.514975167 - 0.683094161\,i$ \\

        5  & $6.349604208 - 0.687364818\,i$
           & $6.462707950 - 0.710056095\,i$
           & $6.294248972 - 0.674829724\,i$
           & $6.303578582 - 0.694396849\,i$
           & $6.313539956 - 0.683977446\,i$ \\

        6  & $7.143304734 - 0.687364818\,i$
           & $7.243584232 - 0.705527613\,i$
           & $7.097374216 - 0.678574820\,i$
           & $7.105704923 - 0.690409951\,i$
           & $7.111078317 - 0.684614671\,i$ \\

        7  & $7.937005260 - 0.687364818\,i$
           & $8.027085338 - 0.702214883\,i$
           & $7.897809828 - 0.680959513\,i$
           & $7.904903600 - 0.688443552\,i$
           & $7.907883580 - 0.685089169\,i$ \\

        8  & $8.730705786 - 0.687364818\,i$
           & $8.812478485 - 0.699723327\,i$
           & $8.696530694 - 0.682541520\,i$
           & $8.702440540 - 0.687468361\,i$
           & $8.704146989 - 0.685451768\,i$ \\

        9  & $9.524406312 - 0.687364818\,i$
           & $9.599280593 - 0.697804783\,i$
           & $9.494103380 - 0.683629540\,i$
           & $9.498989547 - 0.686990041\,i$
           & $9.499998612 - 0.685734918\,i$ \\

        10 & $10.318106838 - 0.687364818\,i$
           & $10.387160326 - 0.696297442\,i$
           & $10.290875184 - 0.684401837\,i$
           & $10.294914844 - 0.686766097\,i$
           & $10.295530005 - 0.685960127\,i$ \\

        \bottomrule
    \end{tabular}%
    }
\end{table}

\begin{table}[H]
    \centering
    \caption{
        Comparison of the fundamental QNMs for $D=10$ and $\mu=1$, obtained using the eikonal
        approximation, first, third, and sixth order WKB methods,
        and the Chebyshev pseudospectral method.
    }
    \label{tab:qnm_D10_n0}

    \resizebox{\textwidth}{!}{%
    \begin{tabular}{c c c c c c}
        \toprule
        $l$
        & $\omega_{\mathrm{eik}}$
        & $\omega_{\mathrm{WKB\,I}}$
        & $\omega_{\mathrm{WKB\,III}}$
        & $\omega_{\mathrm{WKB\,VI}}$
        & $\omega_{\mathrm{Cheb}}$
        \\
        \midrule

        0  & $2.871173746 - 0.767353461\,i$
           & $3.199121475 - 0.901707750\,i$
           & $2.426406007 - 0.541801104\,i$
           & $3.267927152 - 0.691594917\,i$
           & $2.762117945 - 0.739157799\,i$ \\

        1  & $3.691509102 - 0.767353461\,i$
           & $3.942610035 - 0.859809149\,i$
           & $3.427327458 - 0.635290125\,i$
           & $3.591710699 - 0.902232558\,i$
           & $3.603105945 - 0.747898581\,i$ \\

        2  & $4.511844458 - 0.767353461\,i$
           & $4.714827492 - 0.833783665\,i$
           & $4.334850137 - 0.687348246\,i$
           & $4.343445542 - 0.877122490\,i$
           & $4.437532849 - 0.753101734\,i$ \\

        3  & $5.332179814 - 0.767353461\,i$
           & $5.502478292 - 0.816984160\,i$
           & $5.203379394 - 0.716533995\,i$
           & $5.196806436 - 0.830194161\,i$
           & $5.268108782 - 0.756463008\,i$ \\

        4  & $6.152515170 - 0.767353461\,i$
           & $6.299222805 - 0.805668891\,i$
           & $6.052979754 - 0.733502948\,i$
           & $6.052945095 - 0.800389089\,i$
           & $6.096225061 - 0.758763970\,i$ \\

        5  & $6.972850526 - 0.767353461\,i$
           & $7.101739635 - 0.797746026\,i$
           & $6.892425971 - 0.743805593\,i$
           & $6.897735377 - 0.784115197\,i$
           & $6.922672576 - 0.760409157\,i$ \\

        6  & $7.793185882 - 0.767353461\,i$
           & $7.908140558 - 0.792008976\,i$
           & $7.725984945 - 0.750333805\,i$
           & $7.733593901 - 0.775488649\,i$
           & $7.747935241 - 0.761626221\,i$ \\

        7  & $8.613521238 - 0.767353461\,i$
           & $8.717277976 - 0.787733784\,i$
           & $8.555907545 - 0.754636982\,i$
           & $8.563941530 - 0.770913523\,i$
           & $8.572326158 - 0.762551662\,i$ \\

        8  & $9.433856594 - 0.767353461\,i$
           & $9.528415630 - 0.784468934\,i$
           & $9.383463120 - 0.757576081\,i$
           & $9.391039698 - 0.768474309\,i$
           & $9.396057028 - 0.763271523\,i$ \\

        9  & $10.254191950 - 0.767353461\,i$
           & $10.341059824 - 0.781922675\,i$
           & $10.209408758 - 0.759647953\,i$
           & $10.216197241 - 0.767175596\,i$
           & $10.219276180 - 0.763842284\,i$ \\

        10 & $11.074527306 - 0.767353461\,i$
           & $11.154867208 - 0.779900440\,i$
           & $11.034218023 - 0.761149937\,i$
           & $11.040153399 - 0.766496175\,i$
           & $11.042090650 - 0.764302287\,i$ \\

        \bottomrule
    \end{tabular}%
    }
\end{table}

\newpage
\printbibliography
\end{document}